\documentstyle[prd,epsf,aps]{revtex}
\newcommand{\vphi}{\varphi}
\newcommand{\pih}{\frac{\pi}{2}}

\newcommand{\xh}{x_{\rm H}}
\newcommand{\tx}{\tilde{x}}
\newcommand{\txh}{\tilde{x}_{\rm H}}
\newcommand{\DS}{\displaystyle}

\newcounter{fixy}
\begin{document}
\newenvironment{fixy}[1]{\setcounter{figure}{#1}}
{\addtocounter{fixy}{1}}
\renewcommand{\thefixy}{\arabic{fixy}}
\renewcommand{\thefigure}{\thefixy\alph{figure}}
\setcounter{fixy}{1}

\title{AXIALLY SYMMETRIC MONOPOLES AND BLACK HOLES 
IN EINSTEIN-YANG-MILLS-HIGGS THEORY}

\vspace{1.5truecm}
\author{
{\bf Betti Hartmann}\\
Fachbereich Physik, Universit\"at Oldenburg, Postfach 2503\\
D-26111 Oldenburg, Germany\\
{\bf Burkhard Kleihaus}\\
Department of Mathematical Physics, University College, Dublin,\\
Belfield, Dublin 4, Ireland\\
{\bf Jutta Kunz}\\
Fachbereich Physik, Universit\"at Oldenburg, Postfach 2503\\
D-26111 Oldenburg, Germany}

\vspace{1.5truecm}

\date{\today}

\maketitle
\vspace{1.0truecm}

\begin{abstract}
We investigate static axially symmetric monopole 
and black hole solutions with magnetic charge $n \ge 1$
in Einstein-Yang-Mills-Higgs theory.

For vanishing and small Higgs selfcoupling, 
multimonopole solutions are gravitationally bound.
Their mass {\it per unit charge} is lower than the mass of the
$n=1$ monopole. For large Higgs selfcoupling only a repulsive phase exists.

The static axially symmetric hairy black hole solutions 
possess a deformed horizon with constant surface gravity.
We consider their properties in the isolated horizon framework,
interpreting them as bound states of monopoles and black holes.
Representing counterexamples to the ``no-hair'' conjecture,
these black holes are neither uniquely characterized 
by their horizon area and horizon charge.
\end{abstract}
\vfill\eject

\section{Introduction}

Magnetic monopoles \cite{mono}, multimonopoles \cite{multi,rr,kkt,mon}
and monopole-antimonopole pair solutions \cite{taubes,map}
are globally regular solutions of
SU(2) Yang-Mills-Higgs (YMH) theory, with Higgs field in the
triplet representation.
Since their magnetic charge is proportional to their topological charge, 
the monopoles and multimonopoles reside in
topologically non-trivial sectors of the theory, whereas the 
monopole-antimonopole pair solutions are topologically trivial.

In the  Bogomol'nyi-Prasad-Sommerfield (BPS) limit \cite{B,PS}, 
where the strength of the Higgs self-interaction potential vanishes,
the mass of the monopole and multimonopole solutions saturates
its lower bound, the Bogomol'nyi bound. In particular,
the mass {\it per unit charge} of an $n>1$ monopole is precisely
equal to that of the $n=1$ monopole.
The massless Higgs field mediates a long range attractive force 
which exactly cancels the long range repulsive
magnetic force of the $U(1)$ field
\cite{M,N}.

For finite Higgs selfcoupling, however,
the Higgs field is massive and therefore decays exponentially.
Consequently the long range magnetic field dominates at large distances,
leading to the repulsion of like monopoles \cite{Gold}.
In particular, 
as verified numerically for $n=2$ and $n=3$ monopoles \cite{kkt},
the mass {\it per unit charge} 
of an $n>1$ monopole is higher than the mass of the $n=1$ monopole.
Thus for finite Higgs selfcoupling
there is only a repulsive phase between like monopoles.

Let us now consider the effect of gravity on the monopole and
multimonopole solutions.
When gravity is coupled to YMH theory,
a branch of gravitating monopole solutions
emerges from the flat space monopole solution
\cite{gmono}.
With increasing gravitational strength, 
the mass of the gravitating monopole solutions decreases monotonically.
The branch of monopole solutions extends up to some maximal value of the
gravitational strength, beyond which
the size of the soliton core would be smaller than
the Schwarzschild radius of the solution \cite{gmono}.
The same holds true for static axially symmetric gravitating
multimonopole solutions \cite{hkk}.
All these solutions are asymptotically flat.

The inclusion of gravity
allows for an attractive phase of like monopoles 
not present in flat space \cite{hkk}.
There arises a region of parameter space,
where the mass {\it per unit charge} 
of the gravitating multimonopole solutions is lower than 
the mass of the gravitating $n=1$ monopole.
Here the multimonopole solutions are gravitationally bound.

To every regular monopole solution
there exists a corresponding family of black hole solutions
with regular event horizon
and horizon radius  $0 < x_{\Delta} \le x_{\Delta, \rm max}$ 
\cite{gmono}.
Likewise, to every regular axially symmetric multimonopole solution
there exists a corresponding family of black hole solutions
with regular event horizon \cite{hkk}.
Outside their event horizon
these black hole solutions possess non-trivial non-abelian fields.
Therefore they represent counterexamples to the
``no-hair'' conjecture.
The axially symmetric black hole solutions additionally show,
that static black hole solutions need not be spherically symmetric,
i.~e.~Israel's theorem cannot be generalized to non-abelian theories
either \cite{ewein,kk2,kk4}.

Considering the non-abelian black hole solutions 
in the isolated horizon framework \cite{iso,iso1,iso2}, they can be
interpreted as bound states of monopoles and black holes \cite{iso2}.
In particular,
the isolated horizon framework yields an intriguing relation
for the mass of hairy black hole solutions,
representing it as the sum of the monopole mass and the horizon mass
of the black hole solutions \cite{iso1}.
Having shown previously, that
this relation is also valid for black holes inbetween
monopole-antimonopole pair solutions \cite{map3},
we here verify this relation for the magnetically charged
hairy black hole solutions.
The isolated horizon formalism has furthermore led to new conjectures
for black holes. In particular, a ``quasilocal uniqueness conjecture''
has been proposed, stating
that static black holes are uniquely characterized by 
their horizon area and horizon charge(s) \cite{iso1}. 
We investigate the validity of this conjecture for the EYMH 
black hole solutions.

This paper presents a detailed account of the 
static axially symmetric multimonopole and black hole solutions,
reported in \cite{hkk}.
In section II we present the action, the axially
symmetric ansatz in isotropic spherical coordinates
and the boundary conditions. 
In section III we recall the spherically symmetric solutions,
presenting them in isotropic coordinates.
In section IV we discuss
the properties of the axially symmetric regular multimonopole solutions,
and in section V those of the black hole solutions.
We present our conclusions in section V.
In Appendix A some details of the quantities $F_{\mu\nu}$, $D_\mu \Phi$ and 
$T_{\mu\nu}$ are shown,
and in Appendix B the numerical technique is briefly described.

\section{\bf Einstein-Yang-Mills-Higgs Equations of Motion}

\subsection{\bf Einstein-Yang-Mills-Higgs action}

We consider the SU(2) Einstein-Yang-Mills-Higgs action
\begin{equation}
S=\int \left ( \frac{R}{16\pi G} 
- \frac{1}{2} {\rm Tr} (F_{\mu\nu} F^{\mu\nu})
-\frac{1}{4}{\rm Tr}(D_\mu \Phi D^\mu \Phi)
-\frac{1}{8}\lambda{\rm Tr}(\Phi^2 - \eta^2)^2
  \right ) \sqrt{-g} d^4x , 
\ \label{action} \end{equation}
with field strength tensor
\begin{equation}
F_{\mu \nu} = 
\partial_\mu A_\nu -\partial_\nu A_\mu + i e \left[A_\mu , A_\nu \right] 
\ , \label{fmn} \end{equation}
of the gauge field
\begin{equation}
A_{\mu} = \frac{1}{2} \tau^a A_\mu^a
\ , \label{amu} \end{equation}
and with covariant derivative
\begin{eqnarray}
D_\mu \Phi & = & \partial_\mu \Phi + i e \left[ A_\mu, \Phi \right] 
\label{DPdef} \ , \end{eqnarray}
of the Higgs field in the adjoint representation
\begin{eqnarray}
\Phi = \tau^a \phi^a
\ . \label{Higgs} \end{eqnarray}
Here $g$ denotes the determinant of the metric.
The constants in the action represent
Newton's constant $G$,
the Yang-Mills coupling constant $e$,
the Higgs self-coupling constant $\lambda$
and the vacuum expectation value of the Higgs field $\eta$.

Variation of the action (\ref{action}) with respect to the metric
$g^{\mu\nu}$ leads to the Einstein equations
\begin{equation}
G_{\mu\nu}= R_{\mu\nu}-\frac{1}{2}g_{\mu\nu}R = 8\pi G T_{\mu\nu}
\  \label{ee} \end{equation}
with stress-energy tensor
\begin{eqnarray}
T_{\mu\nu} &=& g_{\mu\nu}L_M -2 \frac{\partial L_M}{\partial g^{\mu\nu}} 
 \nonumber \\
  &=& {\rm Tr}(\frac{1}{2}D_\mu \Phi D_\nu \Phi 
    -\frac{1}{4} g_{\mu\nu} D_\alpha \Phi D^\alpha \Phi)
    + 2{\rm Tr} 
    ( F_{\mu\alpha} F_{\nu\beta} g^{\alpha\beta}
   -\frac{1}{4} g_{\mu\nu} F_{\alpha\beta} F^{\alpha\beta})
   -\frac{1}{8}g_{\mu\nu}\lambda{\rm Tr}(\Phi^2 - \eta^2)^2 
\ . 
\end{eqnarray}
Variation with respect to the gauge field $A_\mu$ 
and the Higgs field $\Phi$ 
leads to the matter field equations,
\begin{eqnarray}
& &\frac{1}{\sqrt{-g}} D_\mu(\sqrt{-g} F^{\mu\nu})
   -\frac{1}{4} i e [\Phi, D^\nu \Phi ] = 0 \ ,
\label{feqA} \\
& & \frac{1}{\sqrt{-g}} D_\mu(\sqrt{-g} D^\mu \Phi)
+\lambda (\Phi^2 -\eta^2) \Phi  = 0 \ ,
\label{feqPhi} 
\end{eqnarray}
respectively.
\subsection{\bf Static axially symmetric ansatz}

Instead of the Schwarzschild-like coordinates, used for the
spherically symmetric EYM and EYMH solutions \cite{bm,su2,gmono}
(see section III),
we adopt isotropic coordinates
as in \cite{kk1,kk2,kk3,kk4,map2,map3,hkk},
to construct static axially symmetric solutions.
In terms of the spherical coordinates $r$, $\theta$ and $\vphi$
the isotropic metric reads 
\begin{equation}
ds^2=
  - f dt^2 +  \frac{m}{f} d r^2 + \frac{m r^2}{f} d \theta^2 
           +  \frac{l r^2 \sin^2 \theta}{f} d\vphi^2
\ , \label{metric2} \end{equation}
where the metric functions
$f$, $m$ and $l$ are only functions of 
the coordinates $r$ and $\theta$.
The $z$-axis ($\theta=0$) represents the symmetry axis.
Regularity on this axis requires \cite{book}
\begin{equation}
m|_{\theta=0}=l|_{\theta=0}
\ . \label{lm} \end{equation}

We take a purely magnetic gauge field, $A_0=0$,
and choose for the gauge field the ansatz \cite{rr,kk,kk1,kk2,kk3,kk4,hkk}
\begin{equation}
A_\mu dx^\mu =
\frac{1}{2er} \left[ \tau^n_\vphi 
 \left( H_1 dr + \left(1-H_2\right) r d\theta \right)
 -n \left( \tau^n_r H_3 + \tau^n_\theta \left(1-H_4\right) \right)
  r \sin \theta d\vphi \right]
\ . \label{gf1} \end{equation}
Here the symbols $\tau^n_r$, $\tau^n_\theta$ and $\tau^n_\vphi$
denote the dot products of the cartesian vector
of Pauli matrices, $\vec \tau = ( \tau_x, \tau_y, \tau_z) $,
with the spatial unit vectors
\begin{eqnarray}
\vec e_r^{\, n}      &=& 
(\sin \theta \cos n \vphi, \sin \theta \sin n \vphi, \cos \theta)
\ , \nonumber \\
\vec e_\theta^{\, n} &=& 
(\cos \theta \cos n \vphi, \cos \theta \sin n \vphi,-\sin \theta)
\ , \nonumber \\
\vec e_\vphi^{\, n}   &=& (-\sin n \vphi, \cos n \vphi,0) 
\ , \label{rtp} \end{eqnarray}
respectively.
Since the fields wind $n$ times around, while the
azimuthal angle $\vphi$ covers the full trigonometric circle once,
we refer to the integer $n$ as the winding number of the solutions.
For the Higgs field the corresponding ansatz is \cite{rr,kkt}
\begin{equation}
\Phi= \left(\Phi_1 \tau_r^{n}+\Phi_2 \tau_\theta^{n}\right) \eta
\ . \label{higgs} \end{equation}
The four gauge field functions $H_i$ 
and the two Higgs field functions $\Phi_i$ depend only on 
the coordinates $r$ and $\theta$.
For $n=1$ and $H_1=H_3=\Phi_2=0$, $H_2=H_4=K(r)$ and $\Phi_1=H(r)$,
the spherically symmetric solutions are obtained in isotropic
coordinates.

The ansatz (\ref{gf1})-(\ref{higgs}) is axially symmetric in the sense,
that a rotation around the $z$-axis can be compensated
by a gauge rotation.
The ansatz is form-invariant under the abelian gauge transformation
\cite{rr,kkb,kk,kk1msph}
\begin{equation}
 U= \exp \left({\frac{i}{2} \tau^n_\vphi \Gamma(r,\theta)} \right)
\ .\label{gauge} \end{equation}
The functions $H_1$ and $H_2$ transform inhomogeneously
under this gauge transformation,
\begin{eqnarray}
  H_1 & \rightarrow & H_1 -  r \partial_r \Gamma \ , \nonumber \\
  H_2 & \rightarrow & H_2 +   \partial_\theta \Gamma
\ , \label{gt1} \end{eqnarray}
like a 2-dimensional gauge field.
The functions $H_3$ and $H_4$ combine to form a scalar doublet,
$(H_3+{\rm cot} \theta, H_4)$.
Likewise, the Higgs field functions form a scalar doublet
$(\Phi_1,\Phi_2)$.

We fix the gauge by choosing the gauge condition
as previously \cite{kkb,kk,kk1msph,kk1,kk2,kk3,kk4,hkk}.
In terms of the functions $H_i$ it reads
\begin{equation}
 r \partial_r H_1 - \partial_\theta H_2 = 0 
\ . \label{gc1} \end{equation}
With the ansatz (\ref{metric2})-(\ref{gf1})
and the gauge condition (\ref{gc1}) 
we then obtain the set of EYMH field equations.

The energy density 
of the matter fields $\epsilon =-T_0^0=-L_M$ is given by
\begin{eqnarray}
-T_0^0 &= & 
\frac{\eta^2}{2r^2}f \left[\frac{1}{m}
                            \left( (r\partial_r \Phi_1 +H_1 \Phi_2)^2
                             +(r\partial_r \Phi_2 -H_1 \Phi_1)^2
                             +(\partial_\theta \Phi_1 -H_2 \Phi_2)^2
                             +(\partial_\theta \Phi_2 +H_2 \Phi_1)^2
                             \right) \right.
\nonumber \\
& & 
            \left.    +\frac{n^2}{l}
                              (H_4 \Phi_1 +(H_3+{\rm cot}\theta )\Phi_2)^2
               \right] + 
               \frac{\lambda}{2} \left( \Phi_1^2+\Phi_2^2 - \eta^2\right)^2 
\nonumber \\
& & 
+\frac{f^2}{2 e^2 r^4 m} \left\{
 \frac{1}{m} \left(r \partial_r H_2 + \partial_\theta H_1\right)^2 
 \right.
\nonumber \\
& &  
+  \left.
   \frac{n^2}{l} \left[
  \left(  r \partial_r H_3 - H_1 H_4 \right)^2
+ \left(r \partial_r H_4 + H_1 \left( H_3 + {\rm cot} \theta \right)
    \right)^2 \right. \right.
\nonumber \\
      &  & \left. \left.
 + \left(\partial_\theta H_3 - 1 + {\rm cot} \theta H_3 + H_2 H_4
     \right)^2 +
  \left(\partial_\theta H_4 + {\rm cot} \theta \left( H_4-H_2 \right) 
   - H_2 H_3 \right)^2 \right] \right\}
\ .  \label{edens} \end{eqnarray}
As seen from Eq.~(\ref{edens}),
regularity on the $z$-axis requires 
\begin{equation}
H_2|_{\theta=0}=H_4|_{\theta=0}
\ . \label{h2h4} \end{equation}

\subsection{Boundary conditions}

To obtain asymptotically flat solutions
with the proper symmetries,
which are either globally regular
or possess a regular event horizon,
we must impose appropriate boundary conditions \cite{kk1,kk2,kk3,kk4,hkk}.
Here we are looking for solutions
with parity reflection symmetry.
Therefore we need to consider the solutions only
in the region $0 \le \theta \le \pih$,
imposing boundary conditions along the 
$\rho$- and $z$-axis (i.~e.~for
$\theta=\pih$ and $\theta=0$).

{\sl Boundary conditions at infinity}

Asymptotic flatness imposes
for the metric functions of the regular and black hole solutions 
at infinity ($r=\infty$) the boundary conditions
\begin{equation}
f|_{r=\infty}= m|_{r=\infty}= l|_{r=\infty}=1
\ . \label{bc1a} \end{equation}
For the Higgs field functions we require
\begin{equation}
\Phi_1|_{r=\infty}= 1 \ , \ \ \
\Phi_2|_{r=\infty}=0 
\ , \label{bc1b} \end{equation}
thus the modulus of the Higgs field assumes the vacuum expectation
value $\eta$.
For magnetically charged solutions,
the gauge field functions $H_i$ satisfy
\begin{equation}
H_1|_{r=\infty}=H_2|_{r=\infty}=
H_3|_{r=\infty}=H_4|_{r=\infty}=0
\ . \label{bc1c} \end{equation}

{\sl Boundary conditions at the origin}

Requiring the solutions to be regular at the origin
($r=0$) leads 
to the boundary conditions for the metric functions
\begin{equation}
\partial_r f|_{r=0}= \partial_r m|_{r=0}= \partial_r l|_{r=0}= 0
\ . \label{bc2a} \end{equation}
The Higgs field functions satisfy
\begin{equation}
 \Phi_1|_{r=0} = \Phi_2|_{r=0} = 0 
\ , \label{bc2b} \end{equation}
and the gauge field functions $H_i$ satisfy
\begin{equation}
 H_1|_{r=0}=H_3|_{r=0}=0 \ , \ \ \
 H_2|_{r=0}=H_4|_{r=0}= 1
\ . \label{bc2c} \end{equation}

{\sl Boundary conditions at the horizon}

The event horizon of static black hole solutions
is characterized by $g_{tt}=-f=0$.
In isotropic coordinates $g_{rr}$ is finite at the horizon.
We now impose that the horizon of the black hole solutions
resides at a surface of constant $r$, $r=r_{\rm H}$ \cite{kk2,kk4,hkk}.

Requiring the horizon to be regular, we obtain
the boundary conditions at the horizon $r=r_{\rm H}$.
The metric functions must satisfy
\begin{equation}
f|_{r=r_{\rm H}}=
m|_{r=r_{\rm H}}=
l|_{r=r_{\rm H}}=0 
\ . \label{bh2a} \end{equation}
The boundary conditions for the gauge field functions and the Higgs 
field functions
can be deduced from the equations (\ref{feqA}) and (\ref{feqPhi}), respectively.
\begin{eqnarray}
F_{r\theta}|_{r=r_{\rm H}}= 0 &{\DS \ \ \ \Longleftrightarrow \ \ \ } &
    (r\partial_r H_2 +\partial_\theta H_1)|_{r=r_{\rm H}}= 0  \ , 
\nonumber \\
F_{r\vphi}|_{r=r_{\rm H}}= 0  &{\DS \ \ \ \Longleftrightarrow \ \ \ } &
    (r\partial_r H_3 -H_1 H_4)|_{r=r_{\rm H}}= 0  \ , \ \ 
    (r\partial_r H_4 -H_1 (H_3+{\rm cot}\theta))|_{r=r_{\rm H}}= 0  \ , 
\nonumber \\
D_r\Phi|_{r=r_{\rm H}} = 0 &{\DS \ \ \  \Longleftrightarrow \ \ \ } &
    (r\partial_r \Phi_1 + H_1 \Phi_2)|_{r=r_{\rm H}} = 0  \ , \ \ 
    (r\partial_r \Phi_2 - H_1 \Phi_1)|_{r=r_{\rm H}} = 0 \ . 
\label{bhbc}
\end{eqnarray}         

The equations of motion yield only three boundary conditions
for the four gauge field functions $H_i$;
one gauge field boundary condition is left indeterminate.
However, for the black hole solutions
the gauge condition (\ref{gc1})
still allows non-trivial gauge transformations satisfying
\begin{equation}
r^2 \partial^2_r \Gamma
+r \partial_r \Gamma
+  \partial^2_\theta \Gamma = 0
\ . \label{gfree} \end{equation}
To fix the gauge,
we have chosen the gauge condition \cite{kk2,kk4,hkk}
\begin{equation}
(\partial_\theta H_1) |_{r=r_{\rm H}} = 0 , 
\   \label{gfree2} \end{equation}
which implies ${\DS H_1 |_{r=r_{\rm H}} = 0}$, if we take into account 
the boundary conditions on the axes, ${\DS H_1|_{\theta=0,\pih}=0}$,
(see next paragraph). As a consequence the boundary conditions Eqs.~(\ref{bhbc})
reduce to 
\begin{equation}
\partial_r H_2 |_{r=r_{\rm H}}= 0 , \ \  
\partial_r H_3 |_{r=r_{\rm H}}= 0 , \ \  
\partial_r H_4 |_{r=r_{\rm H}}= 0 , \ \  
\partial_r \Phi_1 |_{r=r_{\rm H}}= 0 , \ \  
\partial_r \Phi_2 |_{r=r_{\rm H}}= 0 . \ \  
\label{bhbc1}
\end{equation}

{\sl Boundary conditions along the axes}

The boundary conditions along the $\rho$- and $z$-axis
($\theta=\pih$ and $\theta=0$) are determined by the 
symmetries.
The metric functions satisfy along the axes
\begin{eqnarray}
& &\partial_\theta f|_{\theta=0} = \partial_\theta m|_{\theta=0} =
   \partial_\theta l|_{\theta=0} =0 \ ,
\nonumber \\
& &\partial_\theta f|_{\theta=\pih} =
   \partial_\theta m|_{\theta=\pih} =
   \partial_\theta l|_{\theta=\pih} =0 \ .
\label{bc4a} 
\end{eqnarray}
Likewise the Higgs field functions satisfy
\begin{eqnarray}
& & \partial_\theta \Phi_1|_{\theta=0} = 0\ , 
    \ \ \ \Phi_2|_{\theta=0} = 0\ , 
\nonumber \\
& & \partial_\theta \Phi_1|_{\theta=\pih} = 0 \ , 
    \ \ \ \Phi_2|_{\theta=\pih} = 0 \ ,
\label{bc4b} 
\end{eqnarray}
along the axes.
For the gauge field functions $H_i$ symmetry considerations
lead to the boundary conditions
\begin{eqnarray}
& & H_1|_{\theta=0}=H_3|_{\theta=0}=0 \ , \ \ \ 
   \partial_\theta H_2|_{\theta=0} = \partial_\theta H_4|_{\theta=0}  = 0 \ ,
\nonumber \\
& & H_1|_{\theta=\pih}=H_3|_{\theta=\pih}=0 \ , \ \ \ 
    \partial_\theta H_2|_{\theta=\pih} = 
    \partial_\theta H_4|_{\theta=\pih} = 0 \ ,
\label{bc4c} 
\end{eqnarray}
along the axes.
In addition, regularity on the $z$-axis requires condition (\ref{lm})
for the metric functions to be satisfied,
and condition (\ref{h2h4}) for the gauge field functions.

\subsection{Mass, temperature and entropy}

Let us introduce the dimensionless coordinate $x$,
\begin{equation}
x= \eta e r
\ . \label{dimx} \end{equation}
The equations then
depend only on two dimensionless  coupling constants, $\alpha$ and $\beta$,
\begin{equation}
\alpha^2 = 4\pi G\eta^2 \ , \ \ \ \beta^2 = \frac{\lambda}{e^{2}}
\ . \end{equation}

The mass $M$ of the solutions can be obtained directly from
the total energy-momentum ``tensor'' $\tau^{\mu\nu}$
of matter and gravitation \cite{wein},
\begin{equation}
M=\int \tau^{00} d^3r
\ . \label{Mass1} \end{equation}
It is related to the dimensionless mass $\mu/\alpha^2$, via
\begin{equation}
\mu/\alpha^2 = \frac{e}{4\pi\eta} M
\ , \label{Mass2} \end{equation}
where $\mu$ is determined by the derivative of the metric function $f$
at infinity \cite{kk1,kk2,kk3,kk4}
\begin{equation}
\mu = \frac{1}{2} \lim_{x \rightarrow \infty} x^2 \partial_x  f
\ . \label{mass} \end{equation}

The surface gravity $\kappa_{\rm sg}$ 
of static black hole solutions is given by
\cite{wald,ewein}
\begin{equation}
\kappa^2_{\rm sg}=-(1/4)g^{tt}g^{ij}(\partial_i g_{tt})(\partial_j g_{tt})
\ . \label{sg} \end{equation}
To evaluate $\kappa_{\rm sg}$, we need to
consider the metric functions at the horizon.
Expanding the equations in the vicinity of the horizon
in powers of the dimensionless coordinate $(x-x_{\rm H})/x_{\rm H}$,
we observe, that the metric functions are quadratic
in $x-x_{\rm H}$,
\begin{equation}
f(x,\theta)=f_2(\theta)\left(\frac{x-x_{\rm H}}{x_{\rm H}}\right)^2 
 \left( 1 -   \frac{x-x_{\rm H}}{x_{\rm H}} \right)
 + O\left(\frac{x-x_{\rm H}}{x_{\rm H}}\right)^4
\ , \label{expanf} \end{equation}
\begin{equation}
m(x,\theta)=m_2(\theta)\left(\frac{x-x_{\rm H}}{x_{\rm H}}\right)^2 
 \left( 1 - 3 \frac{x-x_{\rm H}}{x_{\rm H}} \right)
 + O\left(\frac{x-x_{\rm H}}{x_{\rm H}}\right)^4
\ , \label{expanm} \end{equation}
with $l(x,\theta)$ like $m(x,\theta)$, Eq.~(\ref{expanm}).
We then obtain
the dimensionless surface gravity $\kappa = \kappa_{\rm sg} / e \eta$
\begin{equation}
\kappa =\frac{f_2(\theta)}{x_{\rm H} \sqrt{m_2(\theta)} } 
\ . \label{temp} \end{equation}

The zeroth law of black hole physics states, that 
the surface gravity $\kappa_{\rm sg}$ 
is constant at the horizon of a black hole \cite{wald}.
To show that the zeroth law holds for the hairy black hole solutions
we employ the expansion of the metric functions
(\ref{expanf})-(\ref{expanm})
in the $r \theta$ component of the Einstein equations 
at the horizon \cite{kk4}. This yields the crucial
relation between the expansion coefficients
$f_2(\theta)$ and $m_2(\theta)$,
\begin{eqnarray}
0 & = & \frac{\partial_\theta m_2}{m_2} - 2 \frac{\partial_\theta f_2}{f_2}
\label{exp_horz2}
\ . \end{eqnarray}
The temperature is proportional to the surface gravity
$\kappa_{\rm sg}$ \cite{wald}, 
in particular the dimensionless temperature is given by
\begin{equation}
T=\kappa /(2 \pi)
\ . \label{temp1} \end{equation}

The dimensionless area $A$ 
of the event horizon of the black hole solutions
\begin{equation}
A = 2 \pi \int_0^\pi  d\theta \sin \theta
\frac{\sqrt{l_2 m_2}}{f_2} x_{\rm H}^2
\   \label{area} \end{equation}
is proportional to the dimensionless entropy $S$ \cite{wald},
\begin{equation}
S = \frac{A}{4}
\ , \label{area1} \end{equation}
yielding the product
\begin{equation}
TS = \frac{x_{\rm H}}{4} \int_0^\pi  d\theta \sin \theta
{\sqrt{l_2 }}
\ . \label{area2} \end{equation}

Having defined temperature and entropy, we now derive
a second expression for the mass of the black hole solutions \cite{wald}.
As in EYM theory \cite{kk4}, the equations of motion yield
in EYMH theory the relation
\begin{equation}
\frac{1}{8 \pi G}
\partial_\mu \left( \sqrt{-g} \partial^\mu {\rm ln} f \right) =
  - \sqrt{-g} \left( 2 T_0^{\ 0} - T_\mu^{\ \mu} \right)
\ . \label{area3} \end{equation}
Integrating both sides over $r$, $\theta$ and $\vphi$
from the horizon to infinity, we obtain
\begin{equation}
\frac{1}{4 G}
 \int_0^\pi d\theta \sin \theta  \left.
 \left[ r^2 \sqrt{l} \frac{\partial_r f}{f} \right]
 \right|_{r_{\rm H}}^\infty =
 - \int_0^{2 \pi} \int_0^\pi \int_{r_{\rm H}}^\infty
  d \vphi d \theta d r
  \sqrt{-g} \left( 2 T_0^{\ 0} - T_\mu^{\ \mu} \right)
 = M_{\rm o}
\ . \label{area4} \end{equation}
Changing to dimensionless coordinates, we express the l.h.s.~with help
of the dimensionless mass $\mu$ and the product 
of temperature and entropy, $TS$, obtaining
\begin{equation}
\mu  =\mu_{\rm o} + 2 TS
\ , \label{mass2} \end{equation}
with $\mu_{\rm o}/\alpha^2=  \frac{e}{4\pi\eta} M_{\rm o}$,
in agreement with the general mass formula
for static black hole solutions \cite{wald}.
For regular solutions one simply obtains
\begin{equation}
 M = - \int \left( 2 T_0^{\ 0} - T_\mu^{\ \mu} \right)
   \sqrt{-g} dr d\theta d\vphi 
\ . \label{Mass3} \end{equation}

We finally consider the Kretschmann scalar $K$,
\begin{equation}
 K= R^{\mu\nu\alpha\beta}R_{\mu\nu\alpha\beta}
\ . \label{ks} \end{equation}
In order to show that $K$ is finite at the horizon it is suffient to 
point out, that the expansion of the metric functions 
(\ref{expanf})-(\ref{expanm}) is the same as for the EYM system 
considered in \cite{kk4}. Consequently, it follows from the 
calculations of ref.~\cite{kk4} that the Kretschmann scalar is finite at
the event horizon, if the condition (\ref{exp_horz2}) holds, 
i.~e.~if the temperature is constant at the event horizon.

\subsection{Horizon mass and horizon charge}

In the isolated horizon framework \cite{iso} 
an intriguing relation between the ADM mass $\mu/\alpha^2$
of a black hole with area $A$ and area parameter $x_\Delta$,
\begin{equation}
x_\Delta = \sqrt{A/4\pi} \ ,
\label{IHx} \end{equation}
and the ADM mass $\mu_{\rm reg}/\alpha^2$
of the corresponding regular solution holds \cite{iso1},
\begin{equation}
\mu = \mu_{\rm reg} + \mu_\Delta \ , 
\label{IHmu} \end{equation}
where the horizon mass $\mu_\Delta/\alpha^2$ is defined by
\begin{equation}
\mu_\Delta = \int_0^{x_\Delta} \kappa(x'_\Delta)x'_\Delta d x'_\Delta \ ,
\label{IHmuD}
\end{equation}
with the dimensionless surface gravity $\kappa$ in the integrand.

The isolated horizon formalism then suggests
to interpret a hairy black hole as a bound
state of a regular solution and a Schwarzschild black hole \cite{iso2},
\begin{equation}
\mu = \mu_{\rm reg} + \mu_{\rm S} + \mu_{\rm bind} \ , 
\label{IHbs} \end{equation}
where $\mu_{\rm S}/\alpha^2 = x_\Delta/2\alpha^2$ is the ADM mass 
of the Schwarzschild black hole with horizon radius $x_\Delta$,
and $\mu_{\rm bind}/\alpha^2$ represents the binding energy of the system,
\begin{equation}
\mu_{\rm bind}= \mu_\Delta-\mu_{\rm S} \ . 
\label{IHbind} \end{equation}

Another crucial quantity in the isolated horizon formalism is the
non-abelian magnetic charge of the horizon \cite{iso,iso1}, defined via
the surface integral over the horizon
\begin{equation}
P^{\rm YM}_\Delta = \frac{1}{4\pi} 
\oint \sqrt{\sum_i{\left(F^i_{\theta\vphi}\right)^2}} d\theta d\vphi \ .
\label{Pdelta}
\end{equation}
The non-abelian electric horizon charge is defined analogously
with the dual field strength tensor
\cite{iso,iso1}.
These horizon charges are an important ingredient in a new
``quasilocal uniqueness conjecture''
put forward in ref.~\cite{iso1},
which states that static black holes are uniquely determined
by their horizon area and their horizon electric and magnetic charges.

\section{\bf Spherically Symmetric Solutions}

Previously, spherically symmetric EYMH solutions 
were obtained numerically in Schwarzschild-like coordinates \cite{gmono}.
Since we construct the axially symmetric solutions
in isotropic coordinates, we here discuss the coordinate transformation
between isotropic and Schwarzschild-like coordinates for
spherically symmetric solutions.

We briefly recall the dependence of the monopole solutions on 
the parameters of the theory, $\alpha$ and $\beta$.
In particular, we compare the way the limiting RN solution is
reached in Schwarzschild-like and in isotropic coordinates.

We then turn to the spherically symmetric black hole solutions,
putting particular emphasis on the relations obtained in the
isolated horizon formalism. 
We demonstrate the relation between black hole mass, 
soliton mass and horizon mass, Eq.~(\ref{IHmu}),
we evaluate the binding energy, Eq.~(\ref{IHbs}),
and we discuss the ``quasilocal uniqueness conjecture'' \cite{iso1}.

\subsection{Coordinate transformations}

By requiring $l=m$ and the metric functions $f$ and $m$ to be
only functions of the coordinate $r$,
the axially symmetric isotropic metric (\ref{metric2})
reduces to the spherically symmetric isotropic metric
\begin{equation}
ds^2=
  - f dt^2 +  \frac{m}{f} \left ( d r^2 + r^2 \left( d \theta^2 
           +  \sin^2 \theta d\vphi^2 \right) \right)
\ . \label{metric4} \end{equation}

In Schwarzschild-like coordinates the metric reads \cite{gmono}
\begin{equation}
ds^2=
  - A^2 N dt^2 +  \frac{1}{N} d \tilde r^2 
  + \tilde r^2 \left( d \theta^2 + \sin^2 \theta d\vphi^2 \right)
\ , \label{metric3} \end{equation}
where the metric functions $A$ and $N$
are only functions of the radial coordinate $\tilde r$, and
\begin{equation}
N(\tilde r) = 1 - \frac{2 \tilde M(\tilde r)}{\tilde r}
\ . \label{N} \end{equation}
The spherically symmetric horizon resides at
radial coordinate $\tilde r_{\rm H}$,
and the (dimensionfull) area of the horizon is
\begin{equation}
A = 4 \pi \tilde r_{\rm H}^2 \ , 
\nonumber 
\end{equation}
hence the area parameter $r_\Delta$ is just the Schwarzschild radius  
$\tilde r_{\rm H}$.

Comparison of the metric in Schwarzschild-like coordinates
(\ref{metric3}) with the isotropic metric (\ref{metric4}) yields
for the coordinate transformation
\begin{equation}
\frac{dr}{r} = \frac{1}{\sqrt{N(\tilde r)}} \frac{d \tilde r}{\tilde r}
\ . \label{co2} \end{equation}
The function $N(\tilde r)$ (or equivalently the mass function
$\tilde M(\tilde r)$) is only known numerically.
Therefore the coordinate function $r(\tilde r)$ can only be
obtained numerically from Eq.~(\ref{co2}).

\subsection{Monopole solutions}

Let us briefly recall the dependence of 
the gravitating magnetic monopole solutions
on the parameters $\alpha$ and $\beta$.
When $\alpha$ is increased from zero, while $\beta$ is kept fixed,
a gravitating monopole branch emerges smoothly from the corresponding
flat space monopole solution.
This fundamental monopole branch extends up to
a maximal value $\alpha_{\rm max}$ of the coupling constant $\alpha$.
Beyond this value no monopole solutions exist.
In the BPS limit, $\beta=0$,
and for small values of $\beta$, the fundamental monopole branch 
bends backwards at $\alpha_{\rm max}$,
until a critical coupling constant $\alpha_{\rm cr}$ is reached.
At the critical value $\alpha_{\rm cr}$
the fundamental monopole branch reaches a limiting solution and
bifurcates with the branch of extremal RN solutions of
unit magnetic charge \cite{gmono}.
For larger values of $\beta$ the maximal value of $\alpha$ and the
critical value of $\alpha$ coincide,
$\alpha_{\rm max} = \alpha_{\rm cr}$ \cite{gmono,lw,foot1}.

Let us first recall, how the critical value of $\alpha$ is approached
in Schwarzschild-like coordinates. 
Along the fundamental branch, the metric function $N(\tilde x)$ 
of the monopole solutions develops a minimum, 
which decreases monotonically.
In the limit $\alpha \rightarrow \alpha_{\rm cr}$, 
the minimum approaches zero at 
$ \tilde x_{\rm cr} = \alpha_{\rm cr} $.
The limiting metric function then consists of an inner part, 
$\tilde x \le \tilde x_{\rm cr}$,
and an outer part, $\tilde x \ge \tilde x_{\rm cr}$.
For $\tilde x \ge \tilde x_{\rm cr}$,
the limiting metric function corresponds to the metric function $N(\tilde x)$
of the extremal RN black hole for $\alpha_{\rm cr}$ with unit magnetic charge.

Likewise the gauge and Higgs field functions approach limiting functions,
when $\alpha \rightarrow \alpha_{\rm cr}$.
For $\tilde x \ge \tilde x_{\rm cr}$ they also correspond to those
of the extremal RN black hole for $\alpha_{\rm cr}$ with unit magnetic charge.
The limit $\alpha \rightarrow \alpha_{\rm cr}$ is demonstrated in Figs.~1a-c
for the monopole solutions in the BPS limit
for the metric function $N(\tilde x)$,
the gauge field function $K(\tilde x)$
and the Higgs field function $H(\tilde x)$.

We recall, that a RN solution with magnetic charge $P$ has metric
functions
\begin{equation}
N(\tilde x) = 1 - \frac{2 \mu}{\tilde x} 
+ \frac{\alpha^2 P^2}{\tilde x^2} \ , \ \ \ 
A(\tilde x)=1 
\ . \end{equation}
In the embedded RN solution with unit magnetic charge 
the gauge field function $K(\tilde x)$ and the Higgs
field function $H(\tilde x)$ are constant,
\begin{equation}
K(\tilde x)=0 \ , \ \ \
H(\tilde x)=1 \ ,
\   \end{equation}
implying a Coulomb-type decay for the magnetic field
and a trivial Higgs field, assuming its vacuum expectation value.
In particular, an extremal RN solution has metric function $N(\tilde x)$,
\begin{equation}
N(\tilde x) = \left( 1 - \frac{\alpha P}{\tilde x} \right)^2
\ , \end{equation}
and ADM mass $\mu/\alpha^2$, where
\begin{equation}
\mu = \alpha |P|
\ . \end{equation}

Let us now turn to the monopole solutions in isotropic coordinates.
In Figs.~2a-c we demonstrate the dependence of 
the gravitating magnetic monopole solutions on the parameter $\alpha$ 
along the fundamental branch in isotropic coordinates.
For comparison the same set of parameter values as in Figs.~1a-c
is chosen.
In the limit $\alpha \rightarrow \alpha_{\rm cr}$ 
the fundamental monopole branch bifurcates with the branch of 
extremal RN solutions of unit magnetic charge.
In particular, 
as the critical $\alpha$ is approached, the metric function $f(x)$
develops a zero at the origin, which corresponds to the horizon
of an extremal RN solution in isotropic coordinates.

For a RN solution with charge $P$ and horizon radius $\txh$
the isotropic coordinate $x$ is related to the Schwarzschild-like coordinate 
$\tilde x$ by 
\begin{equation}
x = \frac{\sqrt{\tx^2 -2 \mu \tx + \alpha^2 P^2}+\tx -\mu}{2} 
\end{equation}
in the non-extremal case
${\DS \alpha |P| < \mu = \frac{1}{2\txh} (\txh^2+\alpha^2 P^2)}$, 
and by
\begin{equation}
x = \tx - \alpha|P| 
\end{equation}
in the extremal case $\alpha |P| = \mu = \txh$.

The metric functions $f(x)$ and $m(x)$ of a non-extremal
RN black hole are given by
\begin{eqnarray}
f(x) & = & \frac{(1-\frac{x}{\xh})^2(1+\frac{x}{\xh})^2}
      {\left(1+\frac{2x}{\xh}\sqrt{1+(\frac{\alpha P}{2\xh})^2}
                +(\frac{x}{\xh})^2\right)^2} \ , 
\nonumber \\
m(x) & = & \left(\frac{\xh}{x}\right)^4
                  (1-\frac{x}{\xh})^2(1+\frac{x}{\xh})^2 \ ,
\label{rn2} 
\end{eqnarray}
respectively.                 
The horizon radius $\xh$ in isotropic coordinates is related to the 
horizon radius in Schwarzschild-like coordinates $x_\Delta$ by 
\begin{equation}
\xh = \frac{x_\Delta^2-\alpha^2 P^2}{ 4 x_\Delta^2}\ . 
\nonumber
\end{equation}
Its ADM mass $\mu/\alpha^2$ is obtained from
\begin{equation}
\mu = 2 \xh \sqrt{1+\left(\frac{\alpha P}{2\xh}\right)^2} \ . 
\nonumber
\end{equation}
An extremal RN solution has horizon radius $\xh =0$ and metric
functions
\begin{equation}
f(x) = \left(\frac{x}{x+\alpha |P|}\right)^2 \ , \ \ \ 
m(x) = 1  \ .
\label{rn2x}
\end{equation}

\subsection{Black hole solutions}

Let us now turn to the black hole solutions of the SU(2) EYMH system.
We here limit our discussion to the BPS case, $\beta=0$.

Hairy black hole solutions exist in a limited domain
of the $x_\Delta-\alpha$-plane \cite{gmono}.
For fixed $\alpha < \alpha_{\rm max}$,
hairy black hole solutions emerge from the monopole solution
in the limit $x_\Delta \rightarrow 0$.
They persist up to a maximal value of the horizon radius
$x_{\Delta,\rm max}$, which limits the domain of existence
of hairy black holes.
The domain of existence is shown in Fig.~3,
where the maximal value of the horizon radius
$x_{\Delta,\rm max}$ is shown as a function of $\alpha$.

The domain of existence of hairy black hole solutions
consists of two regions with distinct critical behaviour.
These two regions are separated by the particular value of $\alpha$,
$\hat \alpha = \sqrt{3}/2$ \cite{gmono}.
For $\hat \alpha < \alpha < \alpha_{\rm max}$,
the hairy black hole solutions
bifurcate at a critical value $x_{\Delta, \rm cr}$ 
with an extremal RN solution with unit charge.
In contrast, for $0 < \alpha < \hat \alpha$, the hairy black hole solutions
bifurcate at a critical value $x_{\Delta, \rm cr}$ with a non-extremal
RN solution with unit charge.
In particular,
for small values of $\alpha$, the branch of hairy black hole solutions
extends backwards from the maximal value $x_{\Delta,\rm max}$ to
the critical value $x_{\Delta, \rm cr}$,
whereas for larger values of $\alpha < \hat \alpha$, both values
coincide, $x_{\Delta,\rm max} = x_{\Delta, \rm cr}$.

The dependence of the mass $\mu/\alpha^2$ of the hairy black hole solutions
on the area parameter is demonstrated in Fig.~4 for $\alpha=0.5$ and 1. 
For $\alpha=0.5 < \hat \alpha$,
the mass of the black hole solutions increases monotonically 
with increasing $x_\Delta$, until it
reaches its maximal value at $x_{\Delta,\rm max}$. 
Bending backwards the mass then decreases again, 
until the bifurcation point at $x_{\Delta, \rm cr}$ is reached,
where it coincides with the mass of a non-extremal RN black hole
with unit charge.
Note, that the mass of the hairy black hole solutions exceeds the mass
of the RN solutions in a small region close to $x_{\Delta,\rm max}$.

In contrast for $\alpha=1 > \hat \alpha$, 
the limiting solution reached corresponds 
to an extremal RN solution with unit charge
(for $\tilde x \ge x_{\Delta,\rm cr}$).
Since for $\alpha > \hat \alpha$
the maximal horizon radius $x_{\Delta,\rm max}$ of the hairy black holes
is smaller than the horizon radius
of the corresponding extremal RN solution, a gap between the branch of
hairy black holes and the RN branch arises.
This gap is seen in Fig.~4 for $\alpha=1$ for the mass of the black holes.

In Fig.~5 we exhibit the surface gravity $\kappa$ as function of the
area parameter $x_\Delta$ for the same set of black hole solutions, 
obtained for $\alpha=0.5$ and 1.
The surface gravity of the
hairy black hole solutions decreases monotonically along the
black hole branches. In contrast, the surface gravity of the
corresponding RN branches increases for small horizon radius $x_\Delta$.
For $x_\Delta \rightarrow x_{\Delta,\rm cr}$,
the surface gravity of the hairy black hole solutions 
reaches the surface gravity of the corresponding limiting RN solutions.
In particular, for $\alpha=0.5$ the hairy black hole branch and the RN branch
bifurcate, and the limiting value reached
corresponds to the value of a non-extremal RN solution.
In contrast for $\alpha=1$ 
the surface gravity of the hairy black hole solutions 
reaches the value of an extremal RN solution,
namely zero, even though the hairy
black hole branch is separated from the RN branch by a gap
of the horizon radius.

Let us now turn to the predictions of the isolated horizon formalism,
which have not been considered before for EYMH black hole solutions.
We first consider relation (\ref{IHmu}) between
black hole mass, soliton mass and horizon mass
for the fundamental monopole and black hole solutions.
In Fig.~6 we exhibit the horizon mass $\mu_{\Delta}/\alpha^2$ obtained from
(\ref{IHmuD}). Adding the mass of the corresponding soliton
solution ($\mu_{\rm reg}/\alpha^2(\alpha=0.5)=0.963461$ and 
$\mu_{\rm reg}/\alpha^2(\alpha=1.0)=0.855254$, respectively,) 
precisely gives the masses of the
non-abelian black hole solutions shown in Fig.~4.

Interpreting the hairy black holes as a bound states
of regular solutions and Schwarzschild black holes \cite{iso2},
we can identify the binding energy of these systems
$\mu_{\rm bind}/\alpha^2$, according to Eq.~(\ref{IHbind}).
In Fig.~7 we present the binding energy
of the hairy black hole solutions for $\alpha=0.5$ and $\alpha=1$.
For comparison, we also show the binding energy 
$\mu_{\rm bind}^{(\rm RN)}/\alpha^2$ of the RN solutions,
which we define analogously to Eq.~(\ref{IHbs}) via
\begin{equation}
\mu_{\rm bind}^{\rm RN} = \mu_{\rm RN} - \mu_{\rm reg} - \mu_{\rm S}
\ , \end{equation}
where $\mu_{\rm RN}/\alpha^2$ is the mass of the RN solution with unit charge,
$\mu_{\rm reg}/\alpha^2$ is the mass of the monopole,
and $\mu_{\rm S}/\alpha^2$ is the mass of the Schwarzschild black hole 
with the same horizon area as the RN black hole.
Note that, when the solutions coexist,
the binding energy of the hairy black hole solution
is smaller than the binding energy of the RN solution
(except for a small region close to $x_{\Delta,\rm max}$ (see also Fig.~4)).
This indicates stability of these hairy black hole solutions
(except close to $x_{\Delta,\rm max}$).

Let us finally consider the ``quasilocal uniqueness conjecture''
of ref.\cite{iso1}, which states, that static black holes are uniquely
specified by their horizon area and horizon charges.
The solutions considered here carry no horizon electric charge,
thus they should be uniquely characterized 
by their horizon area and horizon magnetic charge.
In Fig.~8 we exhibit the non-abelian horizon magnetic charge 
for the hairy black hole solutions with $\alpha=0.5$ and 1.
The horizon magnetic charge increases monotonically along the branches of
hairy black hole solutions.
As expected,
for $\alpha=0.5$ the value of the RN solution with unit charge is reached,
when $x_\Delta \rightarrow x_{\Delta,\rm cr}$.
In contrast, for $\alpha=1$ a value smaller than one is reached, 
when $x_\Delta \rightarrow x_{\Delta,\rm cr}$.
In this case
again a gap occurs between the hairy black hole branch and the RN branch.

Concerning the ``quasilocal uniqueness conjecture'' we conclude,
that allowing only for integer values of the magnetic charge
as required for the non-abelian magnetic charge of embedded RN solutions, 
the spherically symmetric black hole solutions are uniquely
characterized by their area parameter and their horizon magnetic charge.

\section{\bf Axially Symmetric Solutions}

We here give a detailed discussion of the properties 
of the regular axially symmetric multimonopole solutions. 
We first investigate the dependence of the monopole solutions on 
the parameter $\alpha$ for fixed $\beta$. In particular, we
demonstrate the convergence of the multimonopole solutions 
with magnetic charge $n$
to limiting extremal RN solutions with magnetic charge $n$,
when $\alpha \rightarrow \alpha_{\rm cr}$.
Focussing on values of $\alpha$ close to
$\alpha_{\rm cr}$, we introduce auxiliary Schwarzschild-like
coordinates to gain better understanding of the limiting solutions
obtained in isotropic coordinates.

We then show, that the inclusion of gravity
allows for an attractive phase of like monopoles 
not present in flat space \cite{hkk}.
There arises a region of parameter space,
where the mass {\it per unit charge} 
of the gravitating multimonopole solutions is lower than 
the mass of the gravitating $n=1$ monopole,
hence the multimonopole solutions are gravitationally bound.

The numerical technique is briefly described in Appendix B.

\subsection{\bf Fundamental multimonopole branch}

Let us first consider the dependence of the gravitating axially symmetric 
multimonopole solutions with magnetic charge $n$
on the parameters $\alpha$ and $\beta$.
Analogously to the monopole solutions,
when $\alpha$ is increased from zero, while $\beta$ is kept fixed,
a branch of gravitating multimonopole solutions with charge $n$
emerges smoothly from the corresponding
flat space multimonopole solution.
The multimonopole branch extends up to
a maximal value $\alpha_{\rm max}(n)$ of the coupling constant $\alpha$,
beyond which no axially symmetric multimonopole solutions 
with charge $n$ exist.
At the maximal value $\alpha_{\rm max}(n)$,
which coincides with the critical value $\alpha_{\rm cr}(n)$ \cite{foot2},
the multimonopole branch reaches a limiting solution 
and bifurcates with the branch of extremal RN solutions 
with magnetic charge $n$ \cite{hkk,foot3}.
The metric functions of this embedded RN solution
are given by Eqs.~(\ref{rn2x}) with $l=m$,
and the gauge field and Higgs field functions are constant, 
$H_i = 0$, $i=1, ..., 4$, $\Phi_1=1$, and $\Phi_2=0$.

We now illustrate the dependence of the multimonopole solutions on $\alpha$,
and in particular the convergence of the non-abelian solutions
to the corresponding RN solution 
in the limit $\alpha \rightarrow \alpha_{\rm cr}$,
for the special case of $n=2$ multimonopole solutions in the BPS limit.
Numerically we find in this case
$\alpha_{\rm cr}\approx \alpha_{\rm max}\approx 1.5$
\cite{hkk,foot2}.
The static axially symmetric solutions depend on two variables, 
the radial coordinate $x$ and the angle $\theta$. 
In the following we present the functions in two-dimensional plots,
exhibiting the $x$-dependence for three fixed angles,
$\theta=0$, $\theta=\pi/4$ and $\theta=\pi/2$.

Let us first discuss the dependence of the metric
functions on the parameter $\alpha$.
In Figs.~9a-c the metric functions $f$, $l$ and $m$ are shown, respectively, 
for $\alpha=1$, $1.2$, $1.4$ and $1.499$. The function $f$ increases 
monotonically with increasing $x$
for all values of $\alpha$. Its value at the origin $f(0)$
decreases with increasing $\alpha$ and tends to zero
as $\alpha$ approaches the critical value $\alpha_{\rm cr}$.
The functions $m$ and $l$ also increase monotonically with increasing $x$. 
As $\alpha$ tends to its critical value, these functions approach the value 
one on an increasing interval. However, at the origin the functions 
$m$ and $l$ assume a value different from one.
Thus, convergence to the limiting RN solution is pointwise.

In Figs.~9d-g we show the gauge fields functions $H_1$-$H_4$, respectively,
for the same values of $\alpha$. 
The function $H_1$ possesses a maximum, whose position decreases
with increasing $\alpha$ and tends to zero, 
when $\alpha$ approaches its critical value. 
In contrast, the height of the maximum depends only weakly on $\alpha$. 
The function $H_2$ decreases monotonically with increasing
$x$ on the $z$-axis ($\theta=0$) and possesses a minimum on the 
$\rho$-axis ($\theta=\pi/2$). The location of the minimum decreases with 
increasing $\alpha$ and tends to zero when $\alpha$ approaches its 
critical value. The function $H_3$ is similar to $H_1$,
and the function $H_4$ to $H_2$, except
that $H_4$ decreases monotonically with increasing $x$ 
for all values of $\theta$.
At the same time the range where the gauge field functions differ 
considerably from zero, decreases with increasing $\alpha$ and vanishes 
as $\alpha$ tends to $\alpha_{\rm cr}$.
However, 
in this limit the gauge field functions are not continuous at the origin. 
Thus, convergence to the gauge field functions of the embedded RN solution,
$H_i=0$, is again pointwise.

The Higgs field functions $\Phi_1$ and $\Phi_2$ are shown in Figs.~9h-i.
$\Phi_1$ increases monotonically with increasing $x$.
$\Phi_2$ is negative and possesses a minimum,
whose position decreases with increasing $\alpha$ and tends to zero, 
as $\alpha$ tends to $\alpha_{\rm cr}$. 
Again the height of the extremum depends only weakly on $\alpha$. 
Since for an embedded RN solution $\Phi_1=1$ and $\Phi_2=0$,
we observe that the Higgs field functions $\Phi_1$ and $\Phi_2$ deviate 
from their respective RN values in a decreasing range, 
as $\alpha$ approaches $\alpha_{\rm cr}$. 
Again, 
in this limit, the Higgs field functions are not continuous at the origin,
thus convergence to the Higgs field function of the embedded RN solution 
is also pointwise. 

\subsection{Coordinate transformation}

Let us focus now on the limit $\alpha \rightarrow \alpha_{\rm cr}$,
where the non-abelian multimonopole solutions approach
the limiting RN solution.
We recall that for $n=1$ in Schwarzschild-like coordinates a degenerate horizon 
forms at $\txh=\alpha_{\rm cr}$, as $\alpha$ tends to $\alpha_{\rm cr}$. 
For $\tx > \txh$ the limiting solution is identical 
to the embedded RN solution,
whereas for $\tx < \txh$ the limiting solution retains its
non-abelian features, and differs from any embedded abelian solution. 
Analogously, for solutions with magnetic charge $n$ one expects that, in 
Schwarzschild-like coordinates, a degenerate horizon should form
at $\txh=n \alpha_{\rm cr}$ as $\alpha$ tends to $\alpha_{\rm cr}$,
and that the limiting solution should retain its non-abelian features
for $\tx < \txh$.

In isotropic coordinates, however, the horizon radius
of an extremal RN black hole is given by $x_{\rm H}=0$. Thus, 
the limiting solution coincides 
in the limit $\alpha \rightarrow \alpha_{\rm cr}$
with the extremal RN solution on the whole interval $0< x < \infty$.
Therefore the question arises, as to
what happens to the region, where the limiting
solution is essentially non-abelian, since 
this region shrinks to zero size in isotropic coordinates
in the limit $\alpha \rightarrow \alpha_{\rm cr}$.

To elucidate this point we
introduce the auxillary Schwarzschild-like coordinate $\hat{x}= x\sqrt{m/f}$,
which coincides with the Schwarzschild-like coordinate $\tilde{x}$
only, if the metric functions are independent of $\theta$. 
As observed above, in the limit $\alpha \rightarrow \alpha_{\rm cr}$,         
the function $f(x)$ tends to zero on an interval of the coordinate $x$, whose 
length tends also to zero. Considered as a function of $\hat{x}$ however, the 
function $f(\hat{x})$ tends to zero on an interval of almost constant length. 
In this sense, the auxillary coordinate                                       
$\hat{x}$ serves as a Schwarzschild-like coordinate.

In Figs.~10a-c we present the $n=2$ solutions in the BPS limit
for $\alpha=1.493$, $1.496$ and $1.499$ on the interval $0\leq \hat{x} \leq 4$,
extending beyond the critical value
$\hat{x}_{\rm H}= 2 \alpha_{\rm cr} \approx 3$.
As $\alpha$ tends to $\alpha_{\rm cr}$, 
the metric function $f$ tends to zero 
on the interval $0\leq \hat{x} \leq 3$, 
and the metric functions $m$ and $l$ tend to nontrivial limiting functions. 
For $\hat{x} > 3$ the metric functions assume 
the form of the extremal RN solution with charge $P=2$ and horizon radius
$\hat{x}_{\rm H}= 2 \alpha_{\rm cr}$ 
for $\alpha \rightarrow \alpha_{\rm cr}$.
The behaviour of the gauge field functions and the Higgs functions is similar. 
For $\hat{x}> 3$ they tend to the functions of the extremal
RN solutions, whereas for $0 \leq \hat{x} \leq 3$ they tend to nontrivial 
limiting functions.

We thus conclude, that there exists a limiting solution
for $\alpha \rightarrow \alpha_{\rm cr}$.
This limiting solution is non-trivial and angle-dependent
on the interior interval $0 \leq \hat{x} \leq 3$.
In Figs.~9b-c this limiting solution has already been reached
in the innermost part of the interval $0 \leq \hat{x} \leq 3$.
For $\hat{x} > 3$, the limiting solution is an extremal RN solution.
Therefore the functions of the limiting solution,
and in particular the metric functions, are spherically symmetric.
Hence, we can identify the coordinate $\hat{x}$
with the Schwarzschild-like coordinate $\tx$ for $\hat{x}>3$, and we conclude
that in Schwarzschild-like coordinates the horizon radius is
$\txh = \alpha_{\rm cr} P$, in accordance with our expectation.

\subsection{Gravitationally bound monopoles}

Let us now consider the mass {\it per unit charge}
of the (multi)monopole solutions, to show, 
that gravitationally bound monopoles exist.
The mass {\it per unit charge}
of the (multi)monopole solutions decreases with increasing $\alpha$
and merges with the mass of the RN solution at $\alpha_{\rm cr}(n)$.
In the BPS limit, for $\alpha=0$ 
the mass {\it per unit charge} 
of the (multi)monopole solutions is precisely
equal to the mass of the $n=1$ monopole.
For $\alpha > 0$, however, we observe that
the mass {\it per unit charge} of the multimonopoles
is smaller than the mass of the $n=1$ monopole.
In particular, the mass {\it per unit charge} decreases
with increasing $n$.
Thus in the BPS limit, there is an attractive phase between like monopoles,
not present in flat space.
Moreover, 
multimonopoles exist for values of the gravitational coupling strength, 
too large for $n=1$ monopoles to exist, since $\alpha_{\rm cr}(n)$
increases with $n$.
The mass of the $n=1$ monopole and the mass {\it per unit charge} 
of $n=2$ and $n=3$ multimonopoles in the BPS limit are shown in Fig.~11.

For finite Higgs selfcoupling, the flat space multimonopoles
have higher mass {\it per unit charge} than
the $n=1$ monopole, allowing only for a repulsive phase
between like monopoles.
By continuity, this repulsive phase persists in the presence of gravity
for small values of $\alpha$, but it
can give way to an attractive phase for larger values of $\alpha$.
Thus the repulsion between like monopoles
can be overcome for small Higgs selfcoupling
by sufficiently strong gravitational attraction.

To mark the region in parameter space,
where an attractive phase exists, 
we introduce the equilibrium value $\alpha_{\rm eq}(n_1=n_2)$,
where the mass {\it per unit charge} of the charge $n_1$ solution
and the mass {\it per unit charge} of the charge $n_2$ solution
equal one another.
Since $\alpha_{\rm max} \approx \alpha_{\rm cr}$ decreases
with increasing Higgs selfcoupling,
and $\alpha_{\rm eq}$ increases
with increasing Higgs selfcoupling,
the region of parameter space, where an attractive phase exists,
decreases with increasing Higgs selfcoupling.
In particular, for large Higgs selfcoupling, only a repulsive phase is left.

We show the equilibrium values $\alpha_{\rm eq}$ in Fig.~12.
Besides the values $\alpha_{\rm eq}(1=2)$ and $\alpha_{\rm eq}(1=3)$, for which
the monopole mass and the mass {\it per unit charge} of the $n=2$ and 
$n=3$ multimonopole equal one another, respectively, 
Fig.~12 also shows the values of $\alpha_{\rm eq}(2=3)$, where 
the mass {\it per unit charge} of the $n=2$ and 
the mass {\it per unit charge} of the $n=3$ multimonopole 
equal one another.
Thus Fig.~12 exhibits the small domain of the $\alpha$-$\beta$-plane
where an attractive phase for like monopoles exists.

While $n=1$ monopole solutions are stable,
stability of the static axially symmetric multimonopole solutions is not
obvious.
We conjecture, that the $n=2$ multimonopole solutions are stable,
as long as their mass {\it per unit charge} is lower than
the mass of the $n=1$ monopole.
For topological number $n \ge 3$, however, solutions with only
discrete symmetry exist in flat space \cite{mon}, which,
by continuity, should also be present in curved space
(at least for small gravitational strength).
For a given topological number $n>2$,
such multimonopole solutions without rotational symmetry
may possess a lower mass than the corresponding axially symmetric solutions.
The axially symmetric solutions may therefore not represent
global minima in their respective topological sectors,
even if their mass {\it per unit charge} is lower than
the mass of the $n=1$ monopole.

\section{Black hole solutions}

Here we present 
the static axially symmetric hairy black hole solutions of EYMH theory.
We describe their properties, starting with the structure of the
energy density of the matter fields 
and the deformation of the regular horizon.
We then discuss the domain of existence of the hairy black hole solutions
and describe the convergence of these solutions to RN solutions
in the limit $x_\Delta \rightarrow x_{\Delta, \rm cr}$.

The hairy black hole solutions are then considered with respect to the
results of the isolated horizon formalism. In particular,
the mass formula is verified, and the bound state interpretation
is investigated. Finally the ``quasilocal uniqueness conjecture''
is addressed.

For the static axially symmetric hairy black hole solutions of EYMH theory
we employ the same numerical technique
as for the globally regular multimonopole solutions (see Appendix B).
The black hole solutions depend on 
the horizon radius $x_{\rm H}$
and on the coupling constants $\alpha$ and $\beta$.
Here we consider only the BPS limit, $\beta=0$.

\subsection{Energy density and horizon}

Let us begin the discussion of the static axially symmetric black hole solutions
by considering the energy density of the matter fields, $\epsilon$. 
The energy density has a pronounced angle-dependence with a maximum 
on the $\rho$-axis.
In particular, the energy density is not constant at the horizon.
Let us now consider two representative examples for the
energy density of the matter fields $\epsilon$. 

In Figs.~13a-d we exhibit the energy density of the matter fields
of the $n=2$ black hole solution with area parameter $x_\Delta=1$ 
for $\alpha=1$.
Fig.~13a shows a 3-dimensional plot of the energy density
as a function of the coordinates $\rho = x \sin \theta$ and
$z= x \cos \theta$ together with a contour plot,
and Figs.~13b-d show surfaces of constant energy density.
For small values of $\epsilon$,
the surfaces of constant energy density appear ellipsoidal,
being flatter at the poles than in the equatorial plane.
With increasing values of $\epsilon$ a toruslike shape appears.

For smaller values of the parameters $x_\Delta$ and $\alpha$, 
the energy density of the matter
fields exhibits a more complicated structure, as seen in Figs.~14a-e, 
where we show the energy density of the matter fields
of the $n=3$ black hole solution, with area parameter $x_\Delta=0.5$ 
for $\alpha=0.5$.
Fig.~14a again shows a 3-dimensional plot of the energy density
as a function of the coordinates $\rho = x \sin \theta$ and
$z= x \cos \theta$ together with a contour plot,
while Figs.~14b-e show surfaces of constant energy density.
Whereas the surfaces of constant energy density still appear ellipsoidal
for small values of $\epsilon$,
here with increasing values of $\epsilon$ the toruslike shape appears
together with two additional ellipsoids covering the poles.
For the largest values of the energy density only the toruslike shape remains.

The $n$-dependence of the energy density of the matter fields is illustrated
in Fig.~15, where we show the energy density of the black hole solutions
with magnetic charges $n=1-3$ 
and area parameter $x_\Delta=0.5$ for $\alpha=0.5$.
With increasing magnetic charge $n$
the absolute maximum of the energy density of the solutions,
residing on the $\rho$-axis, shifts outward
and decreases significantly in height.

Let us now turn to the regular horizon of the hairy black hole solutions, which
resides at a surface of constant radial coordinate $x=x_{\rm H}$.
Even though the radial coordinate is constant at the horizon, 
the horizon is deformed.
The deformation is revealed, when
measuring the circumference of the horizon along the equator, $L_e$, 
and the circumference of the horizon along the poles, $L_p$,
\begin{equation}
L_e = \int_0^{2 \pi} { d \vphi \left.
 \sqrt{ \frac{l}{f}} x \sin\theta
 \right|_{x=x_{\rm H}, \theta=\pi/2} } \ , \ \ \
L_p = 2 \int_0^{ \pi} { d \theta \left.
 \sqrt{ \frac{m  }{f}} x
 \right|_{x=x_{\rm H}, \vphi=const.} }
\ , \label{lelp} \end{equation}
since the hairy black hole solutions have $L_p \ne L_e$ (in general).
The deviation from spherical symmetry is small, though.
For instance, for the solution of Fig.~13 $L_e/L_p=0.99076$,
and for the solution of Fig.~14 $L_e/L_p=0.99977$.

The hairy black holes satisfy the zeroth law of black hole mechanics,
which states that the surface gravity $\kappa$ is constant on the horizon
\cite{wald}. 
This is dictated by the full set of EYMH equations,
as discussed in section II.D.
Numerically the surface gravity is also constant,
as demonstrated in Fig.~16 for the $n=2$ solution with
horizon area parameter $x_\Delta=1$ for $\alpha=1$.
In constrast to the surface gravity itself, the expansion coefficients
$f_2(\theta)$ and $m_2(\theta)$ of the metric functions $f$ and $m$,
which enter into the expression for $\kappa$, Eq.~(\ref{temp}),
possess a non-trivial angular dependence at the horizon,
as seen in Fig.~16.

\boldmath
\subsection{Domain of existence}
\unboldmath

The domain of existence of the axially symmetric non-abelian black holes
is very similar to the domain of existence of the spherically symmetric
black hole solutions.
For a fixed value of the coupling constant $\alpha$,
$\alpha < \alpha_{\rm max}(n)$,
hairy black hole solutions emerge from the globally regular solution
in the limit $x_\Delta \rightarrow 0$.
They persist up to a maximal value of the area parameter
$x_{\Delta,\rm max}$, which depends on $n$ 
and limits the domain of existence of hairy black holes.
The domain of existence of hairy black hole solutions with magnetic
charge $n=2$ is indicated in Fig.~3, where crosses mark the maximal
values $x_{\Delta,\rm max}$ obtained.

The domain of existence of the hairy black hole solutions with $n>1$
also consists of two regions with distinct critical behaviour,
separated by a particular value of $\alpha$, denoted $\hat \alpha(n)$.
For $\hat \alpha(n) < \alpha < \alpha_{\rm max}(n)$,
the hairy black hole solutions
bifurcate at a critical value $x_{\Delta, \rm cr}$ 
with an extremal RN solution with charge $n$,
while for $0 < \alpha < \hat \alpha(n)$, the hairy black hole solutions
bifurcate at a critical value $x_{\Delta, \rm cr}$ with a non-extremal
RN solution with charge $n$.
For small values of $\alpha$, the branch of hairy black hole solutions
extends backwards from the maximal value $x_{\Delta,\rm max}$ to
the critical value $x_{\Delta, \rm cr}$,
whereas for larger values of $\alpha$, both values
coincide, $x_{\Delta,\rm max} = x_{\Delta, \rm cr}$.
This pattern of the $n>1$ hairy black hole solutions is
completely analogously to the pattern observed for the $n=1$ solutions.
Our numerical results, as exhibited in Fig.~3, are consistent with 
the conjecture $\hat \alpha(2) = \sqrt{3}/2$, suggesting that 
$\hat \alpha(n)$ is independent of $n$.

\boldmath
\subsection{Area parameter $x_\Delta$-dependence}
\unboldmath

Before demonstrating the dependence of the hairy black hole 
solutions on the area parameter $x_\Delta$ and the convergence
of the solutions to the limiting RN solution,
we briefly consider the relation between the area parameter
$x_\Delta$ and the horizon radius in isotropic coordinates $x_{\rm H}$,
since $x_{\rm H}$ is one of the parameters employed in the calculations.
 
In the limit $x_{\rm H} \rightarrow 0$,
hairy black hole solutions emerge from the globally regular solution.
With increasing parameter $x_{\rm H}$ the horizon area of the hairy black hole
solutions increases, until a maximal value of the parameter $x_{\rm H}$ 
is reached. This maximal value of the parameter $x_{\rm H}$, however,
does not coincide with the maximal value of the horizon area,
and thus the maximal value of the area parameter $x_\Delta$.
In the further discussion we consider the two regions of the domain of existence
with distinct critical behaviour separately.

For small values of $\alpha$, $0 < \alpha < \hat \alpha(n)$, 
the branch of hairy black hole solutions,
as a function of the horizon radius $x_{\rm H}$,
extends backwards from the maximal value $x_{\rm H, max}$,
until the critical value $x_{\rm H, cr}$ of the bifurcation 
with the non-extremal RN solution is reached.
When $x_{\rm H}$ decreases from $x_{\rm H, max}$, the area parameter
increases further up to its maximal value $x_{\Delta, \rm max}$,
from where it decreases, until it reaches the critical value
$x_{\Delta, \rm cr}$ of the bifurcation.
This is illustrated in Fig.~17 for the $n=2$ black hole solutions
for $\alpha=0.5$. The endpoint of the curve marks the bifurcation
with the non-extremal RN solution at a finite value of the area.

For larger values of $\alpha$, $\hat \alpha(n) < \alpha < \alpha_{\rm max}(n)$, 
the branch of hairy black hole solutions bifurcates with
an extremal RN solution, which has $x_{\rm H}=0$.
As a function of the horizon radius $x_{\rm H}$,
the branch of hairy black hole solutions
also extends backwards from the maximal value $x_{\rm H, max}$, 
but it ends at the critical value $x_{\rm H, cr}=0$, 
where it bifurcates with the extremal RN solution.
The area parameter, in contrast, increases monotonically, and reaches
its maximal value $x_{\Delta, \rm max}=x_{\Delta, \rm cr}$
at the bifurcation,
as illustrated in Fig.~17 for the $n=2$ black hole solutions for $\alpha=1$.

Let us now consider the dependence of the hairy black hole solutions
on the area parameter $x_\Delta$ in more detail, distinguishing
again the regions $0 < \alpha < \hat \alpha(n)$ and
$\hat \alpha(n) < \alpha < \alpha_{\rm max}(n)$.
For $\hat \alpha(n) < \alpha < \alpha_{\rm max}(n)$, the limiting solution
of the branch of hairy black hole solutions is an extremal RN solution. 
Since convergence to an extremal RN solution has been considered in great detail
in section IV, in the following we put more emphasis on the region
$0 < \alpha < \hat \alpha(n)$, where convergence to a non-extremal
RN solution is observed.

For small values of $\alpha$ in the region $0 < \alpha < \hat \alpha(n)$, 
the branch of hairy black hole solutions
extends backwards from the maximal value $x_{\Delta,\rm max}$,
which limits the domain of existence, 
to the critical value $x_{\Delta, \rm cr}$,
where the bifurcation occurs.
In Figs.18a-c we exhibit a set of $n=2$ black hole solutions
for four values of the area parameter along the black hole branch.
Increasing $x_\Delta$ from zero along the branch of hairy black hole solutions,
first the solutions with $x_\Delta=0.46$ and 1.09 are encountered, 
then the solution with $x_\Delta=1.39$ 
close to the maximal value $x_{\Delta, \rm max}$ is passed, 
and finally the solution with $x_\Delta=1.09$ close to the critical value
$x_{\Delta, \rm cr}$ is reached.

The energy density of the matter fields is shown in Fig.~18a,
the metric function $f$ in Fig.~18b,
and the norm of the Higgs field 
$\mid\Phi\mid=\sqrt{\Phi_{1}^{2}+\Phi_{2}^{2}}$
in Fig.~18c.
All functions change drastically along the hairy black hole branch.
In particularly, we observe, that they approach the limiting
functions of the corresponding non-extremal RN solution
for $x_\Delta \rightarrow x_{\Delta,\rm cr}\approx 1.07$.
Thus in this limit the functions become spherically symmetric.

Convergence to the limiting RN solution is nicely seen in
Fig.~19, where $H_2(x_{\rm H})$ and $\Phi_1(x_{\rm H})$ are shown,
the horizon values of the gauge field function $H_2$
and the Higgs field function $\Phi_1$, respectively.
Starting from the unique values $H_2(0)=1$ and $\Phi_1(0)=0$ 
of the regular solution at the origin, 
for finite $x_\Delta$ an angle-dependence arises,
which becomes maximal as the maximal value of the area parameter
$x_{\Delta, \rm max}$ is approached.
On the backward bending part of the hairy black hole branch,
the angle-dependence diminishes and vanishes in the limit
$x_\Delta \rightarrow x_{\Delta,\rm cr}$,
where the horizon values tend to the unique horizon values of the
corresponding RN solution,
$H_2(x_{\rm H})=0$ and $\Phi_1(x_{\rm H})=1$.

For $\alpha > \hat \alpha$,
the limiting solution is reached
at the maximal value of the area parameter 
$x_{\Delta, \rm max}=x_{\Delta, \rm cr}$.
Since $x_{\Delta, \rm cr} < n \alpha$,
a gap between the branch of hairy black hole solutions
and the corresponding RN branch occurs,
similarly as in the case of the globally regular solutions.

Let us now consider the mass, the surface gravity 
and the deformation of the horizon of the hairy black hole solutions.
In Fig.~20 we see the dependence of the mass of the $n=2$ black hole solutions 
on the area parameter $x_\Delta$ for $\alpha=0.5$ and $\alpha=1$.
(Fig.~20 is completely analogous to Fig.~4, where the 
mass of the $n=1$ black hole solutions is shown.)
For $\alpha=0.5 < \hat \alpha$,
the mass increases with increasing $x_\Delta$,
reaching its maximal value at $x_{\Delta,\rm max}$.
Forming a spike, the mass then decreases with decreasing $x_\Delta$,
and reaches the mass of the limiting non-extremal RN solution with charge $n=2$ 
at $x_{\Delta,\rm cr}$. 
For comparison, the corresponding branch of RN solutions is also shown. 
Apart from the critical point $x_{\Delta, \rm cr}$, the RN branch
and the hairy black hole branch also coincide at the crossing point
$x_{\Delta, \rm cross} \approx 1.32$. 
Consequently, the RN solutions have a lower mass than the
non-abelian black hole solutions,
when $x_{\Delta, \rm cross} < x_\Delta < x_{\Delta, \rm cr}$.
For $\alpha=1 > \hat \alpha$,
the mass increases monotonically with increasing $x_\Delta$,
reaching its maximal value at the maximal value
of the area parameter $x_{\Delta, \rm max} = x_{\Delta, \rm cr}$.
Since $x_{\Delta, \rm cr} < 2 \alpha$,
a gap between the branch of hairy black hole solutions
and the corresponding RN branch occurs.
In Fig.~20, the RN branch would begin at $x_\Delta=2$.

In Fig.~21 the dependence of the surface gravity 
of the $n=2$ hairy black hole solutions 
on the area parameter $x_\Delta$ is shown for $\alpha=0.5$ and $\alpha=1$.
(Fig.~21 is completely analogous to Fig.~5, where the 
surface gravity of the $n=1$ black hole solutions is shown.)
Again, convergence the corresponding RN values is observed.

Let us now turn to the shape of the horizon.
In Fig.~22 the dependence of the ratio of circumferences $L_e/L_p$ 
for the $n=2$ black hole solutions 
on the area parameter $x_\Delta$ is shown for $\alpha=0.5$ and $\alpha=1$.
Interestingly, for $\alpha=0.5 < \hat \alpha$ the ratio $L_e/L_p$, 
shows a rather complicated dependence on the area parameter.
With increasing area parameter the ratio $L_e/L_p$ first decreases, 
until it reaches a minimum close to $x_{\Delta, \rm max}$.
Along this part of the branch of hairy black hole solutions $L_e/L_p < 1$.
Continuing along the hairy black hole branch, the ratio then increases,
first with increasing $x_\Delta$ until $x_{\Delta, \rm max}$ is reached,
and then with decreasing $x_\Delta$.
Passing $L_e/L_p = 1$, the ratio increases further with decreasing $x_\Delta$, 
and attains a maximum $L_e/L_p > 1$. 
 From there $L_e/L_p$ decreases with decreasing $x_\Delta$
and reaches the limiting value
$L_e/L_p = 1$ of the spherically symmetric RN solution
for $x_\Delta \rightarrow x_{\Delta, \rm cr}$.
Along the latter part of the branch of hairy black hole solutions,
$L_e/L_p > 1$.
The occurrence of $L_e/L_p > 1$ is a new phenomenon, not seen previously. 
(The EYM solutions had only values of the ratio $L_e/L_p < 1$ \cite{kk4}.)

For $\alpha=1 > \hat \alpha$, the dependence of the ratio $L_e/L_p$
on the area parameter is much simpler.
Again, with increasing area parameter the ratio $L_e/L_p$ decreases, 
until it reaches a minimum not too far from $x_{\Delta, \rm max}$.
The ratio then increases with increasing $x_\Delta$, 
to reach the limiting value
$L_e/L_p = 1$ of the spherically symmetric RN solution
for $x_\Delta \rightarrow x_{\Delta, \rm cr}$.
Thus for $\alpha=1$ the hairy black hole solutions always have $L_e/L_p < 1$.
We also note that the magnitude of the deformation of the horizon
of the $\alpha=1$ solutions is considerably greater than
the magnitude of the deformation of the horizon of the
$\alpha=0.5$ solutions.

Let us finally consider the dependence of the deformation of the horizon
on the magnetic charge $n$.
For the solutions with charge $n>2$ 
we observe the same features of the ratio $L_e/L_p$ as
for the solutions with charge $n=2$, described above.
This is seen in Fig.~22, where the dependence of the
ratio of circumferences $L_e/L_p$ for the $n=3$ black hole solutions 
on the area parameter $x_\Delta$ for $\alpha=0.5 < \hat \alpha$
is also shown.
We observe,
that the maximal deformation of the horizon increases with $n$.

\boldmath
\subsection{Isolated horizon results}
\unboldmath

Let us now address the predictions of the isolated horizon formalism
for the axially symmetric hairy black hole solutions with $n>1$.
We first consider relation (\ref{IHmu}) between
black hole mass, soliton mass and horizon mass
for the $n>1$ solutions.
Our numerical calculations confirm this relation for the
axially symmetric hairy black hole solutions of EYMH theory.
In Fig.~23 we show the dependence of the horizon mass {\it per unit charge}
$\mu_{\Delta}/(\alpha^2 n)$, obtained according to Eq.~(\ref{IHmuD}),
on the area parameter $x_\Delta$ 
for the $n=2$ black hole solutions for $\alpha=0.5$ and $\alpha=1$.
Adding the mass {\it per unit charge}
of the corresponding multimonopole solutions
($\mu_{\rm reg}/(2 \alpha^2)(\alpha=0.5)=0.961105$ and 
$\mu_{\rm reg}/(2 \alpha^2)(\alpha=1.0)=0.847943$, respectively,)
precisely gives the values of the mass {\it per unit charge}
of the hairy black hole solutions in Fig.~20.

Next we consider the interpretation of the hairy black holes as a bound states
of regular solutions and Schwarzschild black holes \cite{iso2}.
We therefore identify the binding energy of these systems
$\mu_{\rm bind}/\alpha^2$, according to Eq.~(\ref{IHbind}).
In Fig.~24 we present the dependence of 
the binding energy $\mu_{\rm bind}/\alpha^2$
on the area parameter $x_\Delta$ 
for the $n=2$ black hole solutions for $\alpha=0.5$ and $\alpha=1$.
For comparison, the binding energy of the 
corresponding RN solutions is also given.
Fig.~24 is completely analogous to Fig.~7, where the 
binding energy of the $n=1$ black hole solutions is shown.

However, for the $n>1$ hairy black hole solutions,
we can also consider other compound systems with magnetic charge $n$,
consisting of a soliton and a black hole.
Therefore the notion of binding energy is no longer unique,
and the binding energy of the compound system 
of a $n=2$ multimonopole and a Schwarzschild black hole
represents only one particular case.

In the following we restrict the discussion to hairy black hole solutions
with magnetic charge $n=2$.
Let us define
the binding energy $\mu_{\rm bind}^1/\alpha^2$ of the compound system 
of a $n=2$ multimonopole and a Schwarzschild black hole by
\begin{equation}
\mu^{(n=2)}(x_\Delta)=
\mu_{\rm reg}^{(n=2)} + \mu_{\rm S}(x_\Delta)
+ \mu_{\rm bind}^1
= \mu^1 + \mu_{\rm bind}^1
\ , \label{bind1} \end{equation}
and compare it to 
the binding energy $\mu_{\rm bind}^2/\alpha^2$ of the compound system 
of two $n=1$ monopoles and a Schwarzschild black hole,
\begin{equation}
\mu^{(n=2)}(x_\Delta)=
 2 \mu_{\rm reg}^{(n=1)} + \mu_{\rm S}(x_\Delta)
+ \mu_{\rm bind}^2
= \mu^2 + \mu_{\rm bind}^2
\ . \label{bind2} \end{equation}
Since the binding energy $\mu_{\rm bind}^2/\alpha^2$ differs from
the binding energy $\mu_{\rm bind}^1/\alpha^2$ 
by the difference between the mass of the $n=2$ multimonopole 
and the mass of two $n=1$ monopoles,
the binding energy $\mu_{\rm bind}^2/\alpha^2$ is smaller 
than the binding energy $\mu_{\rm bind}^1/\alpha^2$, shown in Fig.~24.

Alternatively we can consider a $n=2$ hairy black hole as
a compound system of a $n=1$ monopole and a charged black hole.
For the compound system of a $n=1$ monopole and a $n=1$ hairy black hole,
the binding energy $\mu_{\rm bind}^3/\alpha^2$ is given by
\begin{equation}
\mu^{(n=2)}(x_\Delta)=
\mu_{\rm reg}^{(n=1)} + \mu^{(n=1)}(x_\Delta)
+ \mu_{\rm bind}^3
= \mu^3 + \mu_{\rm bind}^3
\ , \label{bind3} \end{equation}
in the parameter range, where both black hole solutions coexist. 
Likewise, 
for the compound system of a $n=1$ monopole and a RN black hole with unit charge,
the binding energy $\mu_{\rm bind}^4/\alpha^2$ is given by
\begin{equation}
\mu^{(n=2)}(x_\Delta)=
\mu_{\rm reg}^{(n=1)} + \mu_{\rm RN}(x_\Delta)
+ \mu_{\rm bind}^4
= \mu^4 + \mu_{\rm bind}^4
\ , \label{bind4} \end{equation}
in the parameter range, where both solutions coexist. 

Let us now illustrate these possibilities for the binding energy
by considering the masses of the corresponding compound systems.
In Figs.~25a-b we show the dependence
the masses $\mu^1/\alpha^2-\mu^4/\alpha^2$, Eqs.~(\ref{bind1})-(\ref{bind4}),
on the area parameter $x_\Delta$,
and compare it to the mass of the $n=2$ hairy black hole solutions,
$\mu^{(n=2)}(x_\Delta)/\alpha^2$, for $\alpha=0.5$ and $\alpha=1$, respectively.

The binding energy $\mu_{\rm bind}^1/\alpha^2$
and the binding energy $\mu_{\rm bind}^2/\alpha^2$ are always negative
for the $n=2$ hairy black hole solutions.
In contrast, the binding energy $\mu_{\rm bind}^3/\alpha^2$,
representing the binding energy of
the compound system of a $n=1$ monopole and a $n=1$ hairy black hole,
changes sign.
For $\alpha=0.5$, $\mu_{\rm bind}^3/\alpha^2$ 
is positive beyond $x_\Delta \approx 0.34$,
and for $\alpha=1$, it is positive beyond $x_\Delta \approx 0.52$.
The binding energy $\mu_{\rm bind}^4/\alpha^2$, representing the binding energy of
the compound system of a $n=1$ monopole and a RN black hole with unit charge,
changes sign at $x_\Delta \approx 0.63$ for $\alpha=0.5$,
whereas it is always positive for $\alpha=1$.

Thus for $\alpha=0.5$,
the $n=2$ hairy black holes possess a mass lower than the masses $\mu^1-\mu^4$
of the compound systems only for $x_\Delta < 0.34$. 
For $x_\Delta > 0.34$, the $n=2$ hairy black hole is either heavier than 
the compound system of a $n=1$ monopole and a $n=1$ hairy black hole,
or (when the $n=1$ hairy black hole branch ceases to exist)
it is heavier than 
the compound system of a $n=1$ monopole and a RN black hole with unit charge.

For $\alpha=1$
the $n=2$ hairy black hole is energetically favourable
for $x_\Delta < 0.52$, where the
compound system of a $n=1$ monopole and a $n=1$ hairy black hole is heavier,
and for $x_{\Delta, \rm cr}(n=1) < x_\Delta < 1$,
since below $x_\Delta=1$ no RN solutions exist.
($x_{\Delta, \rm cr}(n=1) < x_\Delta < 1$ represents the gap between the
$n=1$ hairy black hole branch and the RN branch with unit charge.)

Let us finally consider the ``quasilocal uniqueness conjecture'',
which states that static black holes are uniquely
specified by their horizon area and horizon charges.
In Fig.~26 we show the dependence of the non-abelian horizon magnetic charge
on the area parameter $x_\Delta$
for the $n=2$ hairy black hole solutions for $\alpha=0.5$ and 1.
The horizon charge increases monotonically along the branches of
$n=2$ hairy black hole solutions, analogously to the
horizon charge of the $n=1$ hairy black hole solutions, shown in Fig.~8.
In the limit $x_\Delta \rightarrow x_{\Delta, \rm cr}$,
for $\alpha=0.5$ the value of the RN solution with charge $n=2$ is reached,
as expected, whereas for $\alpha=1$ the limiting value is smaller than two
(reflecting the gap between the hairy black hole branch and the RN branch).
The horizon electric charge vanishes.

To address the ``quasilocal uniqueness conjecture'', we now also 
consider the branches of embedded RN solutions 
with integer values of the magnetic charge,
beginning at $x_\Delta = \alpha n$.
The lowest branch of embedded RN solutions has unit magnetic charge,
and thus unit horizon magnetic charge.
For $\alpha=0.5$, this branch begins at $x_\Delta = 0.5$,
and crosses the corresponding ($\alpha=0.5$) branch 
of $n=2$ hairy black hole solutions.
Consequently, at the crossing point
there exist two distinct black hole solutions, 
one with hair and the other without hair, with the same horizon area and
the same non-abelian horizon magnetic charge, 
representing a counterexample to the ``quasilocal uniqueness conjecture''.

\section{Conclusions}

We have considered static axially symmetric
multimonopole and black hole solutions in EYMH theory.
We have presented these solutions in detail
and discussed their properties.
Our particular interests were the investigation of an attractive
phase between like monopoles and the study of monopole and black hole
properties predicted by the isolated horizon formalism.

Concerning the multimonopole solutions we observe, 
that the mass {\it per unit charge}
of the (multi)monopole solutions decreases with increasing $\alpha$.
In the BPS limit, for $\alpha=0$ 
the mass {\it per unit charge} is precisely
equal to the mass of the $n=1$ monopole.
For $\alpha > 0$, however, we observe that
the mass {\it per unit charge} of the multimonopoles
is smaller than the mass of the $n=1$ monopole.
In particular, the mass {\it per unit charge} decreases
with increasing $n$.
Thus in the BPS limit, there is an attractive phase between like monopoles,
not present in flat space.
Moreover, multimonopoles exist for gravitational coupling strength, 
too large for $n=1$ monopoles to exist.

For finite Higgs selfcoupling, the flat space multimonopoles
have higher mass {\it per unit charge} than
the $n=1$ monopole, allowing only for a repulsive phase
between like monopoles.
By continuity, this repulsive phase persists in the presence of gravity
for small values of $\alpha$, but it
can give way to an attractive phase for larger values of $\alpha$.
Thus the repulsion between like monopoles
can be overcome for small Higgs selfcoupling
by sufficiently strong gravitational attraction.
At the equilibrium value $\alpha_{\rm eq}$,
multimonopole mass {\it per unit charge} und monopole mass
equal one another.
The equilibrium value
$\alpha_{\rm eq}$ increases with increasing Higgs selfcoupling,
yielding a decreasing region in parameter space for the attractive phase.
For large Higgs selfcoupling, only a repulsive phase is left.

While singly charged monopole solutions are stable,
stability of the static axially symmetric multimonopole solutions is not
obvious.
We conjecture, that the $n=2$ multimonopole solutions are stable,
as long as their mass {\it per unit charge} is lower than
the mass of the $n=1$ monopole.
For topological number $n \ge 3$, however, solutions with only
discrete symmetry exist in flat space \cite{mon}, which,
by continuity, should also be present in curved space
(at least for small gravitational strength).
For a given topological number $n>2$,
such multimonopole solutions without rotational symmetry
may possess a lower mass than the corresponding axially symmetric solutions,
as suggested by analogy from the multiskyrmions in flat space \cite{skyr}.
The axially symmetric solutions may therefore not represent
global minima in their respective topological sectors,
even if their mass {\it per unit charge} is lower than
the mass of the $n=1$ monopole.

Let us now turn to the black hole solutions of SU(2) EYMH theory.
Besides embedded abelian black hole solutions, 
SU(2) EYMH theory also possesses genuine non-abelian black hole solutions
\cite{gmono}.
The static SU(2) EYMH black hole solutions are no longer uniquely
determined by their mass and charge alone. Indeed, in a certain
region of the domain of existence of hairy black hole solutions,
also embedded RN solutions with the same mass and charge exist.
The non-abelian black hole solutions therefore
represent counterexamples to the ``no-hair'' conjecture \cite{gmono}.

While static spherically symmetric ($n=1$) EYMH black holes were studied
in great detail \cite{gmono},
non-abelian black hole solutions with magnetic charge
$n>1$ were previously only considered perturbatively \cite{ewein}.
We have obtained static axially symmetric black hole solutions 
with integer magnetic charge $n>1$ numerically.
These black hole solutions are asymptotically flat,
and they possess a regular deformed horizon.
Being static and not spherically symmetric,
these black hole solutions represent further examples, 
showing that Israel's theorem cannot be generalized to EYM or EYMH theory.
While previous (non-perturbative) counterexamples \cite{kk,map3} 
were classically unstable, hairy EYMH black holes 
should provide classically stable counterexamples \cite{ewein}.

Considering the static axially symmetric solutions
from the isolated horizon formalism point of view, 
we have verified the mass relation between the monopole and the black hole
solutions,
showing that the black hole mass is given by the sum of the soliton mass
and the horizon mass. 
Interpreting the hairy black holes as bound states of solitons
and Schwarzschild black holes \cite{iso2},
we have studied the binding energy of these bound systems.
We have furthermore considered the binding energy with respect to
various other compound systems, such as a $n-1$ soliton and a
$n=1$ hairy black hole or a $n-1$ soliton and a RN black hole
with unit charge.
The ``quasilocal uniqueness conjecture'' claims,
that black holes are uniquely specified by their horizon area
and their horizon electric and magnetic charge.
Since we have constructed a counterexample to this conjecture,
the need for a new formulation of the uniqueness conjecture arises.

The hairy black hole solutions studied here
represent only the simplest type of non-spherical black hole solutions.
Indeed, there are gravitating black hole solutions
with much more complex shapes and only discrete symmetries left.
In curved space such black holes without rotational symmetry,
have been considered so far only perturbatively \cite{ewein}.
It remains a challenge to construct such solutions non-perturbatively
and to find out, whether such black hole solutions 
without rotational symmetry are stable.

{\it Acknowledgement}
We would like to thank the RRZN in Hannover for computing time.

\vfill\eject

\clearpage

\section{Appendix A}

\boldmath
\subsection{Tensors $F_{\mu\nu}$, $D_{\mu}\Phi$, $T_{\mu\nu}$} 
\unboldmath

We expand the field strength tensor and the covariant derivative
with respect to the Pauli matrices 
$\tau^{n}_{\lambda}$, ($\lambda=r$, $\theta$, $\vphi$,
see Eq.~(\ref{rtp}))
\begin{eqnarray*}
F_{\mu \nu} & = & F^{(\lambda)}_{\mu \nu} \ \frac{\tau^{n}_{\lambda}}{2}
 \  \\
\end{eqnarray*}
and
\begin{eqnarray*}
D_{\mu}\Phi & = & D^{(\lambda)}_{\mu}\Phi \ {\tau^{n}_{\lambda}}
 \ .\\
\end{eqnarray*}
Inserting ansatz (\ref{gf1}) for the gauge field,
we obtain the non-vanishing coefficients 
$F_{\mu\nu}^{(\lambda)}$ and $D_{\mu}\Phi^{(\lambda)}$
\begin{eqnarray}
F_{r\theta }^{(\vphi )}
&  = & 
 - \frac{1}{r}\left( H_{1,\theta} + r H_{2,r} \right) 
\ , \nonumber \\
F_{r\vphi }^{(r)} 
& = & 
-n \frac{ \sin \theta}{r}\left( r H_{3,r} - H_1  H_4  \right) 
\ , \nonumber \\
F_{r\vphi }^{(\theta )} 
& = & 
n \frac{ \sin \theta}{r}\left( r H_{4,r} + H_1 H_3   
+ {\rm cot} \theta H_1 \right) 
\ , \nonumber \\
F_{\theta\vphi }^{(r)} 
& = & 
-n \sin \theta  \left( H_{3,\theta} - 1 +  H_2 H_4  
+ {\rm cot} \theta H_3  \right) 
\ , \nonumber \\
F_{\theta\vphi }^{(\theta )} 
& = &  
n \sin \theta \left( H_{4,\theta} - H_2 H_3   
 - {\rm cot} \theta \left( H_2 - H_4 \right)  \right) 
\ , \nonumber \\
\end{eqnarray}
and $F^{(\lambda)}_{\mu \nu} = -F^{(\lambda)}_{\nu \mu}$,

\begin{eqnarray}
D_{r}\Phi^{(r )}
&  = & 
   \frac{1}{r}\left( H_{1}\Phi_{2} + r \Phi_{1,r} \right) 
\ , \nonumber \\
D_{r}\Phi^{(\theta )} 
& = & 
 - \frac{1}{r}\left( H_{1}\Phi_{1} - r \Phi_{2,r} \right) 
\ , \nonumber \\
D_{\theta}\Phi^{(r)} 
& = & 
 \left( \Phi_{1,\theta} -  H_2\Phi_2  \right) 
\ , \nonumber \\
D_{\theta}\Phi^{(\theta)} 
& = & 
 \left( \Phi_{2,\theta} +  H_2\Phi_1  \right) 
\ , \nonumber \\ 
D_{\vphi}\Phi^{(\vphi )} 
& = &  
n \sin \theta \left( H_{4}\Phi_1 + H_3\Phi_2   
 + {\rm cot} \theta  \Phi_2   \right) 
\ . \nonumber \\
\end{eqnarray}

It is convenient to define
\begin{eqnarray}
F^2_{r \theta} 
& = & 
\left( F_{r\theta }^{(\vphi )}\right)^2 
+\frac{1}{r^2} \left( r H_{1,r}-H_{2,\theta}\right)^2 
\ , \nonumber \\
F^2_{r \vphi} 
& = & 
\left( F_{r\vphi }^{(r)}\right)^2 
+\left( F_{r\vphi }^{(\theta )}\right)^2 
\ , \nonumber \\ 
F^2_{\theta \vphi} 
& = & 
\left( F_{\theta\vphi }^{(r)}\right)^2 
+\left(F_{\theta\vphi }^{(\theta )}\right)^2 
\ , \nonumber \\ 
\end{eqnarray}
where the second term in the definition of $F^2_{r \theta}$ 
represents the gauge fixing term, as well as
\begin{eqnarray}
D^2_{r}\Phi 
& = & 
\left( D_{r}^{(r )}\Phi\right)^2
+ \left( D_{r}^{(\theta )}\Phi\right)^2 
\ , \nonumber \\
D^2_{\theta}\Phi 
& = & 
\left( D_{\theta}^{(r )}\Phi\right)^2 
+ \left( D_{\theta}^{(\theta )}\Phi\right)^2 
\ , \nonumber \\
D^2_{\vphi}\Phi 
& = & 
\left( D_{\vphi}^{(\vphi )}\Phi\right)^2 
 \ . \nonumber \\
\end{eqnarray}

With the ansatz for the metric (\ref{metric2}) we obtain the
Lagrange densities
\begin{eqnarray}
L_F      
& = & 
-\frac{f}{2 m} \left( \frac{f}{r^2 m} F^2_{r \theta} 
 + \frac{f}{r^2 \sin^2 \theta l} (F^2_{r \vphi}
+\frac{1}{r^2}F^2_{\theta \vphi}) \right) 
\ , \nonumber \\
L_{\Phi} 
& = & 
-\frac{f}{2 \ m} \left(D^2_{r}\Phi  
+ \frac{1}{r^2}D^2_{\theta}\Phi +\frac{m}{lr^2\sin^2\theta}
D^2_{\vphi}\Phi \right)-\frac{\lambda}{8}Tr(\Phi^2-\eta^2)^2
\ , \nonumber \\
L_M 
& = & 
L_{\Phi}+ L_{F}
\ , \nonumber \\   
\end{eqnarray}
and the non-vanishing components of the stress energy tensor, 
\begin{eqnarray}
T_{00} 
& = & 
\frac{f^2}{2 m} \left[      
    \frac{f}{r^2 m} F^2_{r \theta} 
    + \frac{f}{r^2 \sin^2 \theta l}(F^2_{r \vphi} 
         + \frac{1}{r^2} F^2_{\theta \vphi})
         + D^2_{r}\Phi  
           + \frac{1}{r^2}D^2_{\theta}\Phi +
           \frac{m}{lr^2\sin^2\theta}
            D^2_{\vphi}\Phi \right] 
            + f\frac{\lambda}{8}Tr(\Phi^2-\eta^2)^2 
\ , \nonumber \\
T_{rr} 
& = & 
\frac{1}{2} \left[  
     \frac{f}{r^2 m} F^2_{r \theta} 
     + \frac{f}{r^2 \sin^2 \theta l}(F^2_{r \vphi} 
     - \frac{1}{r^2} F^2_{\theta \vphi})
      + D^2_{r}\Phi  
           - \frac{1}{r^2}D^2_{\theta}\Phi -
           \frac{m}{lr^2\sin^2\theta}
            D^2_{\vphi}\Phi \right] 
            - \frac{m}{f}\frac{\lambda}{8}Tr(\Phi^2-\eta^2)^2   
\ , \nonumber \\
T_{\theta \theta} 
& = & 
\frac{r^2}{2} 
     \left[ \frac{f}{r^2 m} F^2_{r \theta} 
     + \frac{f}{r^2 \sin^2 \theta l}(-F^2_{r \vphi} 
     + \frac{1}{r^2} F^2_{\theta \vphi})
     - D^2_{r}\Phi  
           + \frac{1}{r^2}D^2_{\theta}\Phi -
           \frac{m}{lr^2\sin^2\theta}
            D^2_{\vphi}\Phi \right] 
            - \frac{mr^2}{f}\frac{\lambda}{8}Tr(\Phi^2-\eta^2)^2 
\ , \nonumber \\
T_{\vphi \vphi} 
& = & 
\frac{r^2\sin^2\theta}{2}\frac{l}{m} 
     \left[ -\frac{f}{r^2 m} F^2_{r \theta} 
     + \frac{f}{r^2 \sin^2 \theta l}(F^2_{r \vphi} 
     + \frac{1}{r^2} F^2_{\theta \vphi})
     - D^2_{r}\Phi  
           - \frac{1}{r^2}D^2_{\theta}\Phi +
           \frac{m}{lr^2\sin^2\theta}
            D^2_{\vphi}\Phi \right] 
            - \frac{lr^2\sin^2\theta}{f}\frac{\lambda}{8}Tr(\Phi^2-\eta^2)^2 
\ . \nonumber \\
\end{eqnarray}

\section{Appendix B}

Subject to the corresponding set of boundary conditions,
we solve the system of coupled non-linear partial
differential equations numerically.
For the globally regular solutions, we employ the radial coordinate 
\begin{equation}
z = \frac{x}{1+x}
\   \label{barx} \end{equation}
instead of $x$,
to map the infinite interval of the variable $x$ onto 
the finite interval $[0,1]$ of the variable $z$.
For the derivatives this leads to the substitutions 
\begin{eqnarray}
r F_{,r}     & \longrightarrow & z (1-z) F_{,z} \ \ , 
\nonumber \\ 
r^2 F_{,r,r} & \longrightarrow & 
z^2 \left( (1-z)^2 F_{,z,z} 
  - 2 (1-z) F_{,z} \right) \\ 
\end{eqnarray}
for any function $F$ in the differential equations. 
In this form we have solved the system of differential equations 
numerically.  

To map spatial infinity to the finite value $z=1$,
we employ for the black hole solutions the radial coordinate 
\begin{equation}
z = 1-\frac{x_{\rm H}}{x}
\ . \label{barx2} \end{equation}
For the derivatives this leads to the substitutions 
\begin{eqnarray}
r F_{,r}     & \longrightarrow &  (1-z) F_{,z} \ \ , 
\nonumber \\ 
r^2 F_{,r,r} & \longrightarrow & 
(1-z)^2  F_{,z,z} 
  - 2 (1-z) F_{,z} \\ 
\end{eqnarray}
for any function $F$ in the differential equations. 

For the black hole solutions we furthermore introduce the functions
$\bar{f}(z,\theta)$, $\bar{m}(z,\theta)$ and $\bar{l}(z,\theta)$ \cite{kk2,kk4},
\begin{equation}
\bar{f}(z,\theta)=\frac{f(z,\theta)}{z^2}\ , \
\bar{m}(z,\theta)=\frac{m(z,\theta)}{z^2}\ , \
\bar{l}(z,\theta)=\frac{l(z,\theta)}{z^2}\,
\end{equation}
where $z$ is the compactified coordinate (\ref{barx2}).
Since in the limit $x\rightarrow\infty$, the variable $z$ approaches the value $1$,
the boundary conditions 
for the functions $\bar{f}$, $\bar{m}$ and $\bar{l}$ coincide 
with the boundary conditions for the functions $f$, $m$ and $l$ at infinity.
At the horizon, the boundary conditions of the functions 
$\bar{f}$, $\bar{m}$ and $\bar{l}$ are given by
\begin{equation}
(\bar{f}-\partial_{z}\bar{f})|_{z=0}=0 \ , \ 
(\bar{m}+\partial_{z}\bar{m})|_{z=0}=0 \ , \ 
(\bar{l}+\partial_{z}\bar{l})|_{z=0}=0 \ .
\end{equation}
To satisfy the regularity condition (\ref{lm}) in the numerical
calculations, we have introduced the new function $g(z,\theta)$
\begin{equation}
g(z,\theta)=\frac{\bar{m}(z,\theta)}{\bar{l}(z,\theta)}
\end{equation}
with the boundary conditions on the symmetry axis and at the horizon
\begin{equation}
g|_{\theta=0}=1 \ , \   \partial_{z}g|_{z=0}=0 \ .
\end{equation}

The numerical calculations are performed with help of the program
FIDISOL, which is extensively documented in \cite{schoen}.
The equations are discretized on a non-equidistant
grid in $z$ and  $\theta$.
Typical grids used have sizes $150 \times 30$, 
covering the integration region 
$0\leq z\leq 1$ and $0\leq\theta\leq\pi/2$.

The numerical method is based on the Newton-Raphson
method, an iterative procedure to find a good approximation to
the exact solution. 
Let us put the partial differential equations into the form 
$P(u)=0$, where $u$ denotes the unknown functions 
(and their derivatives). For an approximate solution $u^{(1)}$,
$P(u^{(1)})$ does not vanish. If we could find a small correction 
$\Delta u$, such that $u^{(1)}+\Delta u$ is the exact solution, 
$P(u^{(1)}+\Delta u)=0$ should hold. Approximately the expansion in 
$\Delta u$ gives 
$$
0=P(u^{(1)}+\Delta u) \approx 
P(u^{(1)})+\frac{\partial P}{\partial u }(u^{(1)}) \Delta u \ .
$$
The equation 
$P(u^{(1)})= -\frac{\partial P}{\partial u }(u^{(1)}) \Delta u $
can be solved to determine the correction  $\Delta u^{(1)}= \Delta u$.
$u^{(2)}=u^{(1)}+\Delta u^{(1)}$ will not be the exact solution
but an improved approximate solution. Repeating the calculations 
iteratively, the approximate solutions will converge to the exact solutuion,
provided the initial guess solution is close enough to the exact solution. 
The iteration stops after $i$ steps if the Newton residual $P(u^{(i)})$ 
is smaller than a prescribed tolerance.
Therefore it is essential to have a good
first guess, to start the iteration procedure.
Our strategy therefore is to use a known solution as guess
and then vary some parameter to produce the next solution.

To construct axially symmetric EYMH solutions, 
we have used the known spherically symmetric EYMH solutions
as starting solutions with $n=1$.
We have then increased the `parameter' $n$ slowly, to obtain
the desired axially symmetric solutions at integer values of $n$.

For a numerical solution it is important to have information 
about its quality, i.~e.~to have an error estimate. The error originates 
from the discretization of the system of partial differential equations.
It depends on the number of gridpoints and on the order of consistency of 
the differential formulae for the derivatives. FIDISOL provides
an error estimate for each unknown function, corresponding to 
the maximum of the discretization error divided by the 
maximum of the function. For the solutions presented here  
the estimations of the relative error for the functions are 
on the order of $10^{-3}$ for $n=2$ and $n=3$.

\clearpage

\newpage

\begin{fixy}{0}
\begin{figure}
\centering
\epsfysize=10cm
\mbox{\epsffile{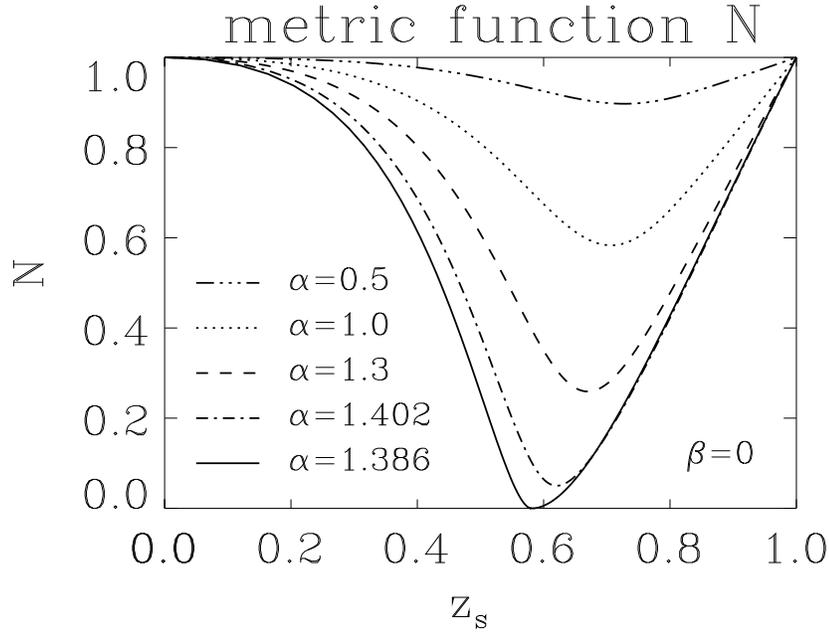}}
\caption{\label{Fig.1a} 
The metric function $N$ of the $n=1$ monopole solution
is shown as a function of the
compactified dimensionless Schwarzschild-like coordinate
$z_{\rm s}=\tilde{x}/(1+\tilde{x})$ in the BPS limit
for five values of $\alpha$ along the monopole branch,
in particular for a value close to the maximal value of $\alpha$,
$\alpha_{\rm max} \approx 1.403$ and
for a value close to the critical value of $\alpha$,
$\alpha_{\rm cr} \approx 1.386$. }
\end{figure}

\begin{figure}
\centering
\epsfysize=10cm
\mbox{\epsffile{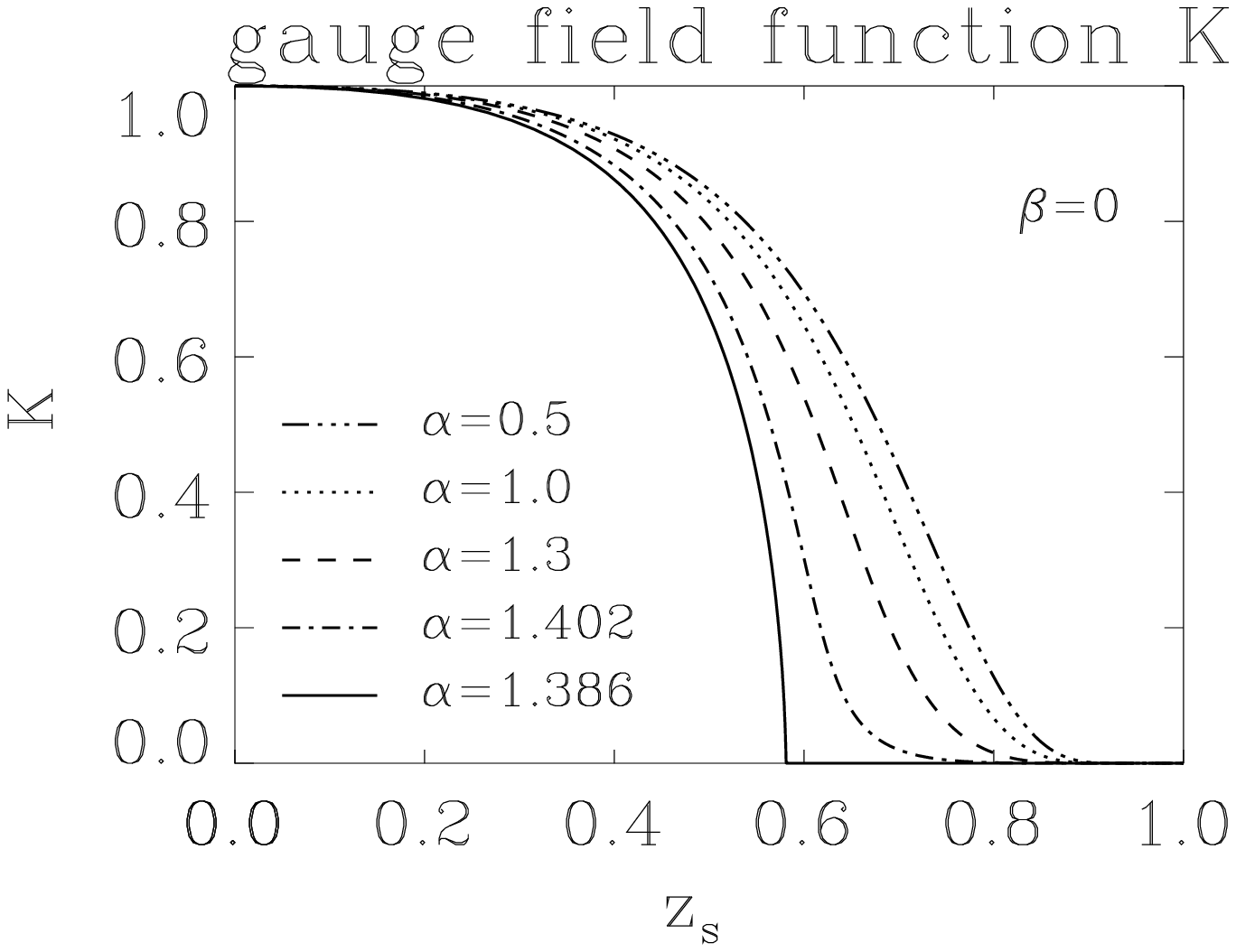}}
\caption{\label{Fig.1b} The same as Fig.~1a for the gauge field function $K$.
}
\end{figure}

\begin{figure}
\centering
\epsfysize=10cm
\mbox{\epsffile{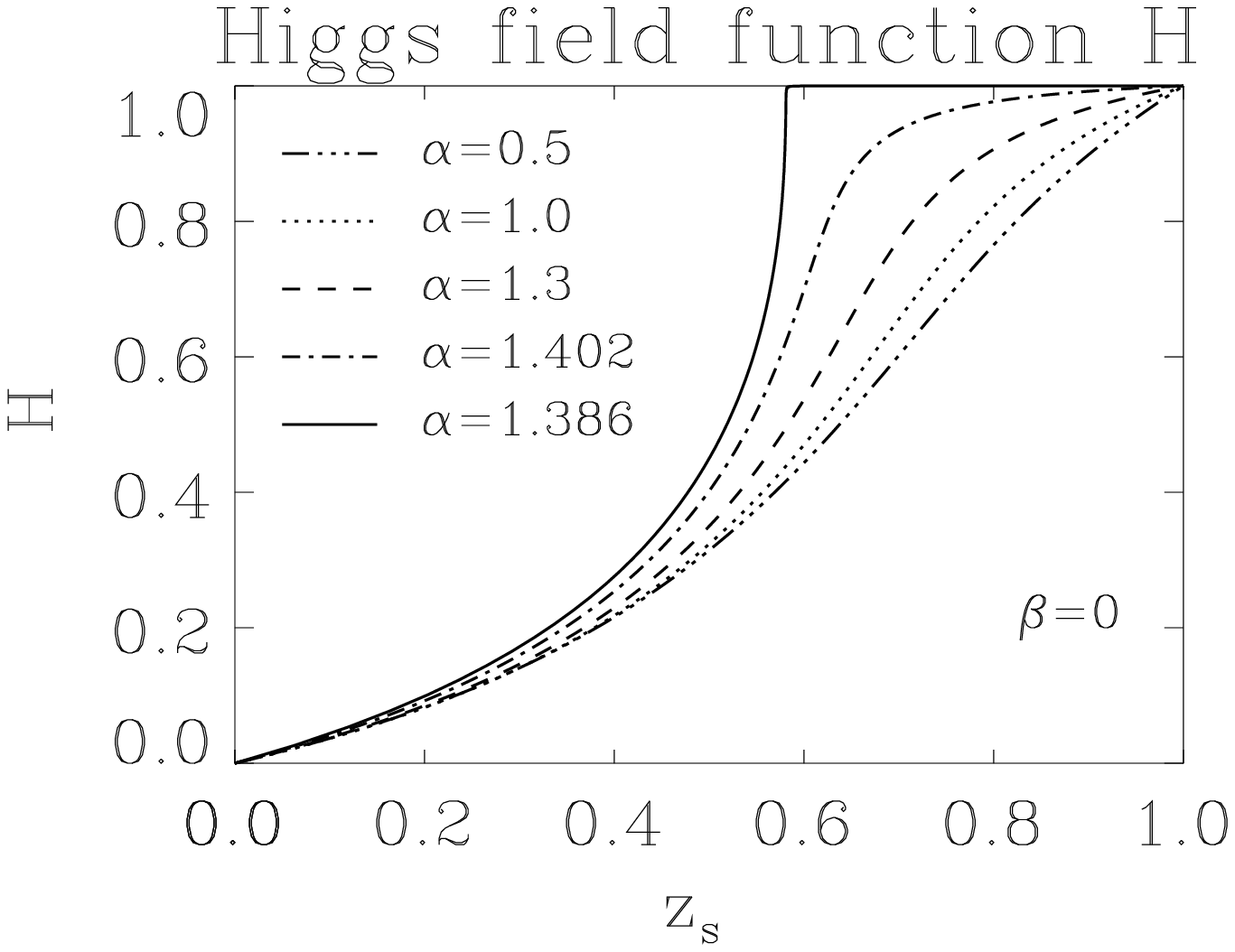}}
\caption{\label{Fig.1c} The same as Fig.~1a for the Higgs field function $H$.}
\end{figure}
\end{fixy}

\begin{fixy}{0}
\begin{figure}
\centering
\epsfysize=10cm
\mbox{\epsffile{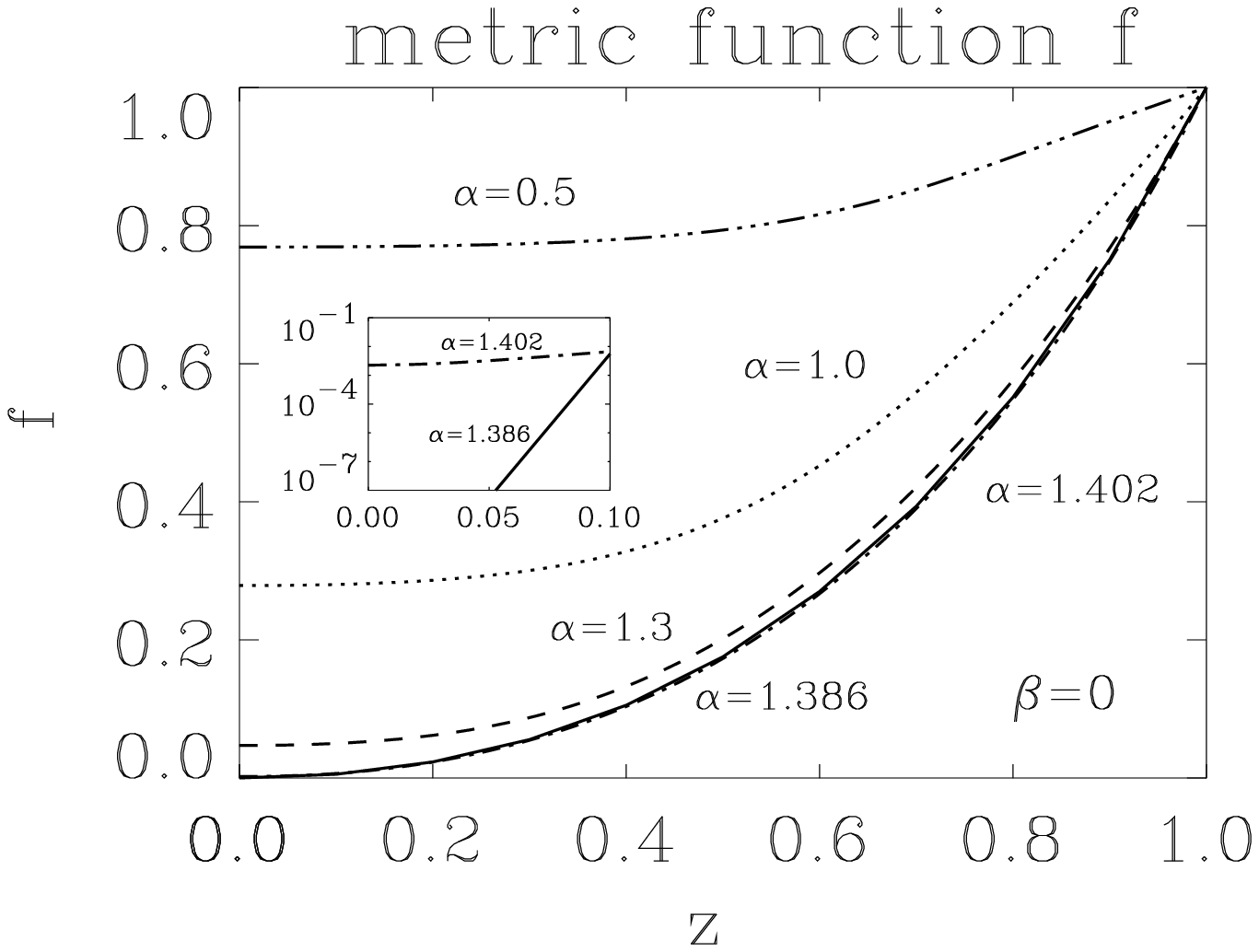}}
\caption{\label{Fig.2a} 
The metric function $f$ of the $n=1$ monopole solution
is shown as a function of the
compactified dimensionless isotropic coordinate
$z=x/(1+x)$ in the BPS limit
for five values of $\alpha$ along the monopole branch,
in particular for a value close to the maximal value of $\alpha$,
$\alpha_{\rm max} \approx 1.403$ and
for a value close to the critical value of $\alpha$,
$\alpha_{\rm cr} \approx 1.386$. }
\end{figure}

\begin{figure}
\centering
\epsfysize=10cm
\mbox{\epsffile{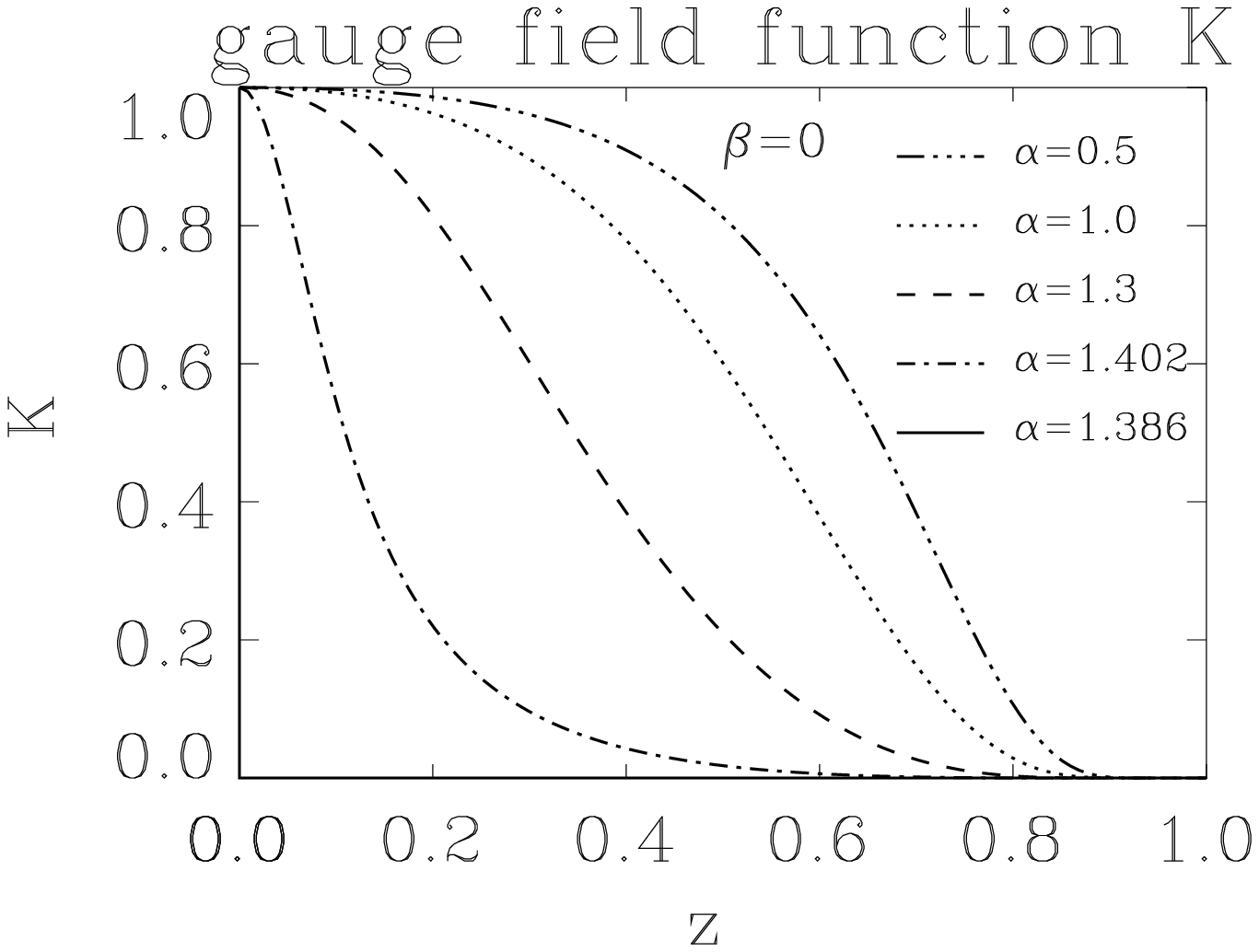}}
\caption{\label{Fig.2b} The same as Fig.~2a for the gauge field function $K$.
}
\end{figure}

\begin{figure}
\centering
\epsfysize=10cm
\mbox{\epsffile{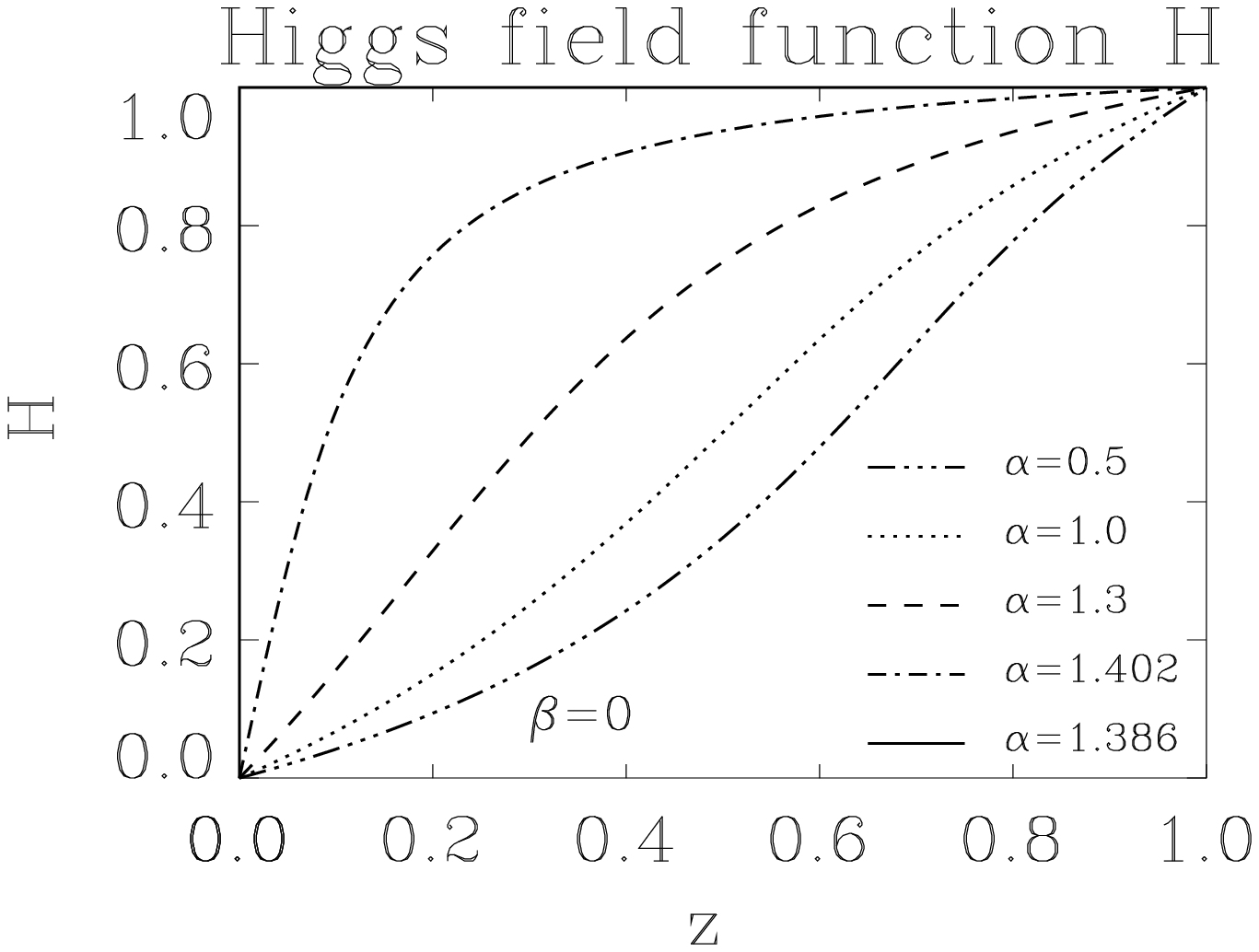}}
\caption{\label{Fig.2c} The same as Fig.~2a for the Higgs field function $H$.}
\end{figure}
\end{fixy}

\begin{fixy}{-1}
\begin{figure}
\centering
\epsfysize=10cm
\mbox{\epsffile{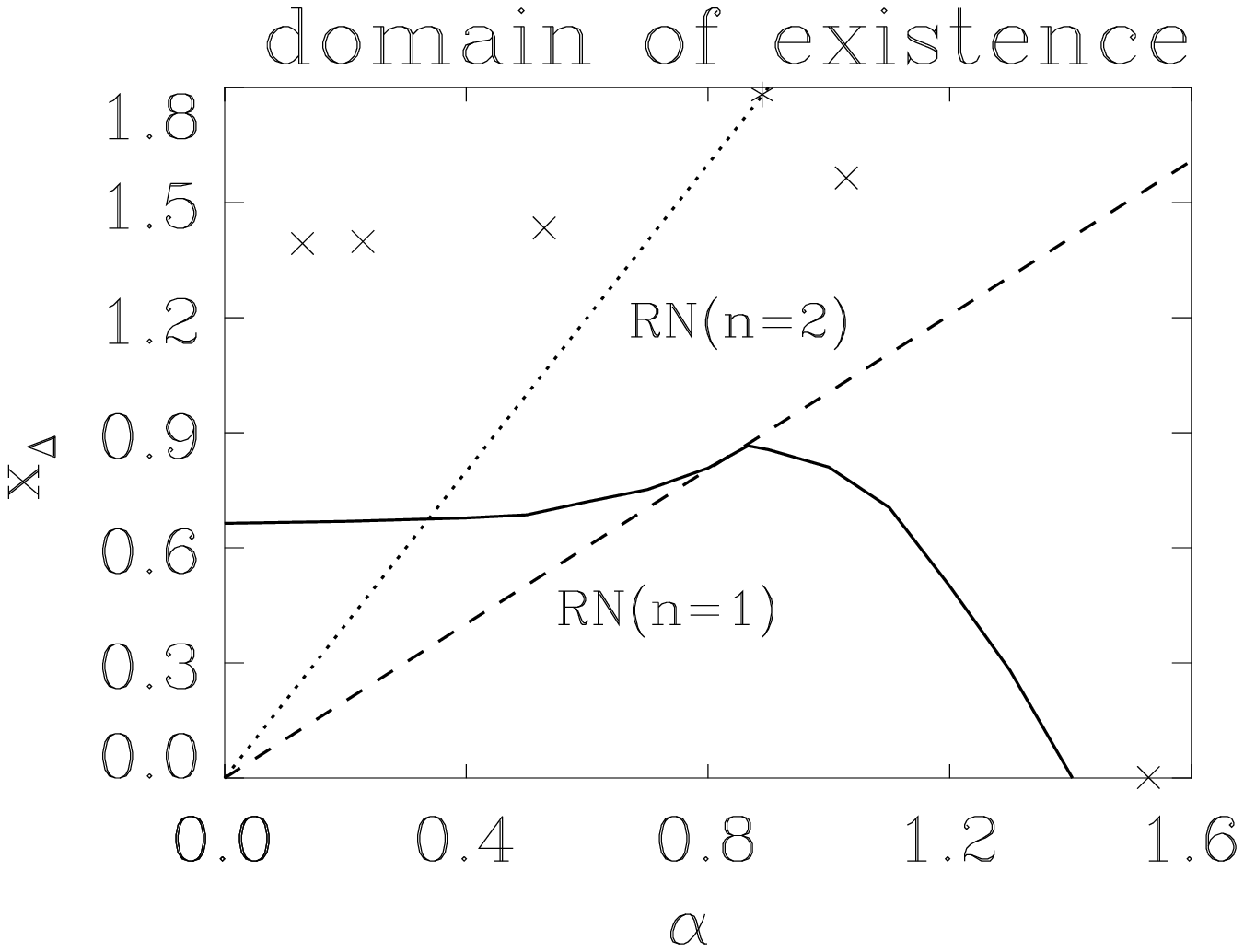}}
\caption{\label{Fig.3} 
The domain of existence of the hairy black hole solutions 
in the BPS limit is shown in the $x_{\Delta}$-$\alpha$-plane.
The solid line shows the maximal values $x_{\Delta, \rm max}$
obtained for the $n=1$ hairy black hole solutions,
while the crosses represent the maximal values $x_{\Delta, \rm max}$
obtained for $n=2$ hairy black hole solutions.
The asterisk marks the value $\hat \alpha(2) = \sqrt{3}/2$,
conjectured to separate the two regions with distinct critical behaviour.
Also shown are the extremal RN solutions with unit charge and charge two.}
\end{figure}
\end{fixy}

\begin{fixy}{-1}
\begin{figure}
\centering
\epsfysize=10cm
\mbox{\epsffile{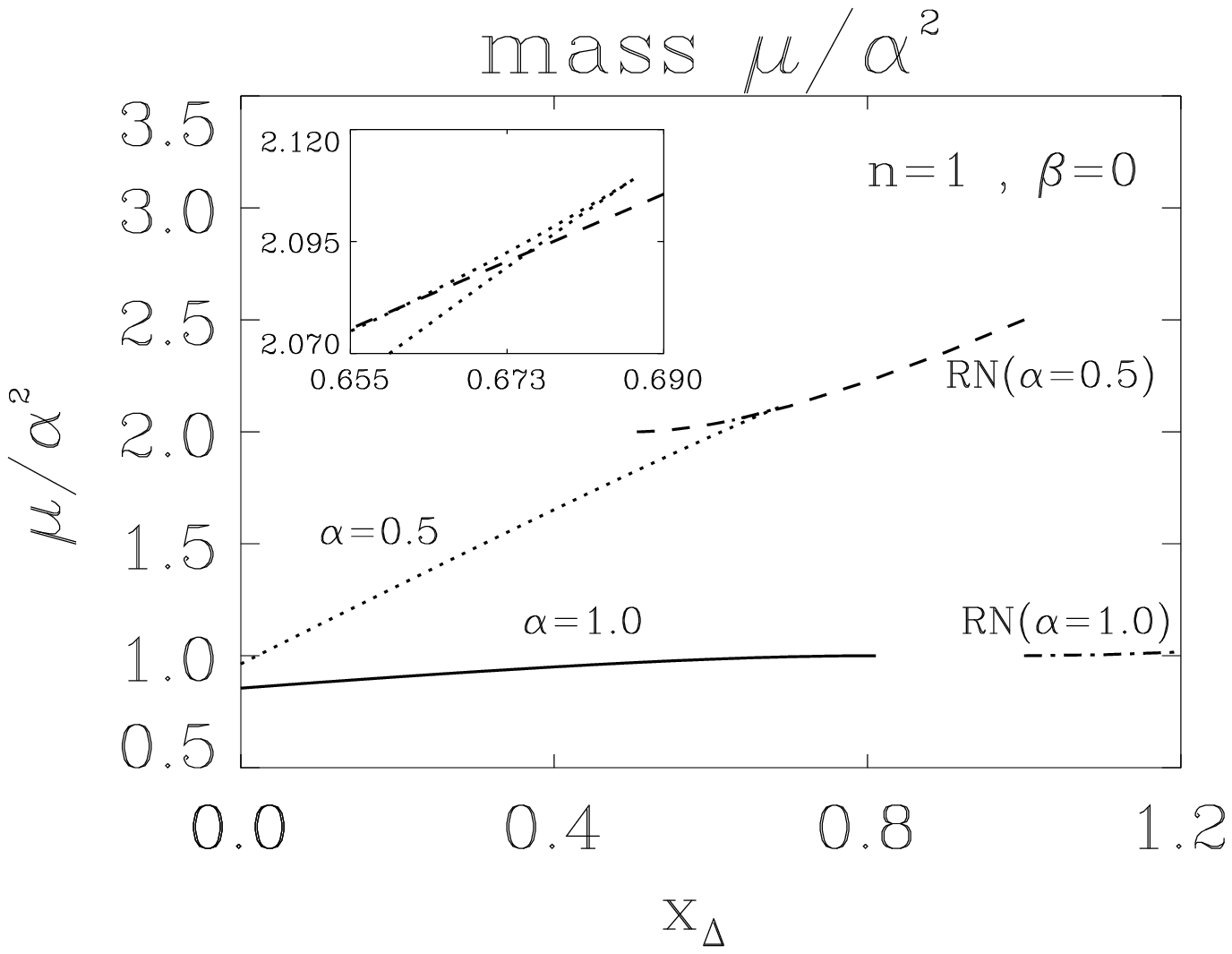}}
\caption{\label{Fig.4} 
The dependence of the mass $\mu/\alpha^2$ 
of the $n=1$ hairy black hole solutions on the area parameter $x_{\Delta}$ 
is shown in the BPS limit for $\alpha=0.5$ and $\alpha=1$.
For comparison, the mass of the corresponding RN solutions is also shown. }
\end{figure}
\end{fixy}

\begin{fixy}{-1}
\begin{figure}
\centering
\epsfysize=11cm
\mbox{\epsffile{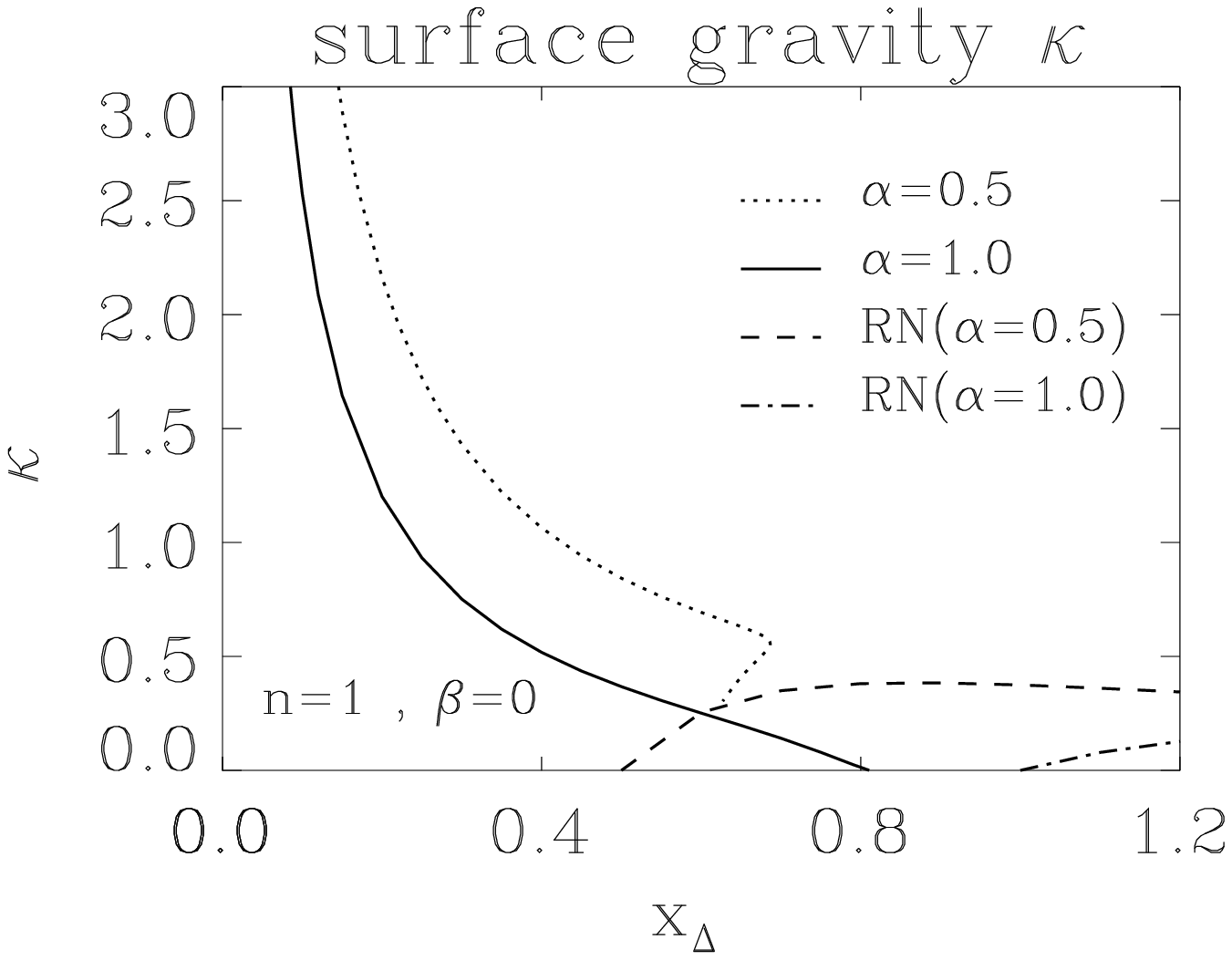}}
\caption{\label{Fig.5} 
The dependence of the surface gravity $\kappa$
of the $n=1$ hairy black hole solutions on the area parameter $x_{\Delta}$ 
is shown in the BPS limit for $\alpha=0.5$ and $\alpha=1$.
For comparison, the surface gravity of the corresponding RN solutions 
is also shown. }
\end{figure}
\end{fixy}

\begin{fixy}{-1}
\begin{figure}
\centering
\epsfysize=10cm
\mbox{\epsffile{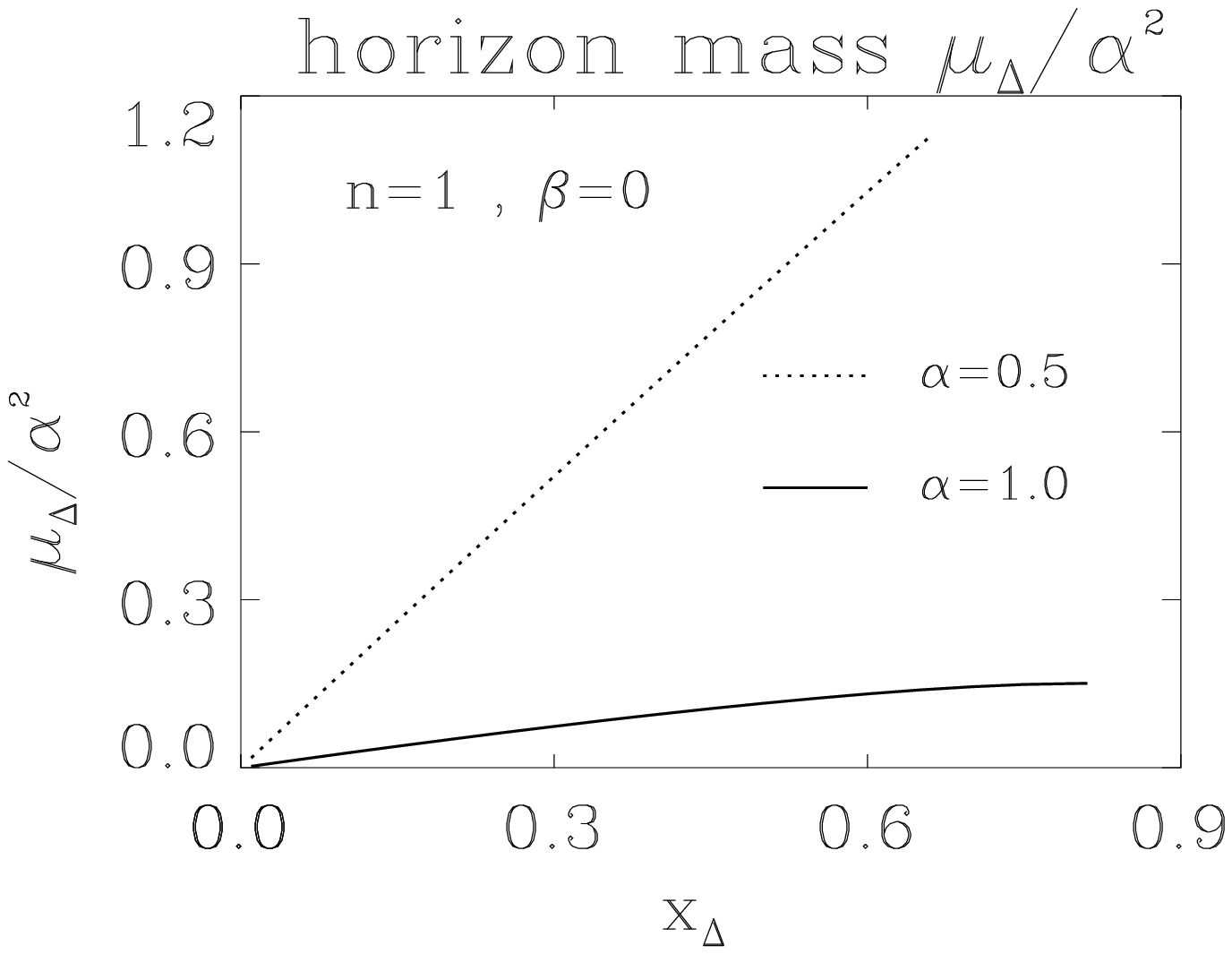}}
\caption{\label{Fig.6} 
The dependence of the horizon mass $\mu_{\Delta}/\alpha^2$ 
of the $n=1$ hairy black hole solutions on the area parameter $x_{\Delta}$ 
is shown in the BPS limit for $\alpha=0.5$ and $\alpha=1$. }
\end{figure}
\end{fixy}

\begin{fixy}{-1}
\begin{figure}
\centering
\epsfysize=10cm
\mbox{\epsffile{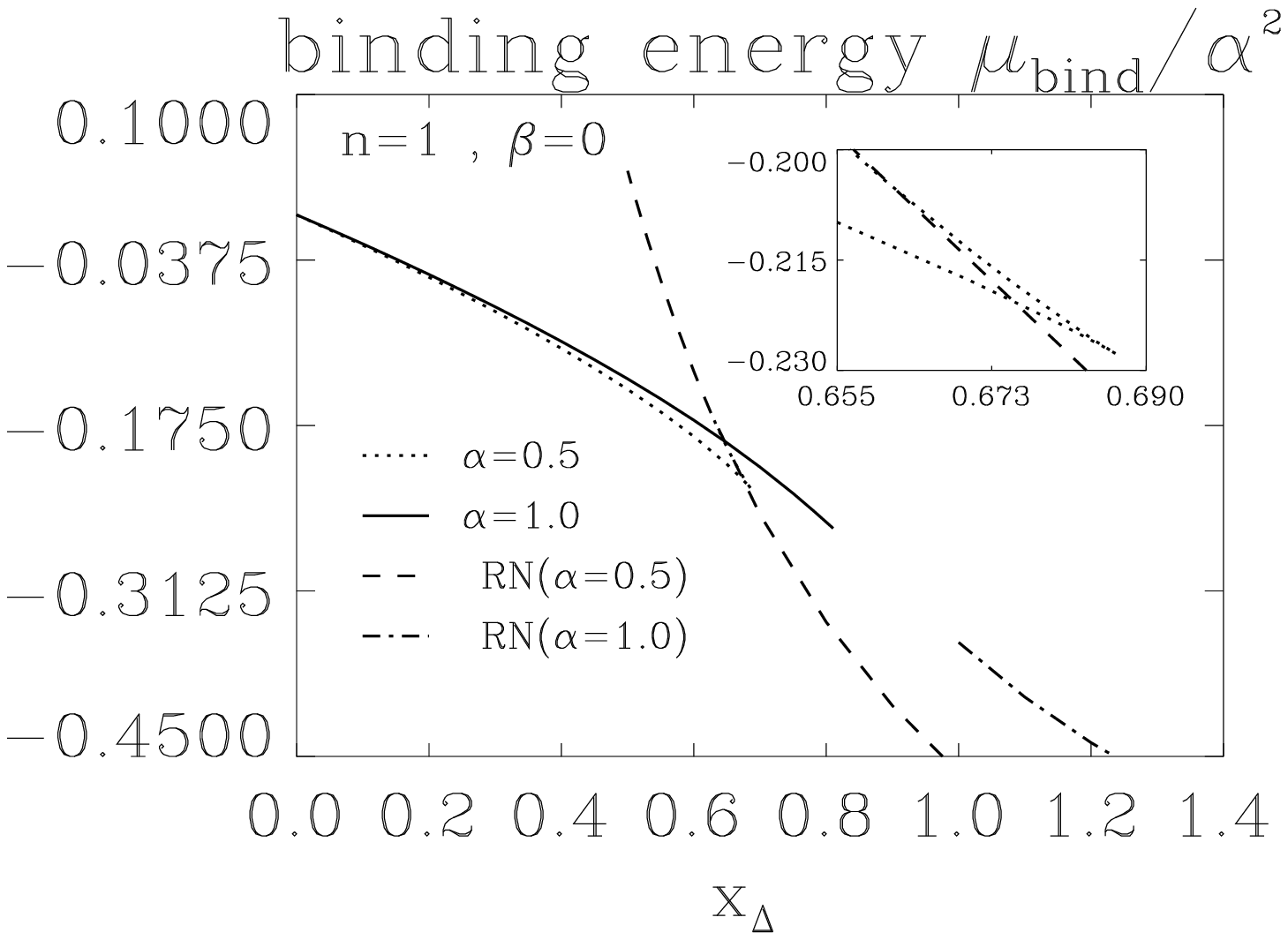}}
\caption{\label{Fig.7} 
The dependence of the binding energy $\mu_{\rm bind}/\alpha^2$
of the $n=1$ hairy black hole solutions on the area parameter $x_{\Delta}$ 
is shown in the BPS limit for $\alpha=0.5$ and $\alpha=1$.
For comparison, the binding energy of the corresponding RN solutions 
is also shown. }
\end{figure}
\end{fixy}

\begin{fixy}{-1}
\begin{figure}
\centering
\epsfysize=10cm
\mbox{\epsffile{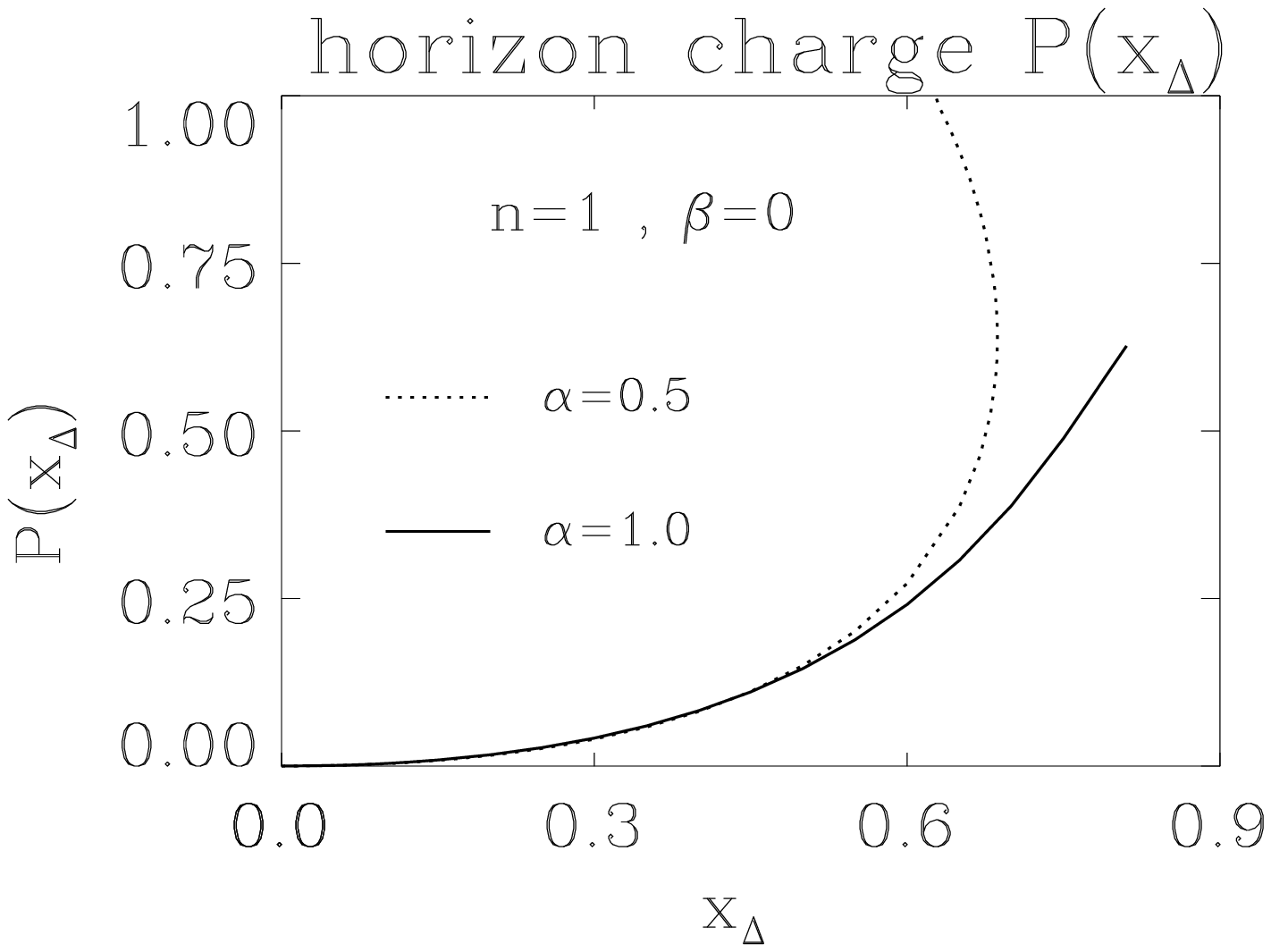}}
\caption{\label{Fig.8} 
The dependence of the non-abelian horizon magnetic charge $P(x_{\Delta})$
of the $n=1$ hairy black hole solutions on the area parameter $x_{\Delta}$ 
is shown in the BPS limit for $\alpha=0.5$ and $\alpha=1$. }
\end{figure}
\end{fixy}

\begin{fixy}{0}
\begin{figure}
\centering
\epsfysize=10cm
\mbox{\epsffile{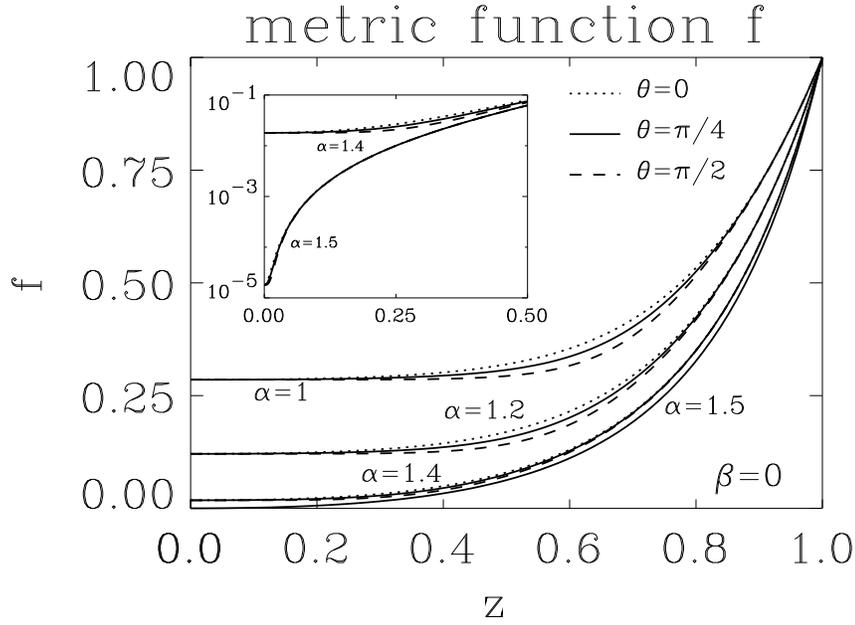}}
\caption{\label{Fig.9a} 
The metric function $f$ of the axially symmetric $n=2$ monopole solution
is shown as a function of the
compactified dimensionless isotropic coordinate
$z=x/(1+x)$ for the angles $\theta=0$, $\theta=\pi/4$ and $\theta=\pi/2$
in the BPS limit
for four values of $\alpha$ along the multimonopole branch,
$\alpha=1$, 1.2, 1.4 and 1.499. }
\end{figure}

\begin{figure}
\centering
\epsfysize=10cm
\mbox{\epsffile{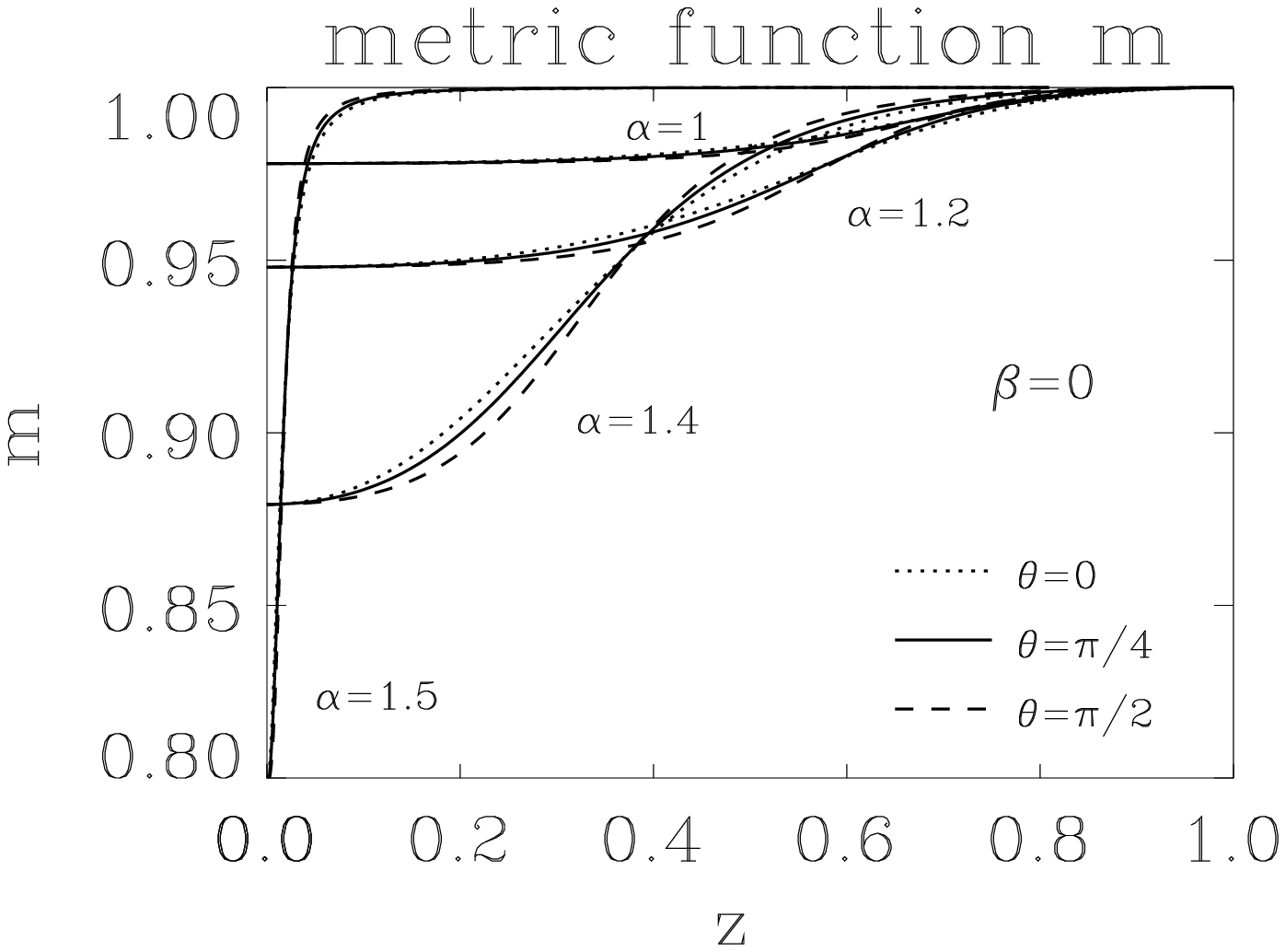}}
\caption{\label{Fig.9b} The same as Fig.~9a for the metric function $m$.}
\end{figure}

\begin{figure}
\centering
\epsfysize=10cm
\mbox{\epsffile{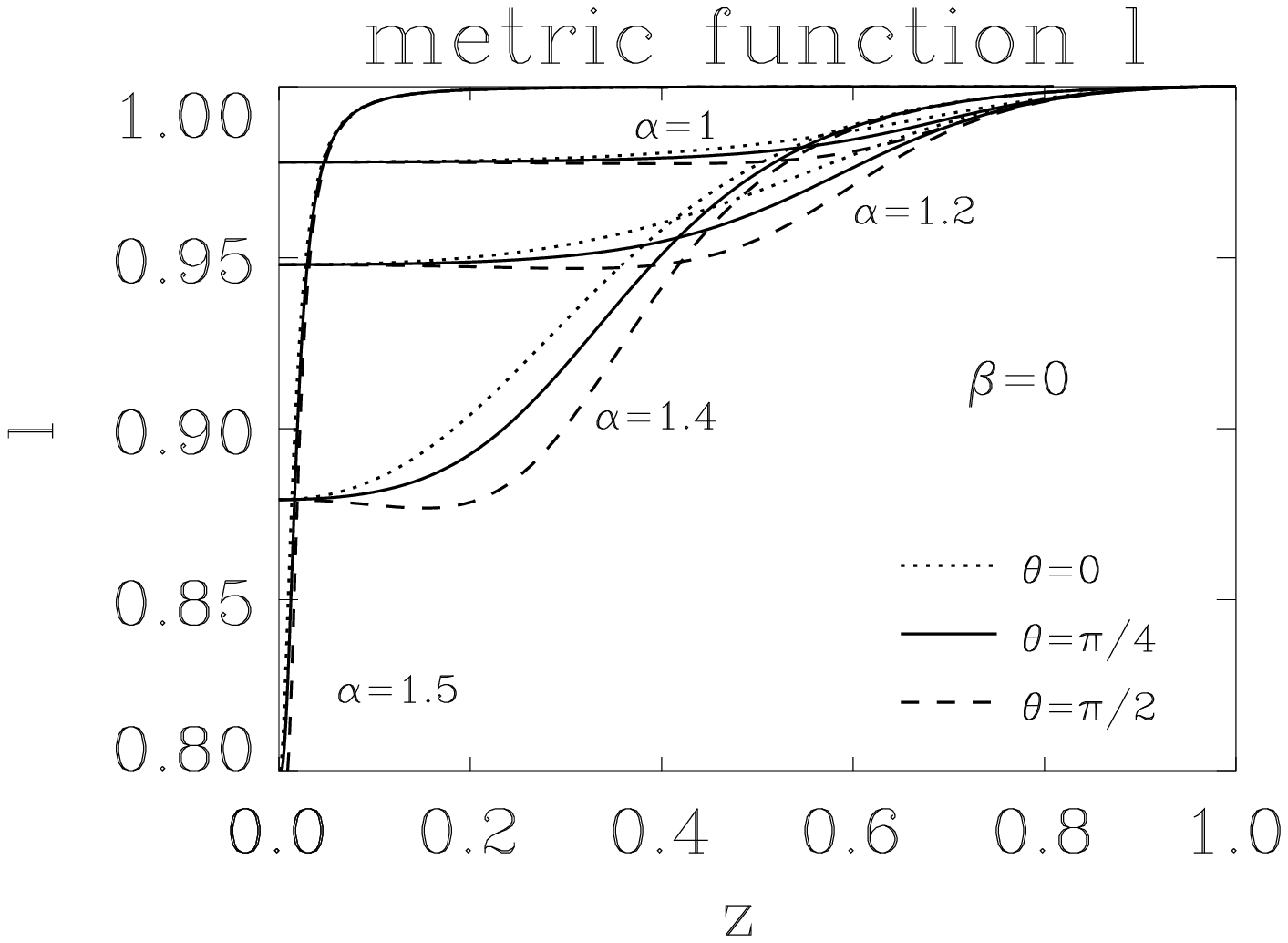}}
\caption{\label{Fig.9c} The same as Fig.~9a for the metric function $l$.
}
\end{figure}

\begin{figure}
\centering
\epsfysize=10cm
\mbox{\epsffile{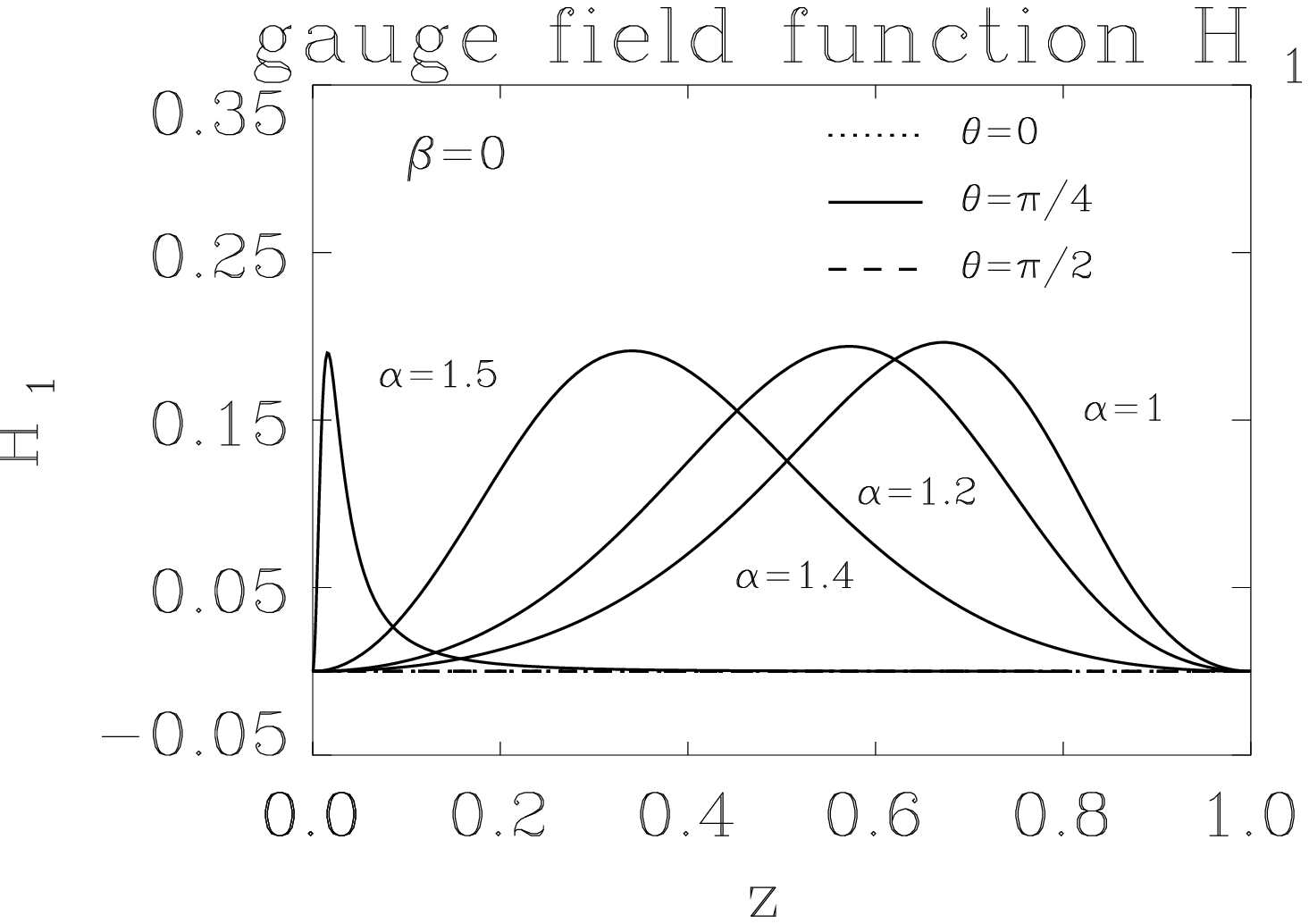}}
\caption{\label{Fig.9d} The same as Fig.~9a for the gauge field function $H_1$.}
\end{figure}

\begin{figure}
\centering
\epsfysize=10cm
\mbox{\epsffile{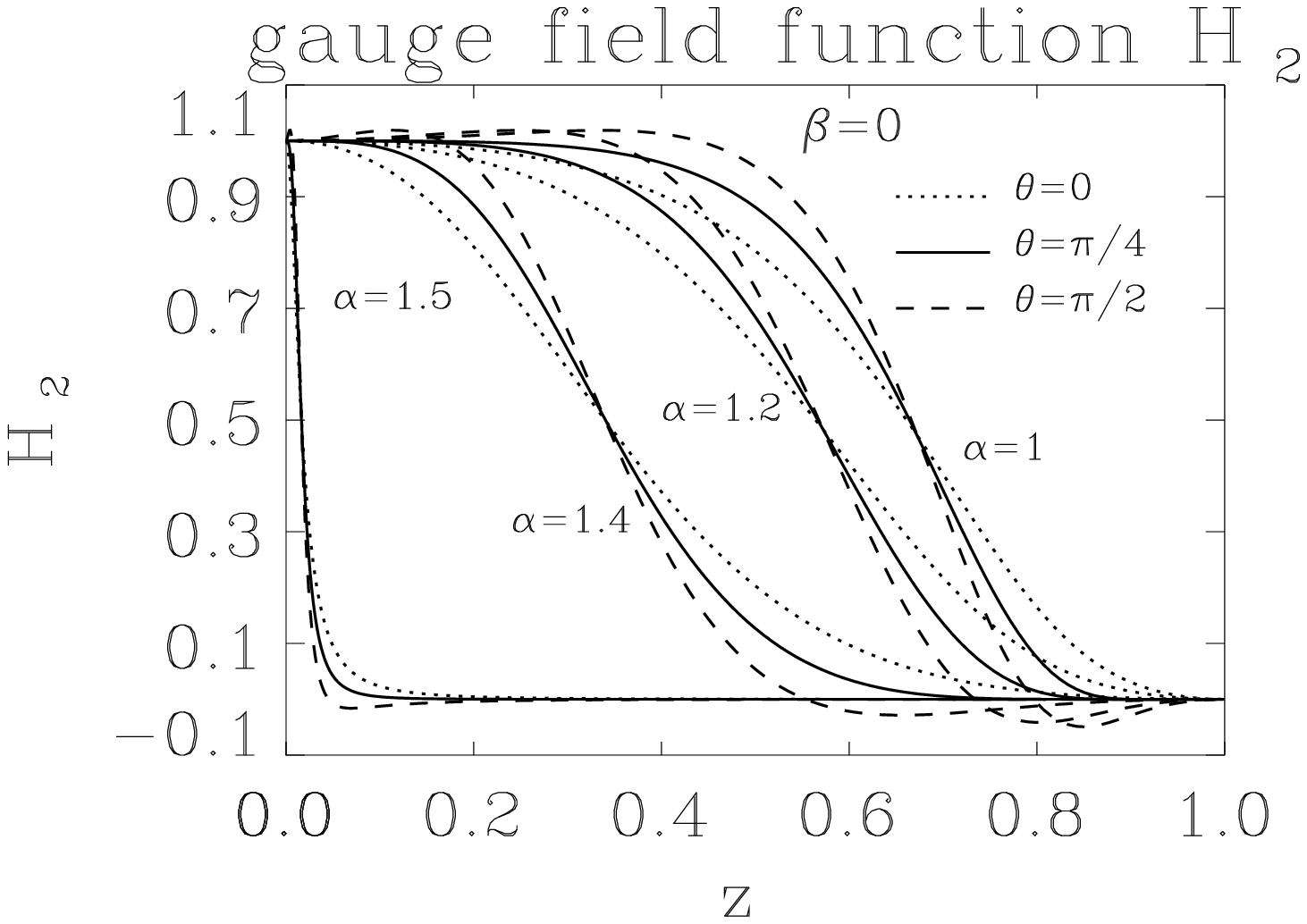}}
\caption{\label{Fig.9e} The same as Fig.~9a for the gauge field function $H_2$.}
\end{figure}

\begin{figure}
\centering
\epsfysize=10cm
\mbox{\epsffile{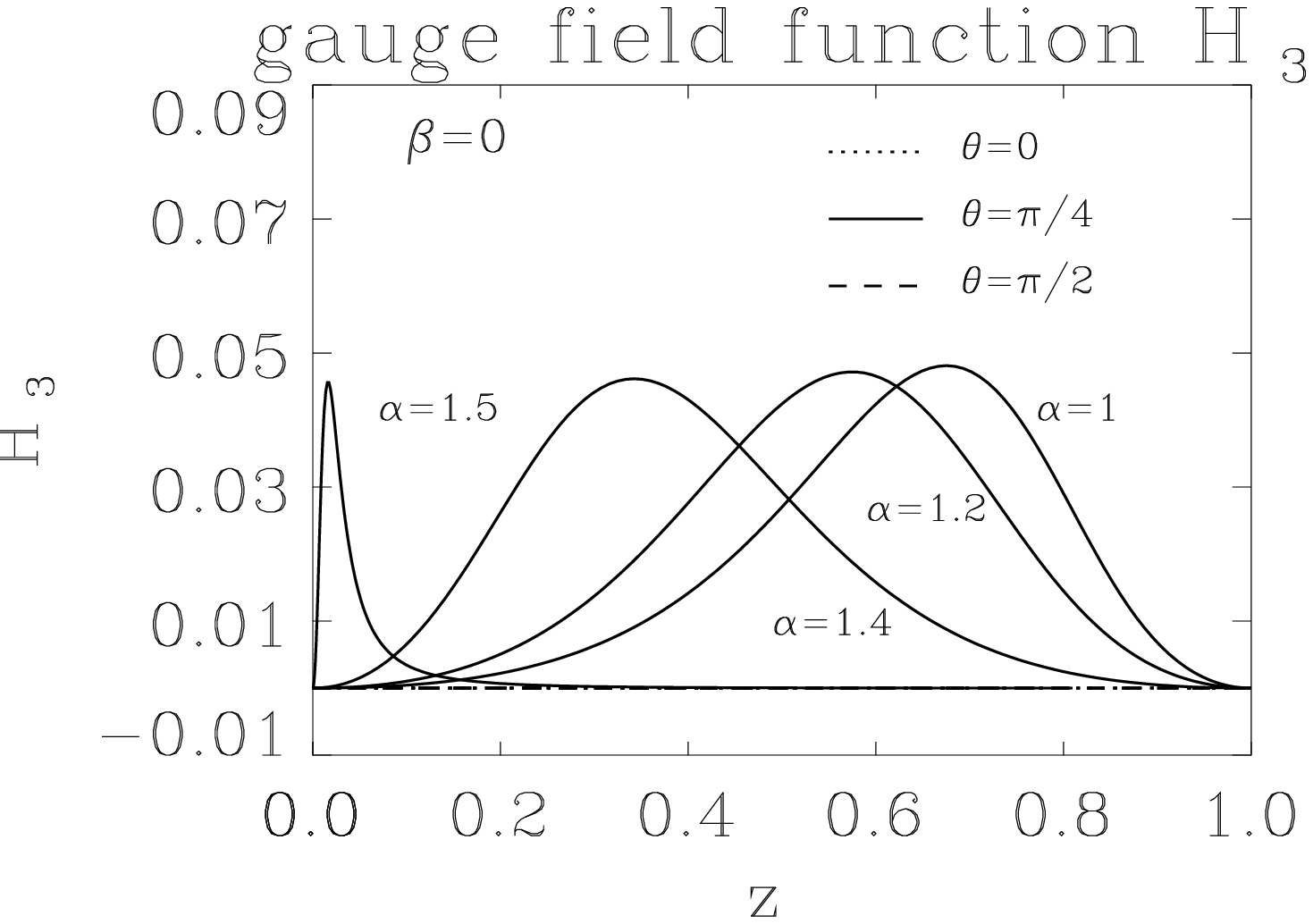}}
\caption{\label{Fig.9f} The same as Fig.~9a for the gauge field function $H_3$.}
\end{figure}

\begin{figure}
\centering
\epsfysize=10cm
\mbox{\epsffile{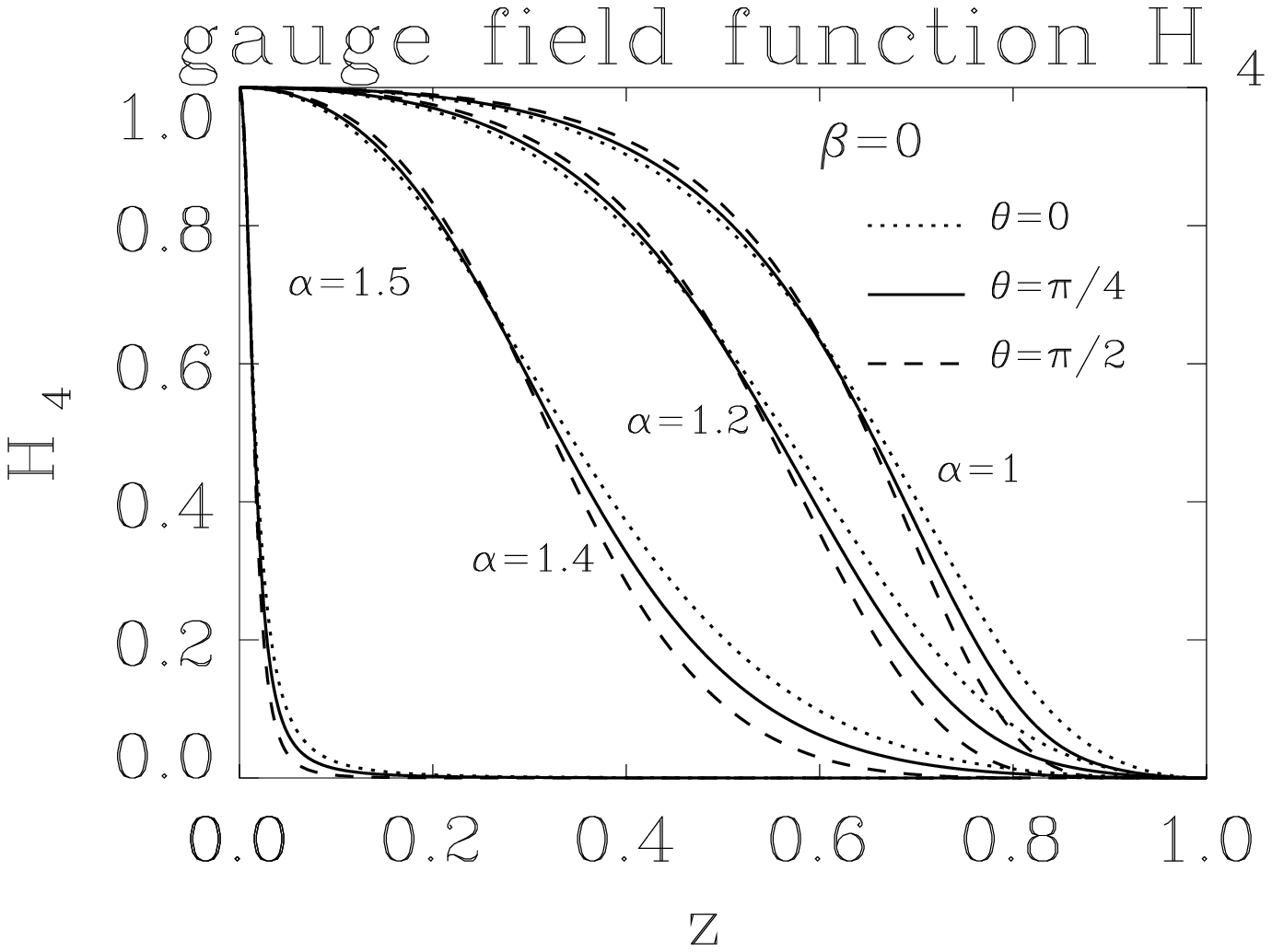}}
\caption{\label{Fig.9g} The same as Fig.~9a for the gauge field function $H_4$.}
\end{figure}

\begin{figure}
\centering
\epsfysize=10cm
\mbox{\epsffile{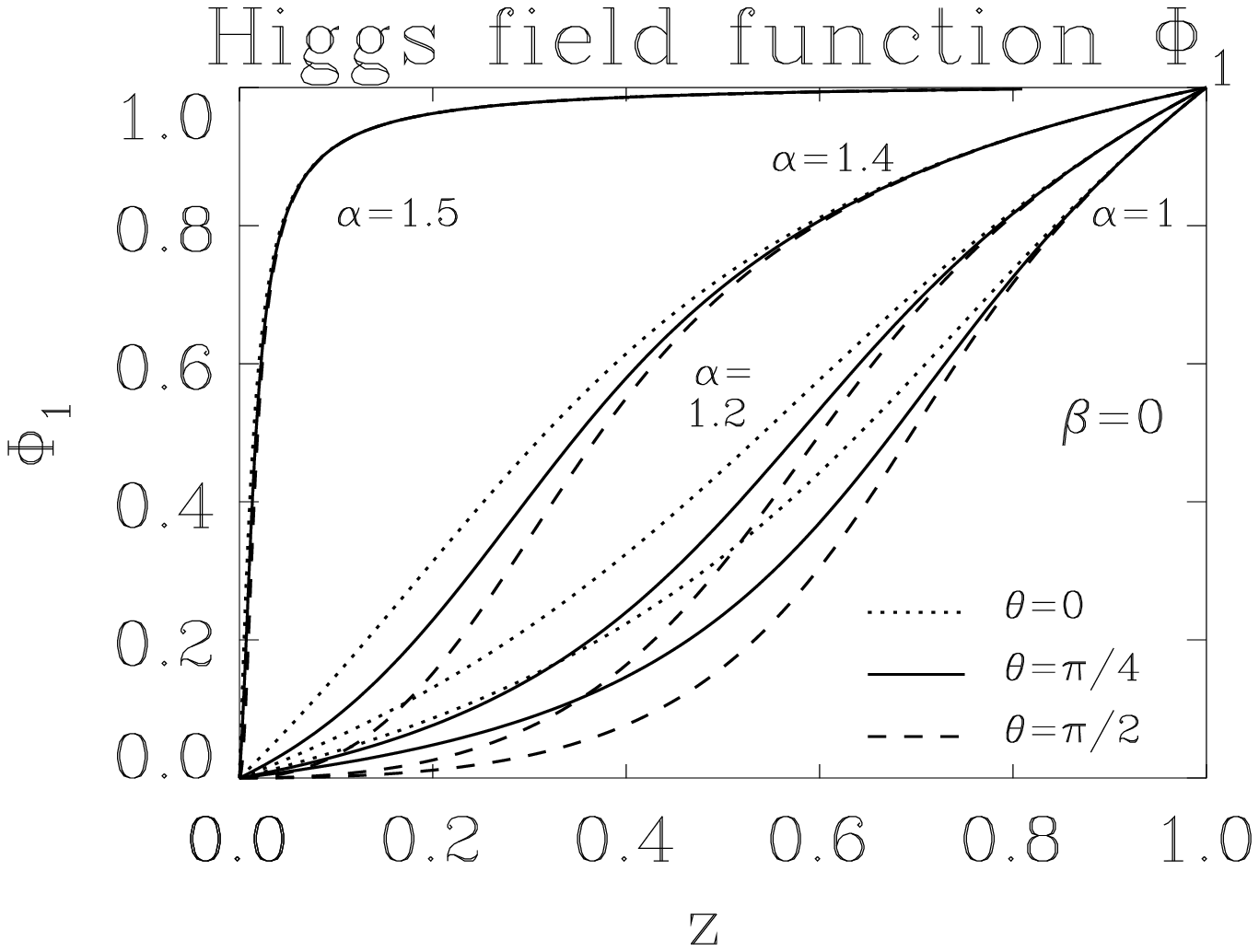}}
\caption{\label{Fig.9h} The same as Fig.~9a for the Higgs field function 
$\Phi_1$.}
\end{figure}

\begin{figure}
\centering
\epsfysize=10cm
\mbox{\epsffile{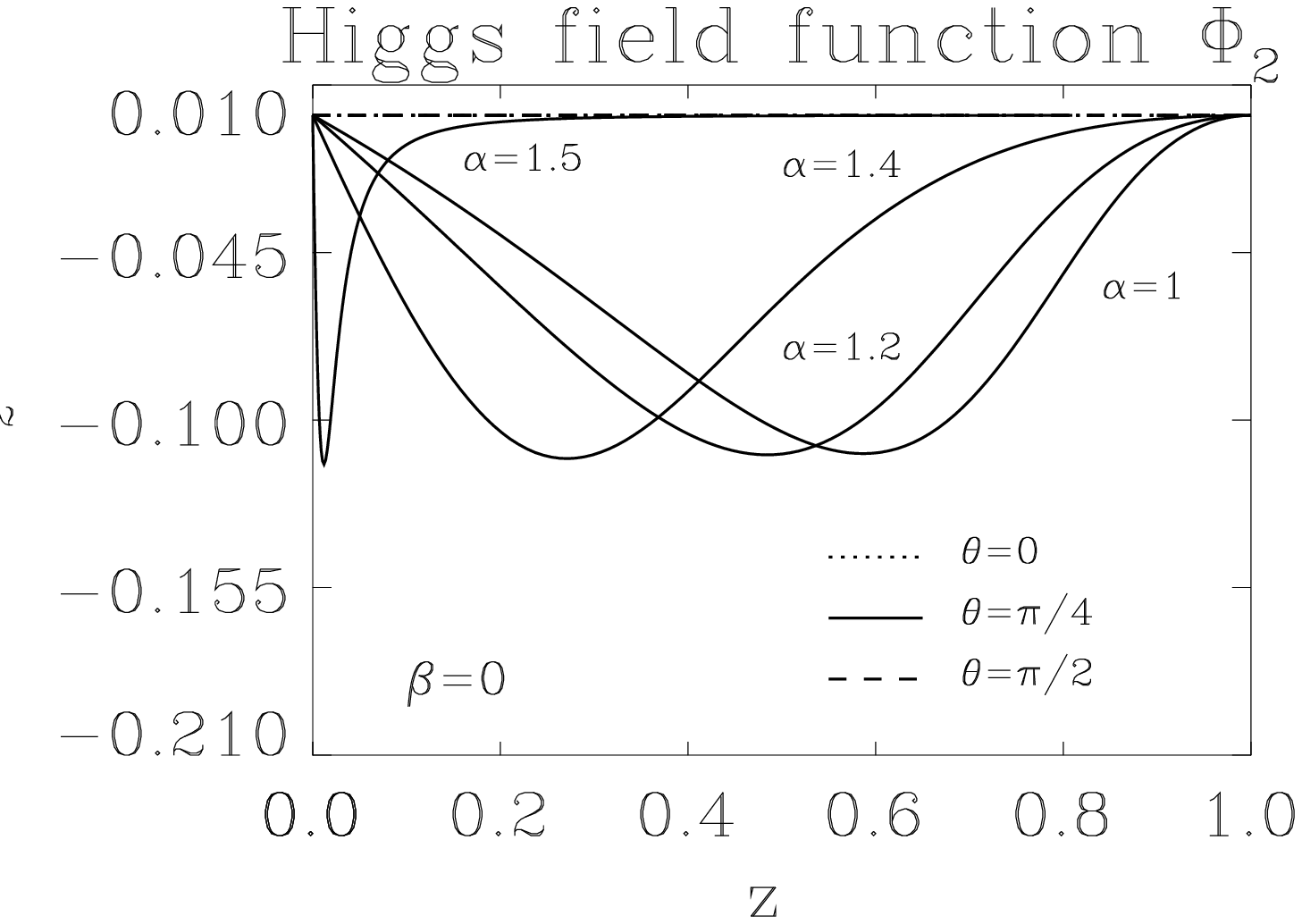}}
\caption{\label{Fig.9i} The same as Fig.~9a for the Higgs field function 
$\Phi_2$.}
\end{figure}
\end{fixy}


\begin{fixy}{0}
\begin{figure}
\centering
\epsfysize=10cm
\mbox{\epsffile{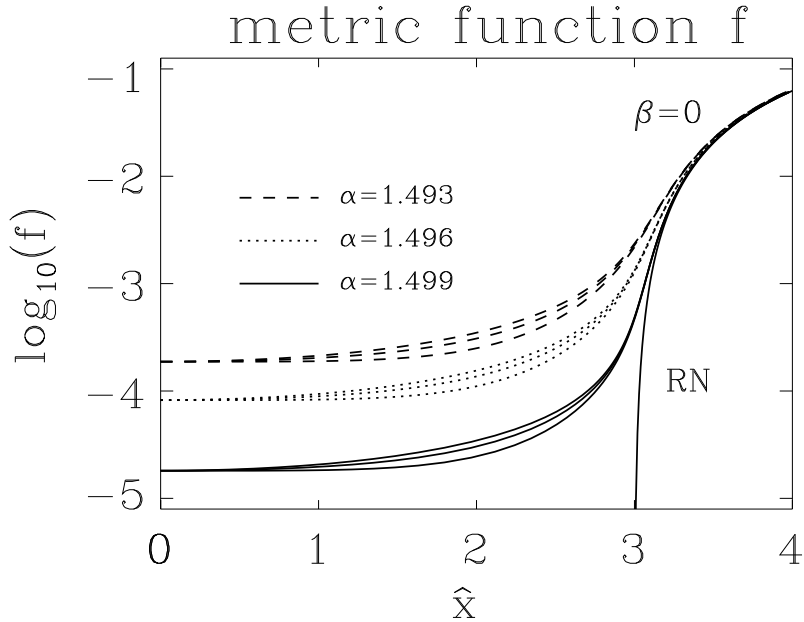}}
\caption{\label{Fig.10a}
The metric function $f$ of the axially symmetric $n=2$ monopole solution
is shown as a function of the
auxillary Schwarzschild-like coordinate $\hat{x}= x\sqrt{m/f}$,
for the angles $\theta=0$, $\theta=\pi/4$ and $\theta=\pi/2$
in the BPS limit
for $\alpha=1.493$, $1.496$ and $1.499$, close to the critical $\alpha$.
Also shown is the metric function $f$ of the RN solution with charge 
$P=2$ and area parameter $x_\Delta=1.5$. }
\end{figure}

\begin{figure}
\centering
\epsfysize=10cm
\mbox{\epsffile{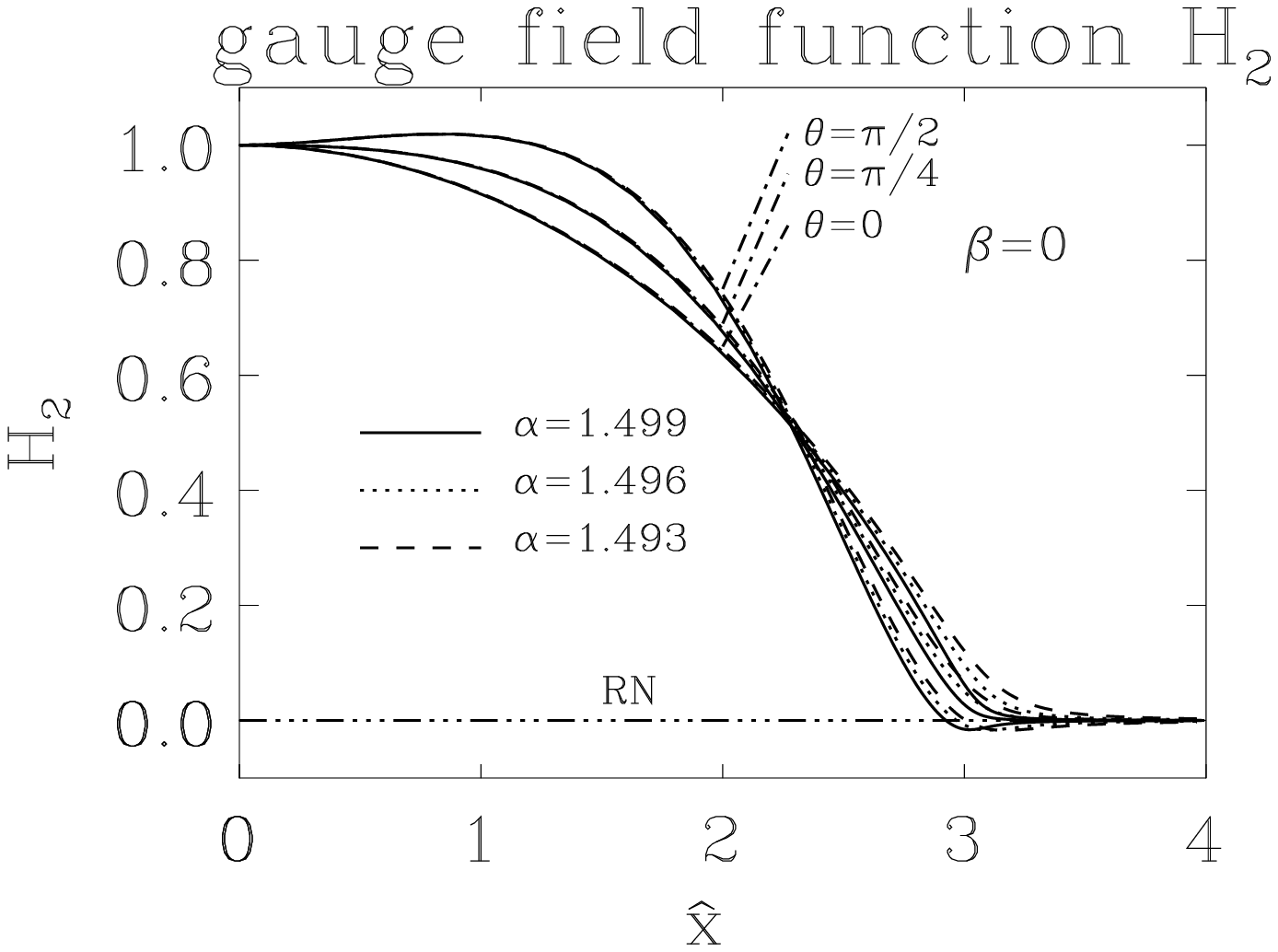}}
\caption{\label{Fig.10b}
The same as Fig. 10a for the gauge field function $H_2$. }
\end{figure}

\begin{figure}
\centering
\epsfysize=10cm
\mbox{\epsffile{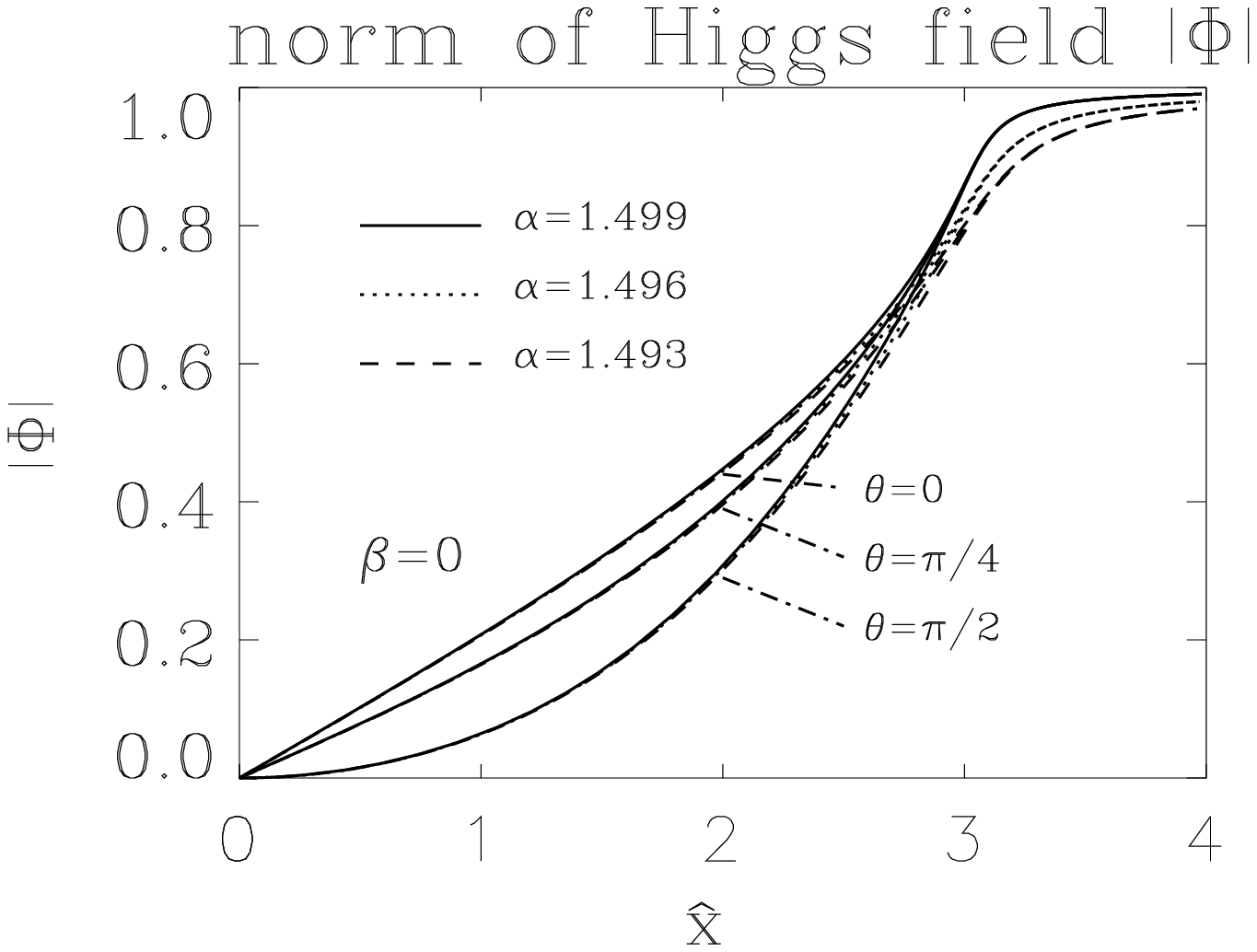}}
\caption{\label{Fig.10c}
The same as Fig. 10a for the norm of the Higgs field 
$|\Phi|= \sqrt{\Phi_1^2+\Phi_2^2}$. }
\end{figure}
\end{fixy}

\begin{fixy}{-1}
\begin{figure}
\centering
\epsfysize=10cm
\mbox{\epsffile{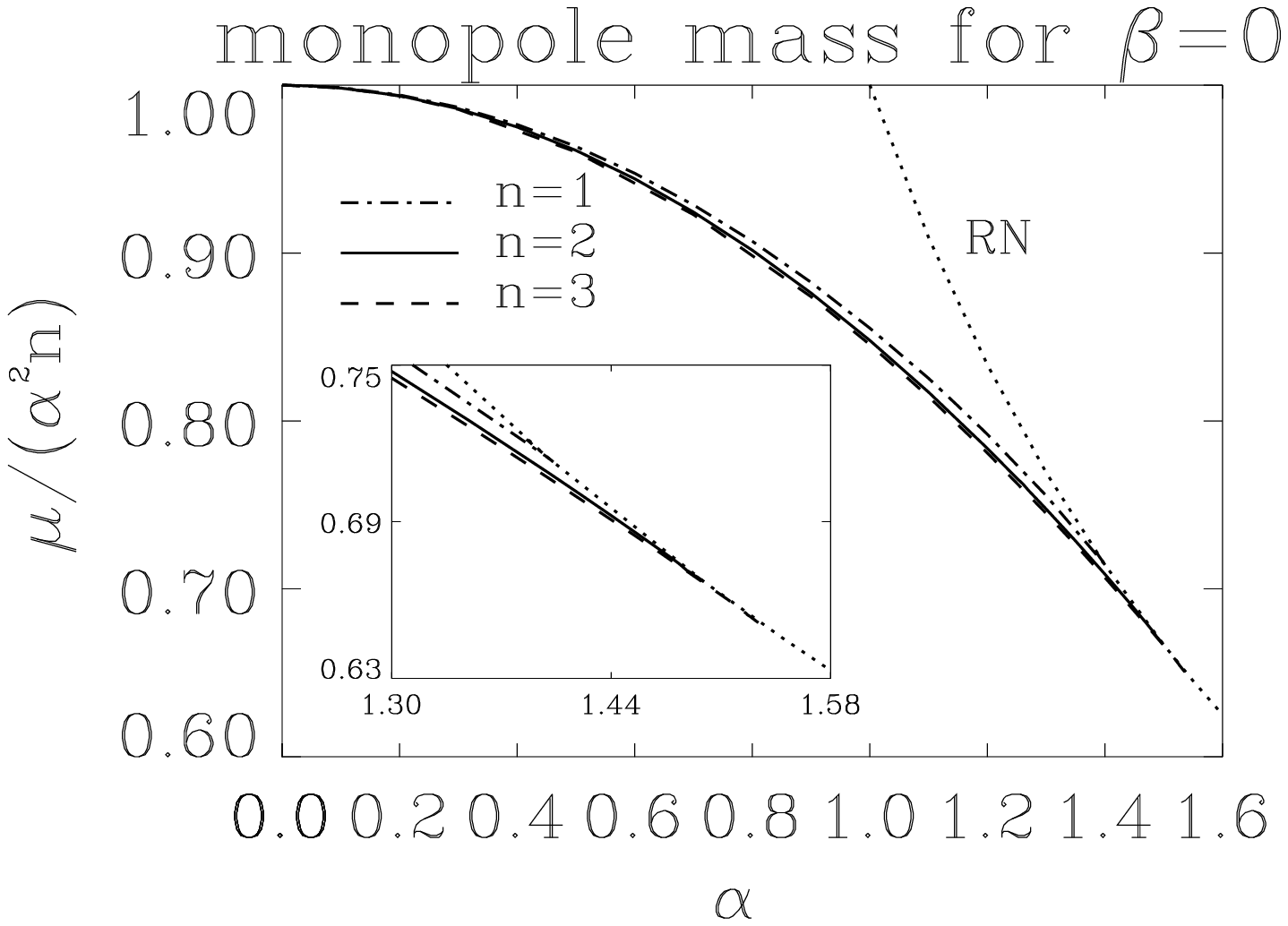}}
\caption{\label{Fig.11} 
The dependence of the mass {\it per unit charge} $\mu/(\alpha^2n)$ 
of the hairy black hole solutions on the parameter $\alpha$
is shown in the BPS limit for magnetic charge $n=1$, $2$ and $3$.
For comparison, the mass {\it per unit charge}
of the extremal RN solutions is also shown. }
\end{figure}
\end{fixy}


\begin{fixy}{-1}
\begin{figure}
\centering
\epsfysize=10cm
\mbox{\epsffile{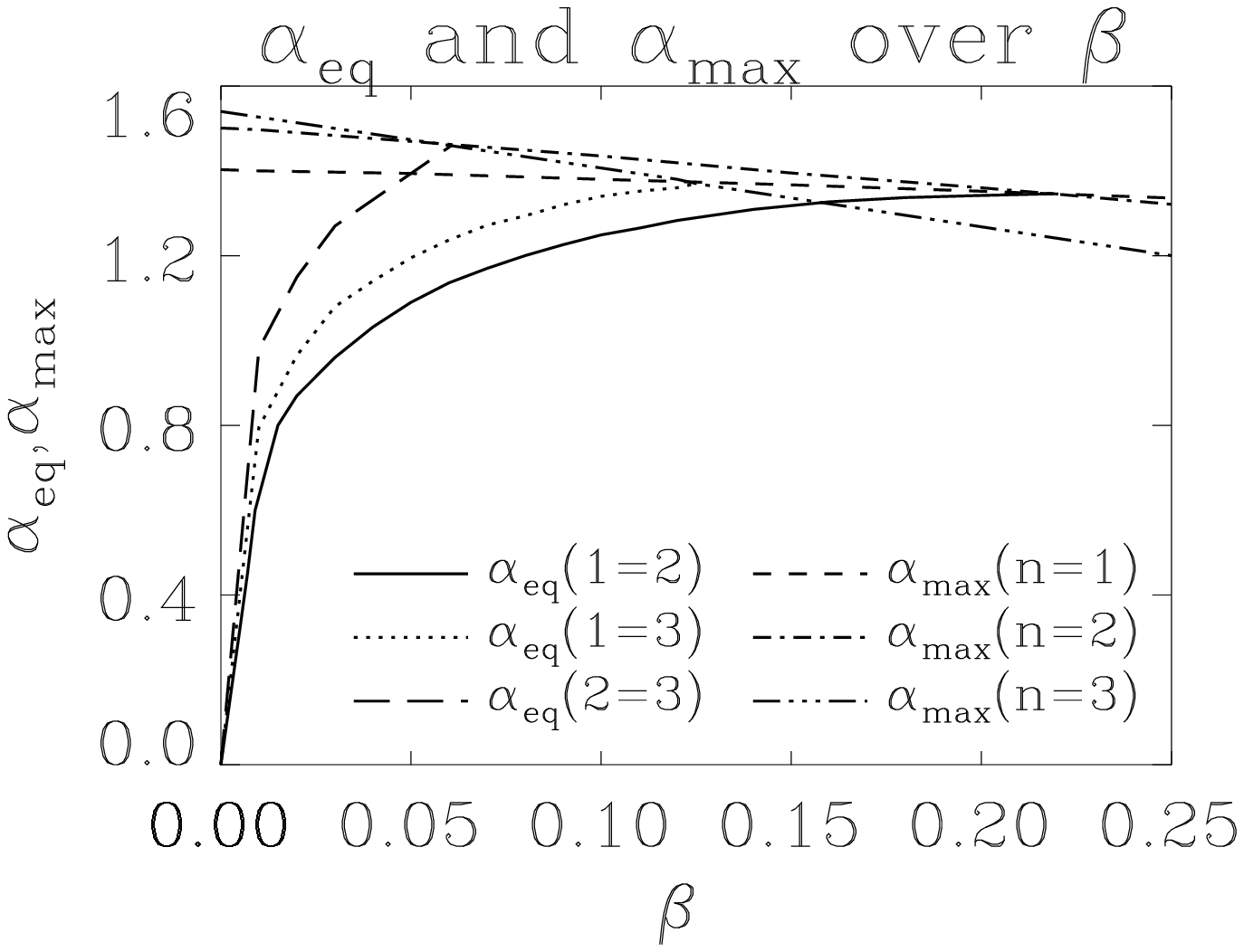}}
\caption{\label{Fig.12} 
The dependence of the maximal values $\alpha_{\rm max}(n)$
of the hairy black hole solutions on the parameter $\beta$ 
is shown for magnetic charge $n=1$, $2$ and $3$.
Also shown are the equilibrium values $\alpha_{\rm eq}(n_1=n_2)$,
in particular $\alpha_{\rm eq}(1=2)$, $\alpha_{\rm eq}(1=3)$ 
and $\alpha_{\rm eq}(2=3)$,
for which the mass {\it per unit charge} of the charge $n_1$ and 
and the charge $n_2$ (multi)monopole equal one another. }
\end{figure}
\end{fixy}

\begin{fixy}{-1}
\begin{figure}
\centering
\epsfysize=19cm
\mbox{\epsffile{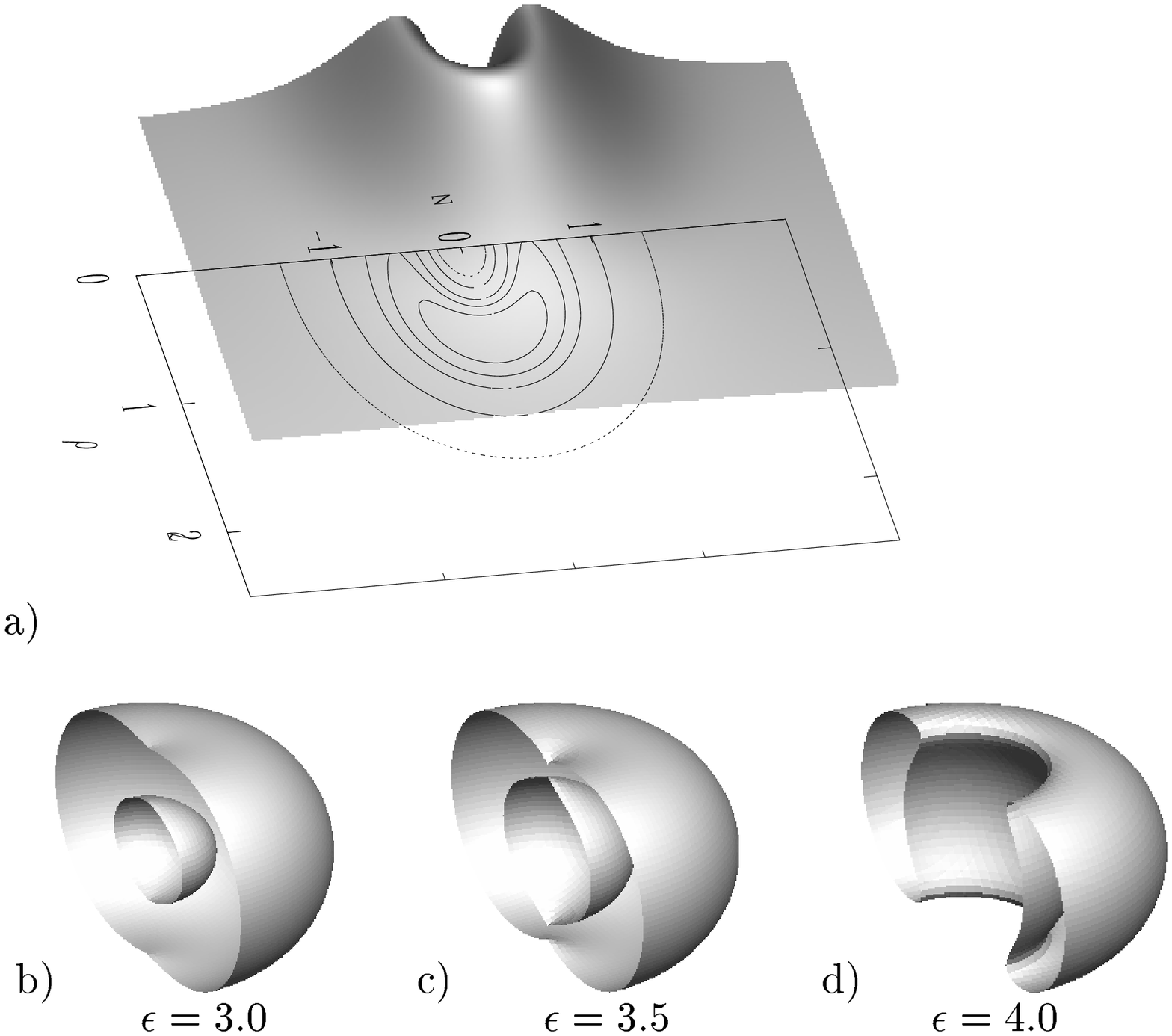}}
\caption{\label{Fig.13}
a) The energy density $\epsilon = - T_0^0$ is shown for 
the hairy black hole solution with magnetic charge $n=2$
and area parameter $x_{\Delta}=1.0$ 
in the BPS limit for $\alpha=1.0$
in a three-dimensional plot and a contour plot with axes $\rho$ and $z$.
b)-d) Surfaces of constant energy density $\epsilon = - T_0^0$ 
are shown for the solution of a). }
\end{figure}
\end{fixy}

\newpage

\begin{fixy}{-1}
\begin{figure}
\centering
\epsfysize=19cm
\mbox{\epsffile{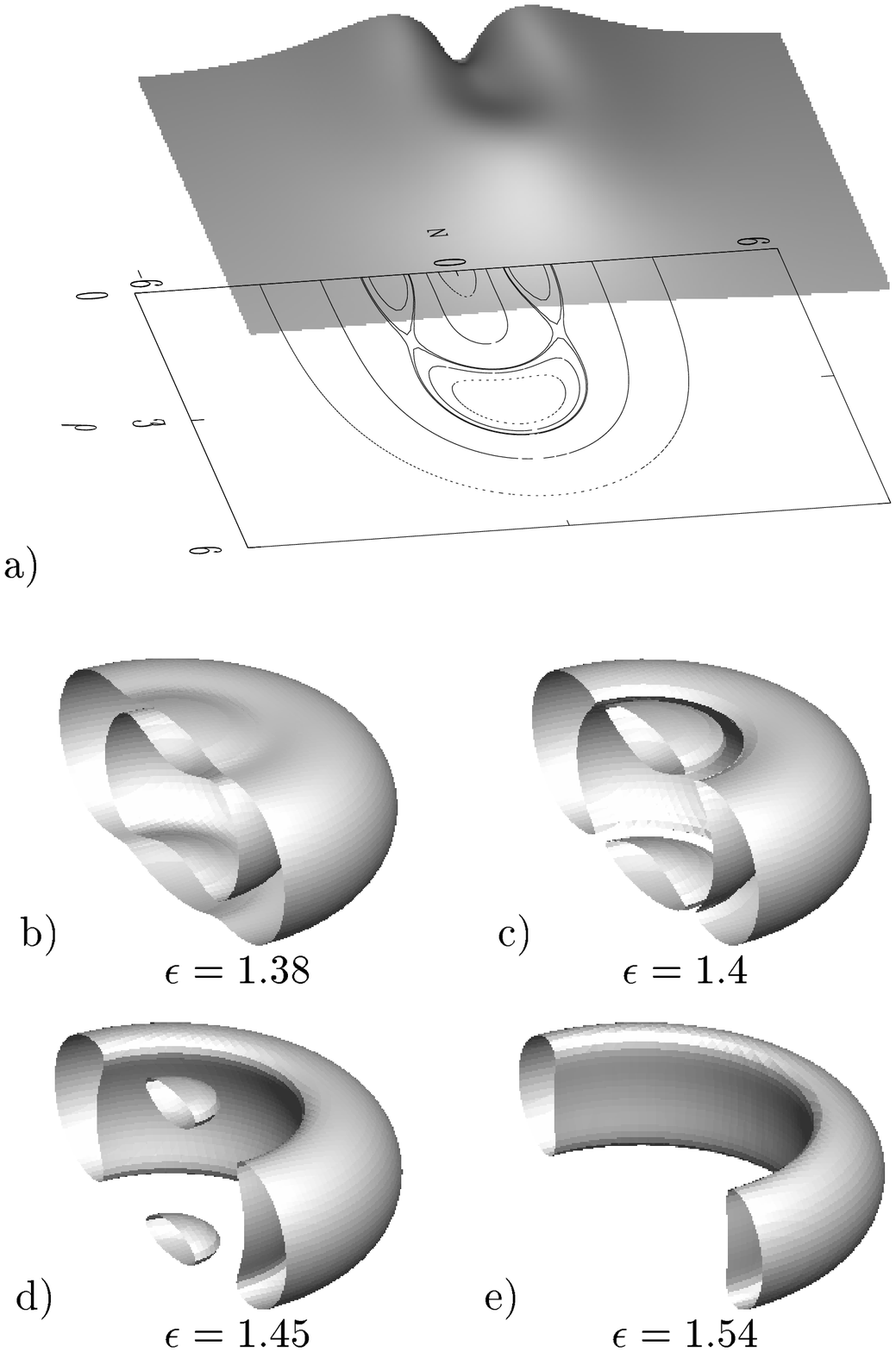}}
\caption{\label{Fig.14} 
The same as Fig.~13 for the hairy black hole solution with
$n=3$, $x_{\Delta}=0.5$ and $\alpha=0.5$. }
\end{figure}
\end{fixy}

\newpage

\begin{fixy}{-1}
\begin{figure}
\centering
\epsfysize=10cm
\mbox{\epsffile{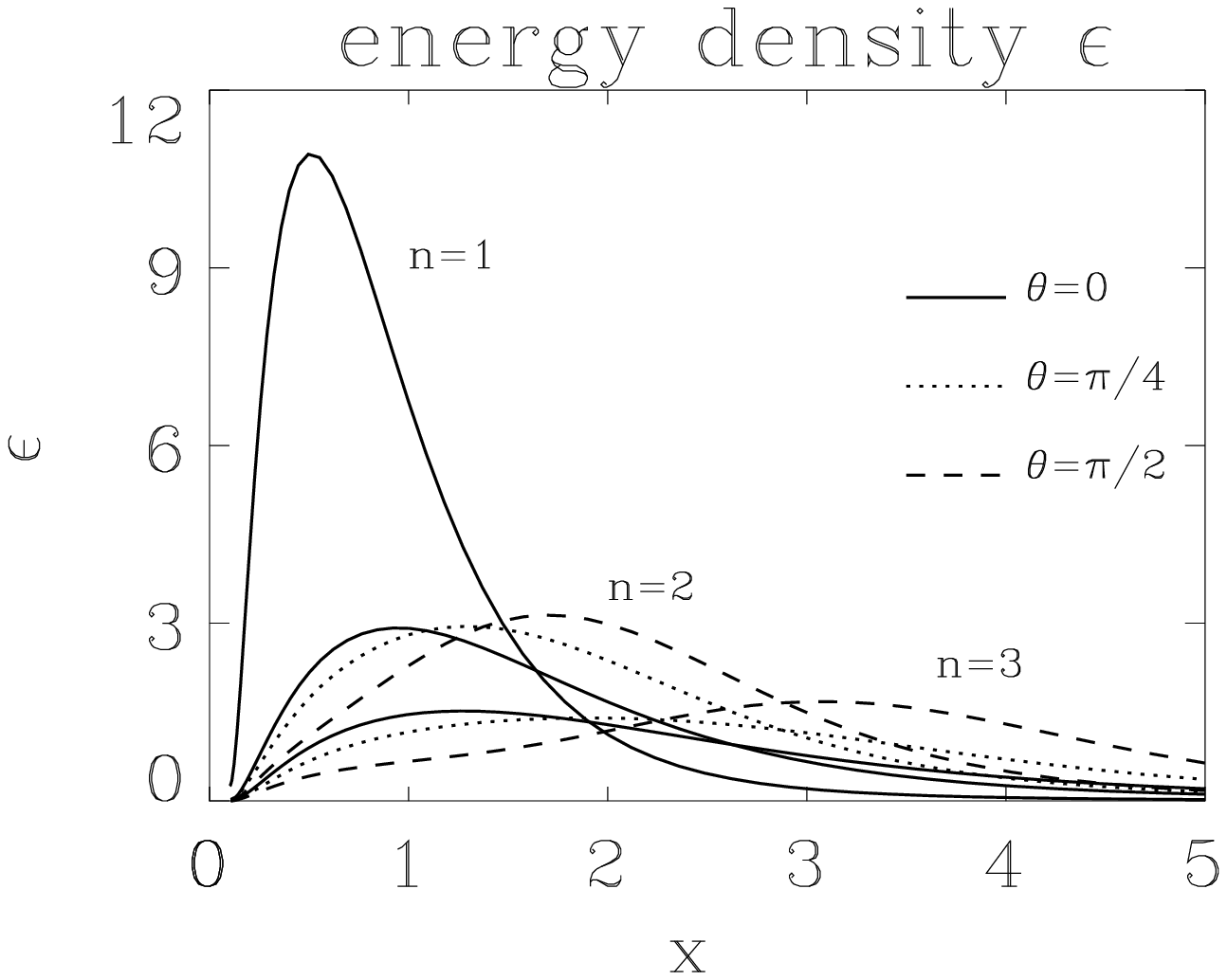}}
\caption{\label{Fig.15} 
The energy density $\epsilon=-T_0^0$
of the hairy black hole solutions
with magnetic charge $n=1$, $2$ and $3$
and area parameter $x_{\Delta}=0.5$
is shown as a function of the
dimensionless isotropic coordinate $x$
for the angles $\theta=0$, $\theta=\pi/4$ and $\theta=\pi/2$
in the BPS limit for $\alpha=0.5$. }
\end{figure}
\end{fixy}

\begin{fixy}{-1}
\begin{figure}
\centering
\epsfysize=10cm
\mbox{\epsffile{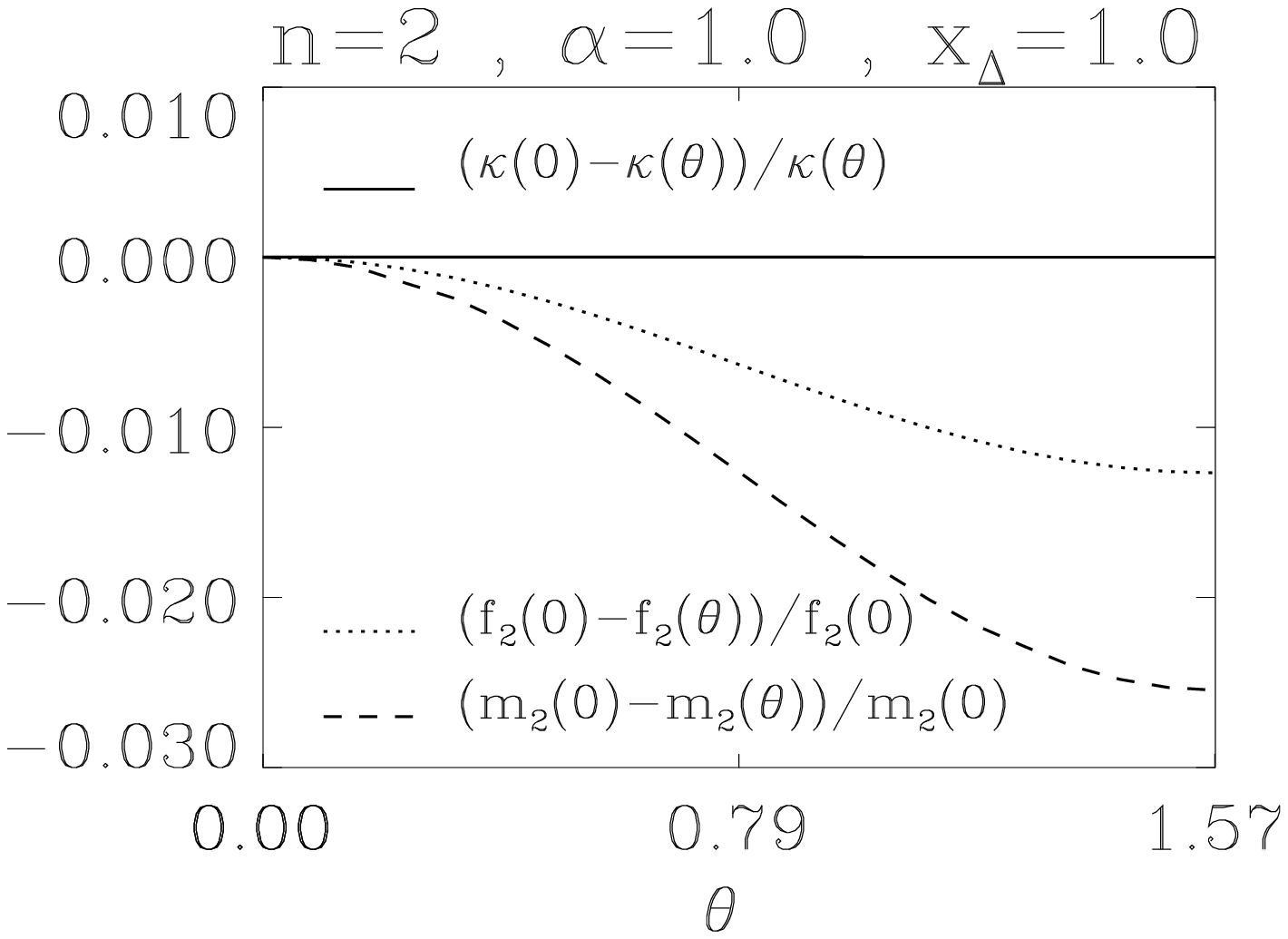}}
\caption{\label{Fig.16} 
The angle-independence of the surface gravity $\kappa$ 
is shown for the hairy black hole solution with magnetic charge
$n=2$ and area parameter $x_{\Delta}=1.0$
in the BPS limit for $\alpha=1.0$.
Also shown is the angle-dependence of the normalized expansion coefficients 
$f_2$ and $m_2$ of the metric functions $f$ and $m$. }
\end{figure}
\end{fixy}

\begin{fixy}{-1}
\begin{figure}
\centering
\epsfysize=10cm
\mbox{\epsffile{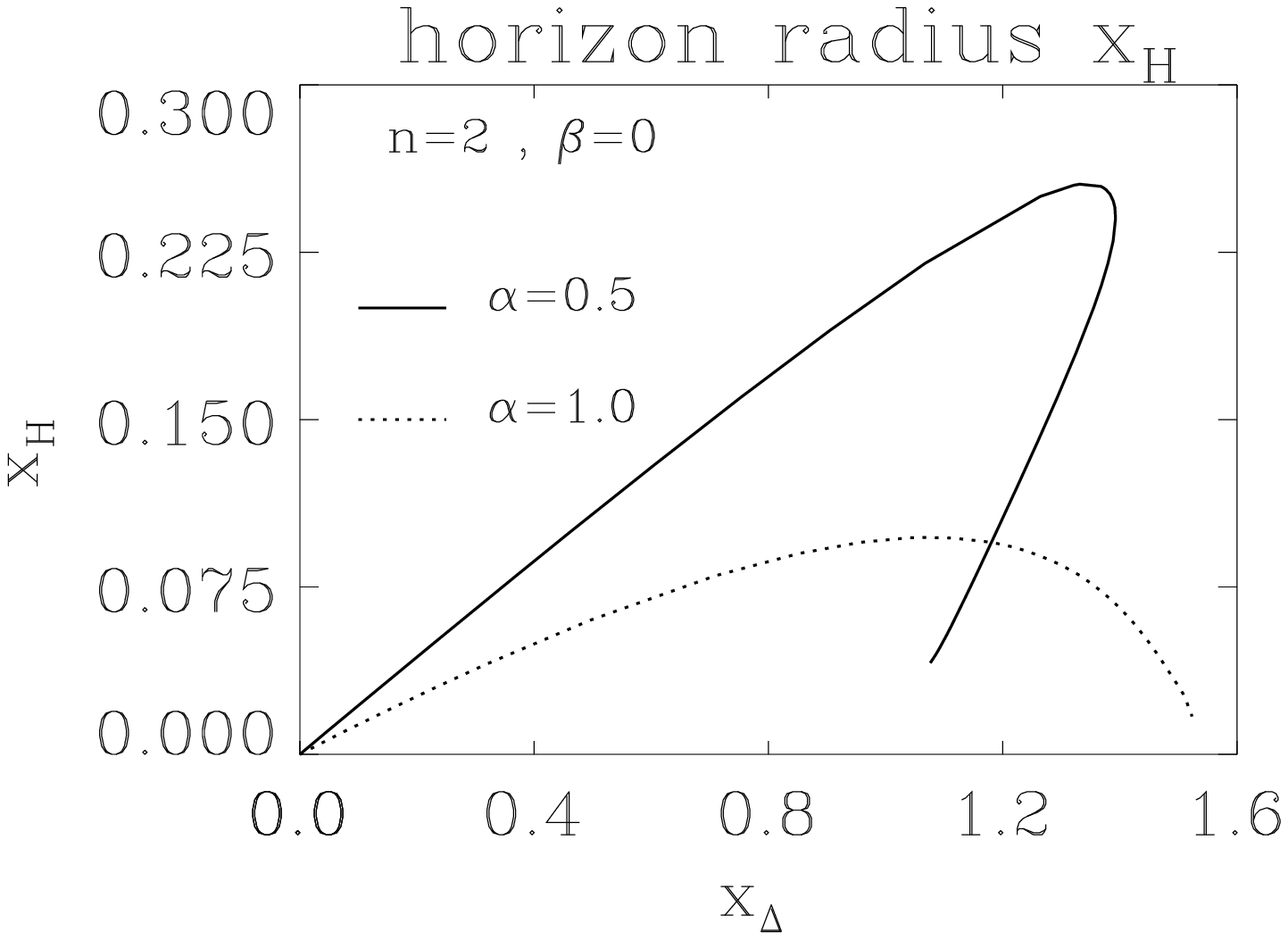}}
\caption{\label{Fig.17} 
The dependence of the horizon radius in isotropic coordinates $x_{\rm H}$ 
of the $n=2$ hairy black hole solutions on the area parameter $x_{\Delta}$ 
is shown in the BPS limit for $\alpha=0.5$ and $\alpha=1$. }
\end{figure}
\end{fixy}


\begin{fixy}{0}
\begin{figure}
\centering
\epsfysize=10cm
\mbox{\epsffile{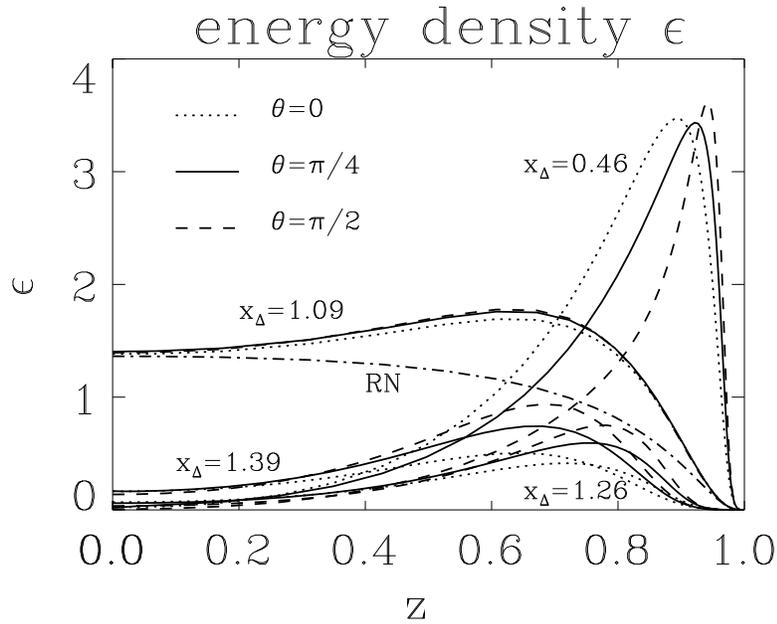}}
\caption{\label{Fig.18a} 
The energy density $\epsilon=-T_0^0$
of the $n=2$ hairy black hole solution
is shown as a function of the dimensionless 
compactified coordinate $z=1-\frac{x_{\rm H}}{x}$
for the angles $\theta=0$, $\theta=\pi/4$ and $\theta=\pi/2$
in the BPS limit
for four values of $\alpha$ along the hairy black hole branch,
in particular for a value close to the maximal value of $x_\Delta$,
$x_{\Delta, \rm max} \approx 1.39$, and
for a value close to the critical value of $x_\Delta$,
$x_{\Delta, \rm cr} \approx 1.07$. }
\end{figure}

\begin{figure}
\centering
\epsfysize=10cm
\mbox{\epsffile{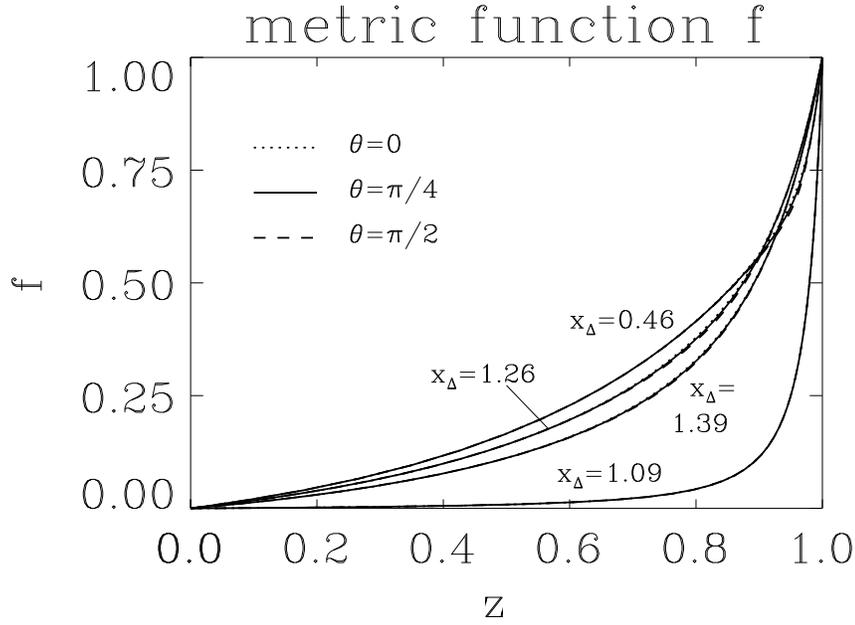}}
\caption{\label{Fig.18b} 
The same as Fig. 18a for the metric function $f$. }
\end{figure}

\begin{figure}
\centering
\epsfysize=10cm
\mbox{\epsffile{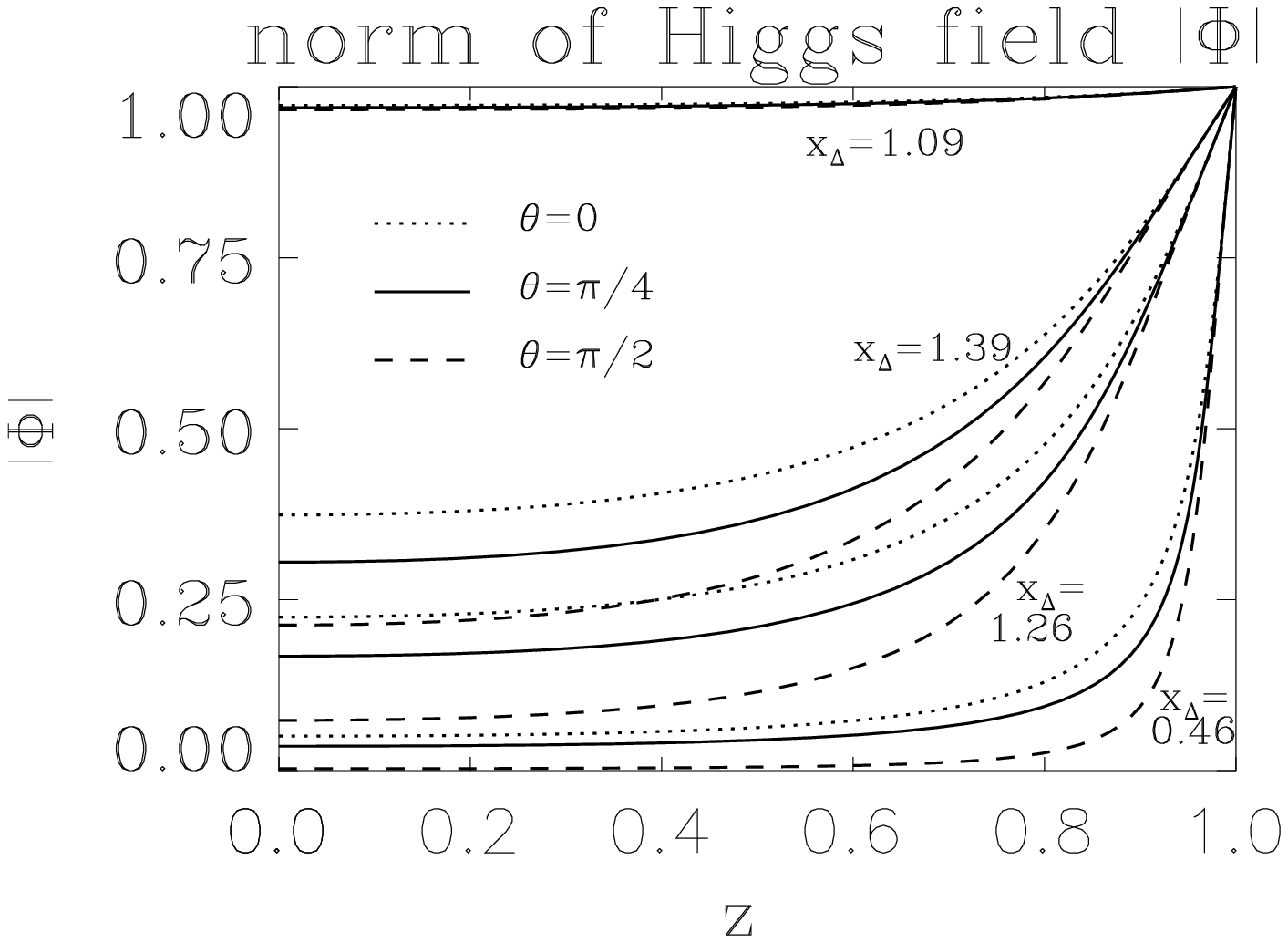}}
\caption{\label{Fig.18c} 
The same as Fig. 18a for the norm of the Higgs field
$\mid\Phi\mid=\sqrt{\Phi_{1}^{2}+\Phi_{2}^{2}}$. }
\end{figure}
\end{fixy}

\begin{fixy}{-1}
\begin{figure}
\centering
\epsfysize=10cm
\mbox{\epsffile{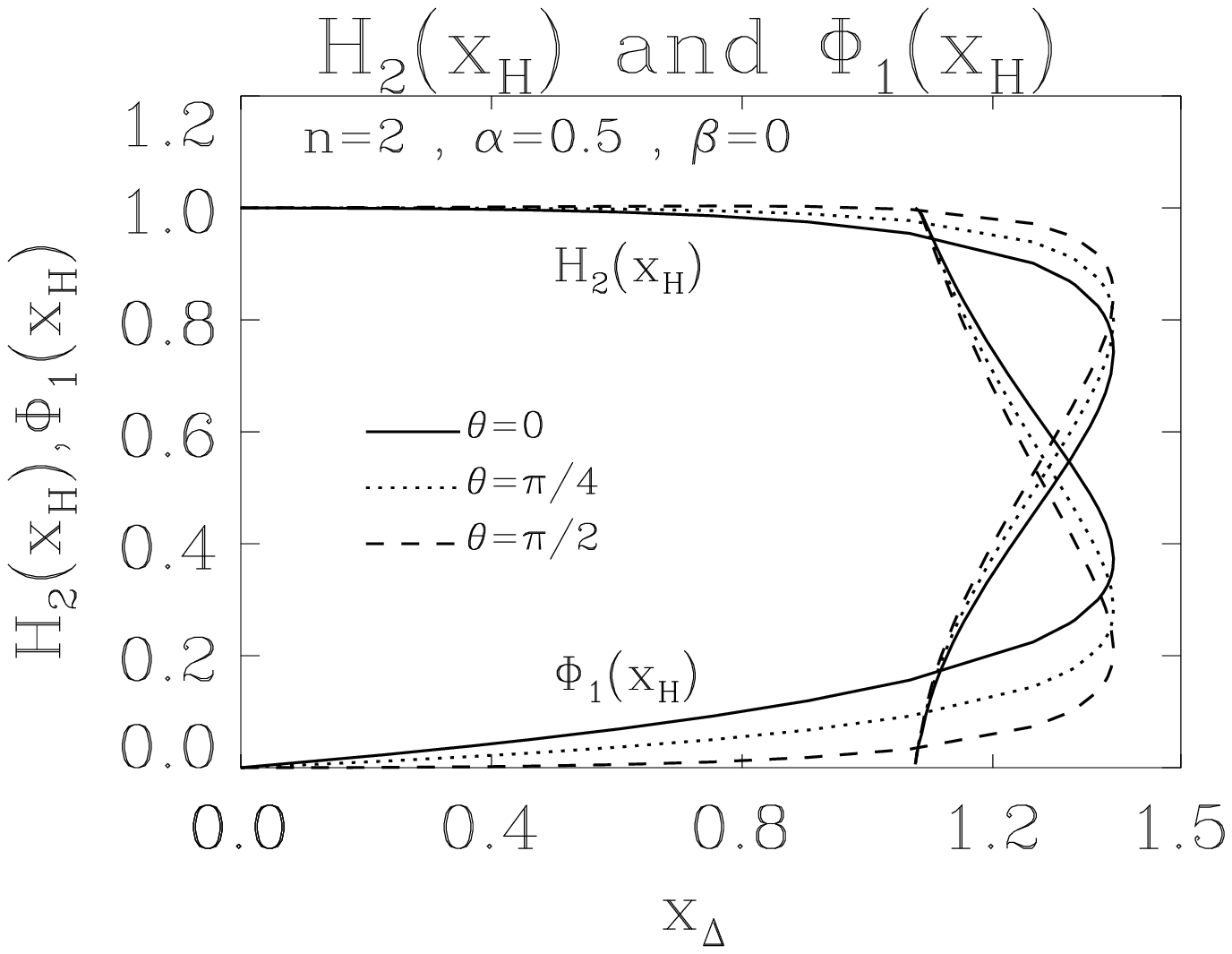}}
\caption{\label{Fig.19} 
The dependence of the horizon value of the gauge field function $H_2(x_{\rm H})$
and the horizon value of the Higgs field function $\Phi_1(x_{\rm H})$
of the $n=2$ hairy black hole solutions on the area parameter $x_{\Delta}$ 
is shown in the BPS limit for $\alpha=0.5$. }
\end{figure}
\end{fixy}

\begin{fixy}{-1}
\begin{figure}
\centering
\epsfysize=10cm
\mbox{\epsffile{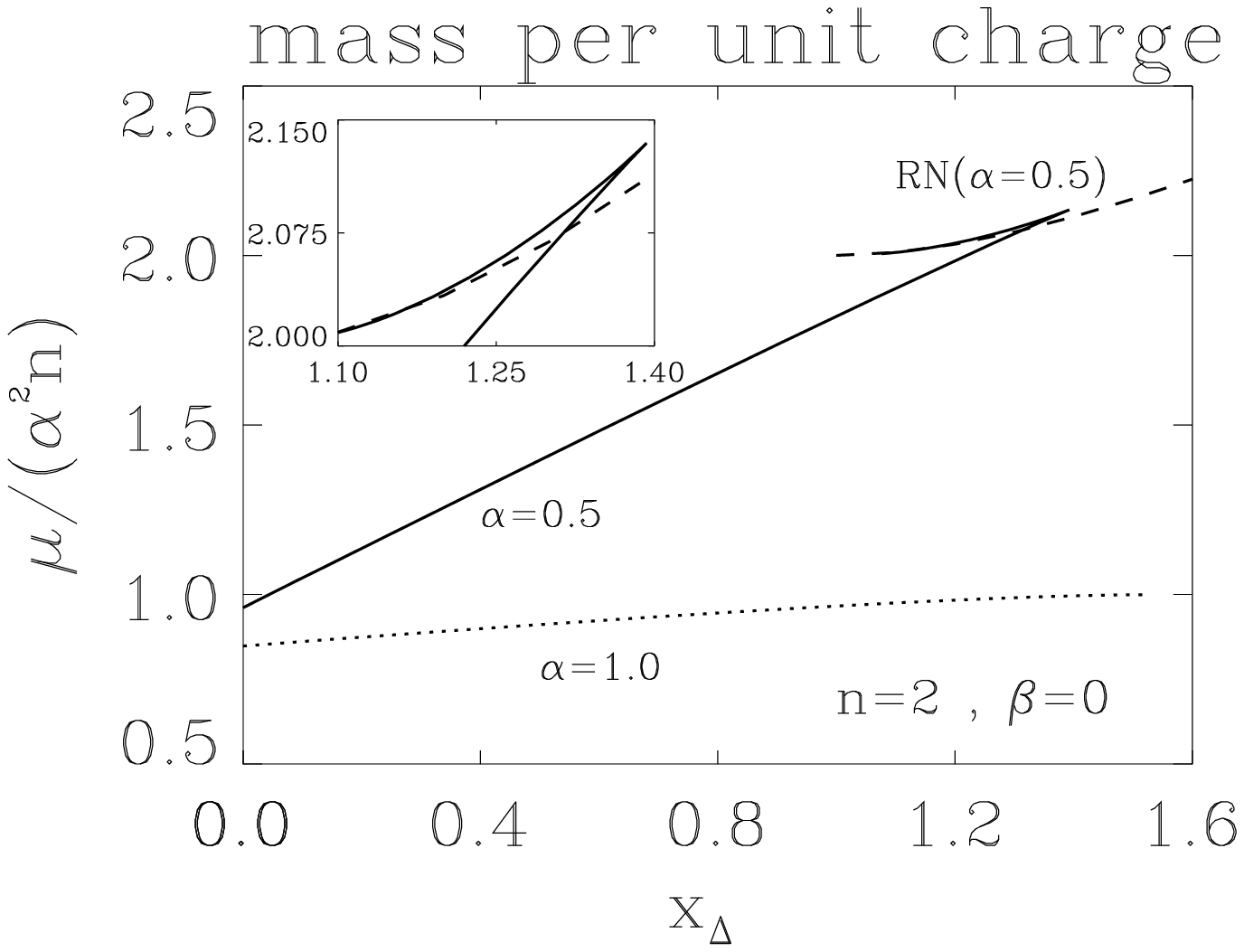}}
\caption{\label{Fig.20} 
The dependence of the mass {\it per unit charge} $\mu/(\alpha^2 n)$
of the $n=2$ hairy black hole solutions on the area parameter $x_{\Delta}$ 
is shown in the BPS limit for $\alpha=0.5$ and $\alpha=1$.
For comparison, also the mass {\it per unit charge} 
of the RN solutions for $\alpha=0.5$ is shown.
For $\alpha$=1.0, RN solutions with charge $P=2$
exist only for $x_{\Delta}\geq 2$. }
\end{figure}
\end{fixy}

\begin{fixy}{-1}
\begin{figure}
\centering
\epsfysize=10cm
\mbox{\epsffile{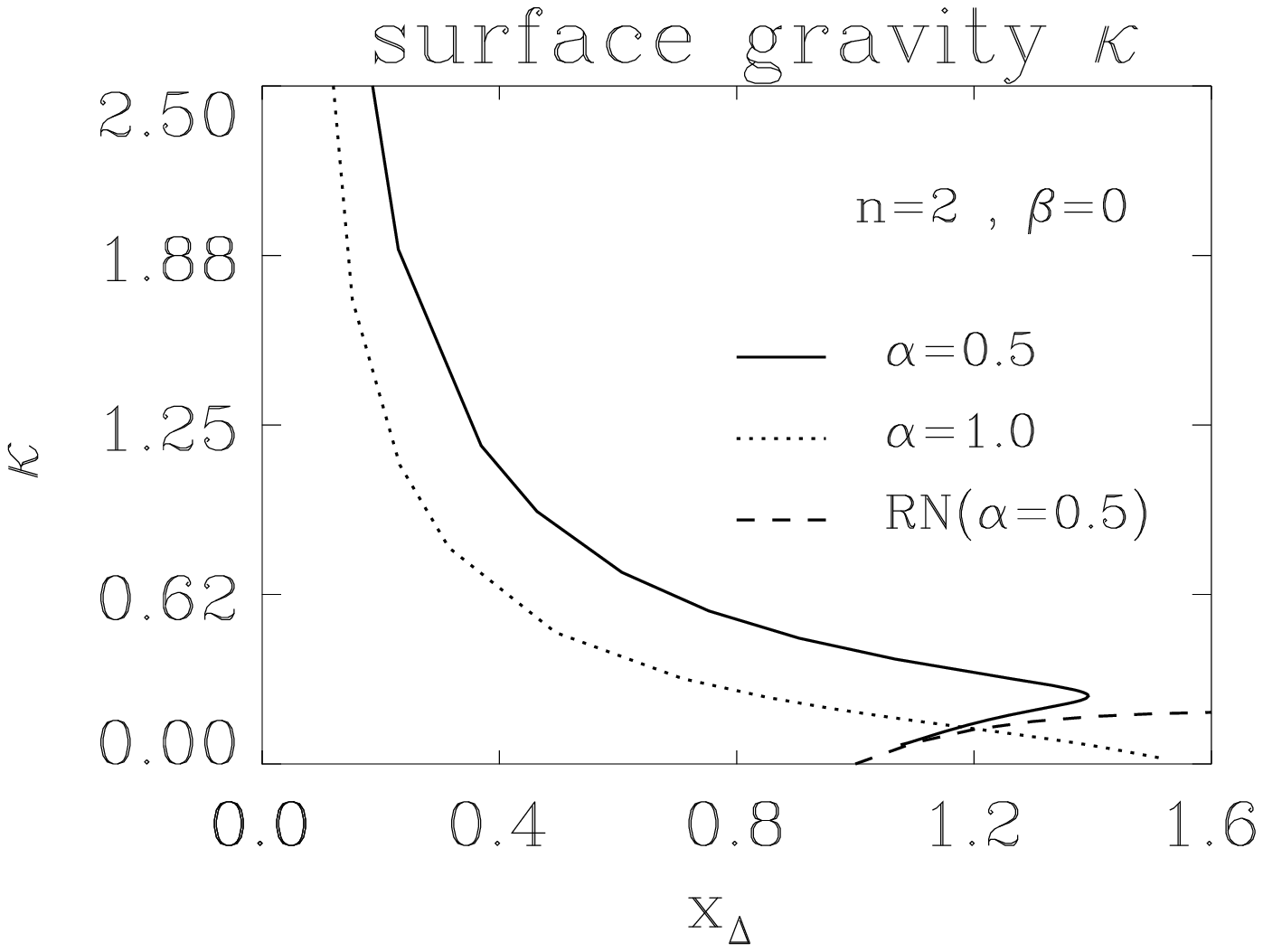}}
\caption{\label{Fig.21} 
The dependence of the surface gravity $\kappa$
of the $n=2$ hairy black hole solutions on the area parameter $x_{\Delta}$ 
is shown in the BPS limit for $\alpha=0.5$ and $\alpha=1$.
For comparison, also the surface gravity
of the RN solutions for $\alpha=0.5$ is shown. }
\end{figure}
\end{fixy}

\begin{fixy}{-1}
\begin{figure}
\centering
\epsfysize=10cm
\mbox{\epsffile{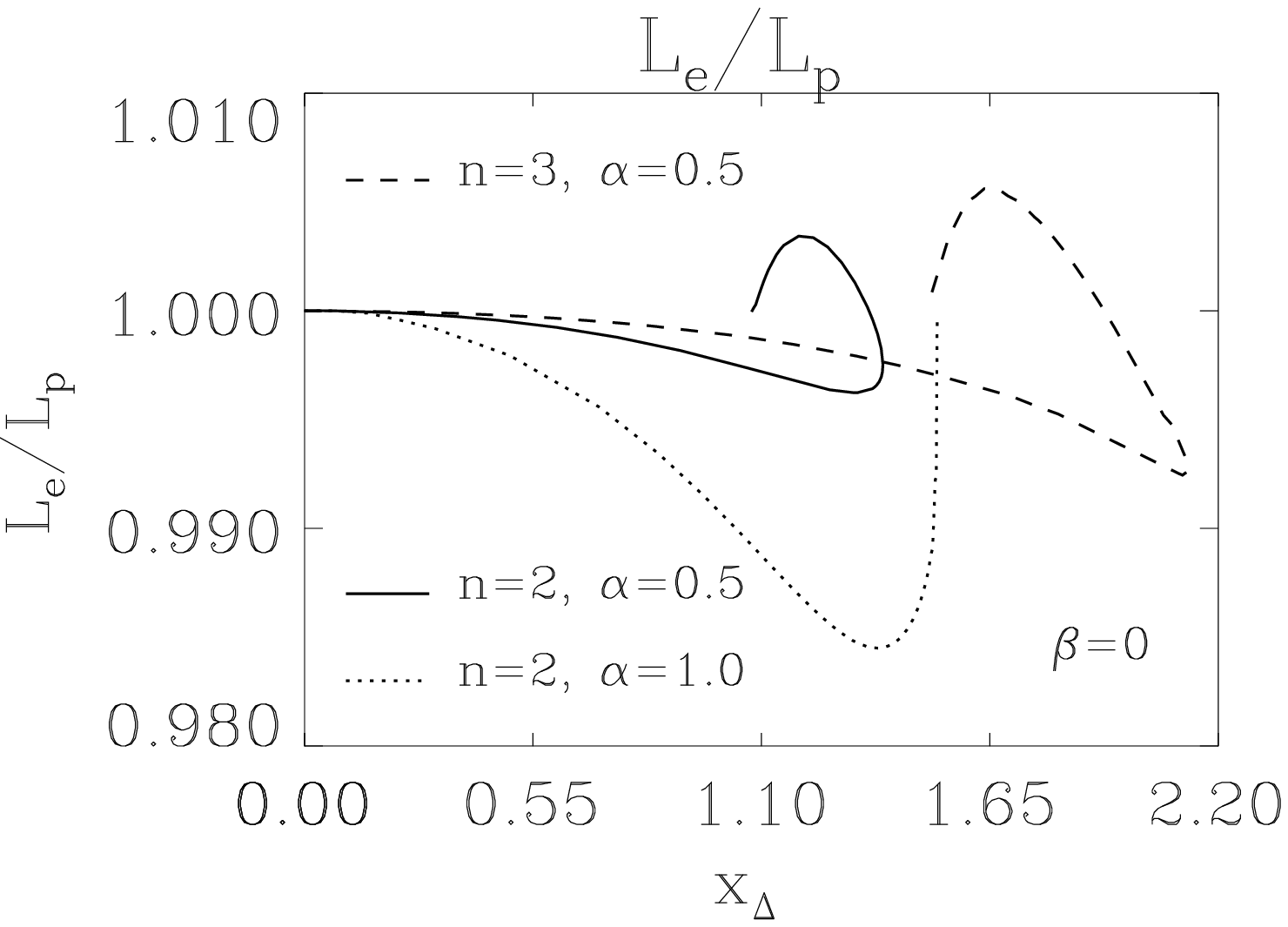}}
\caption{\label{Fig.22} 
The dependence of the ratio of circumferences $L_e/L_p$
of the $n=2$ hairy black hole solutions on the area parameter $x_{\Delta}$ 
is shown in the BPS limit for $\alpha=0.5$ and $\alpha=1$.
Also shown is the ratio of circumferences $L_e/L_p$
of the $n=3$ hairy black hole solutions
for $\alpha=0.5$. }
\end{figure}
\end{fixy}

\begin{fixy}{-1}
\begin{figure}
\centering
\epsfysize=10cm
\mbox{\epsffile{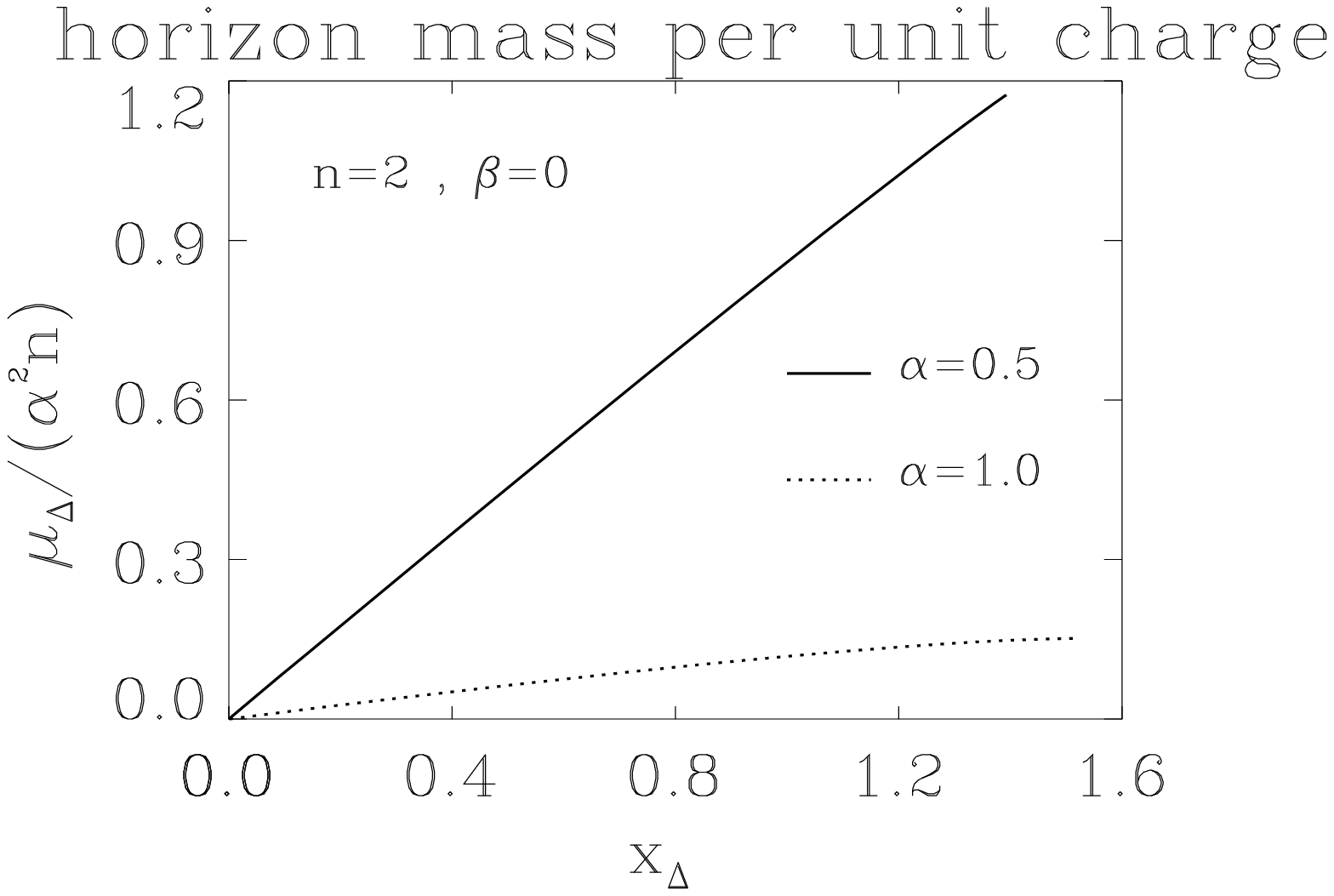}}
\caption{\label{Fig.23} 
The dependence of the horizon mass $\mu_{\Delta}/\alpha^2$
of the $n=2$ hairy black hole solutions on the area parameter $x_{\Delta}$ 
is shown in the BPS limit for $\alpha=0.5$ and $\alpha=1$. }
\end{figure}
\end{fixy}


\begin{fixy}{-1}
\begin{figure}
\centering
\epsfysize=10cm
\mbox{\epsffile{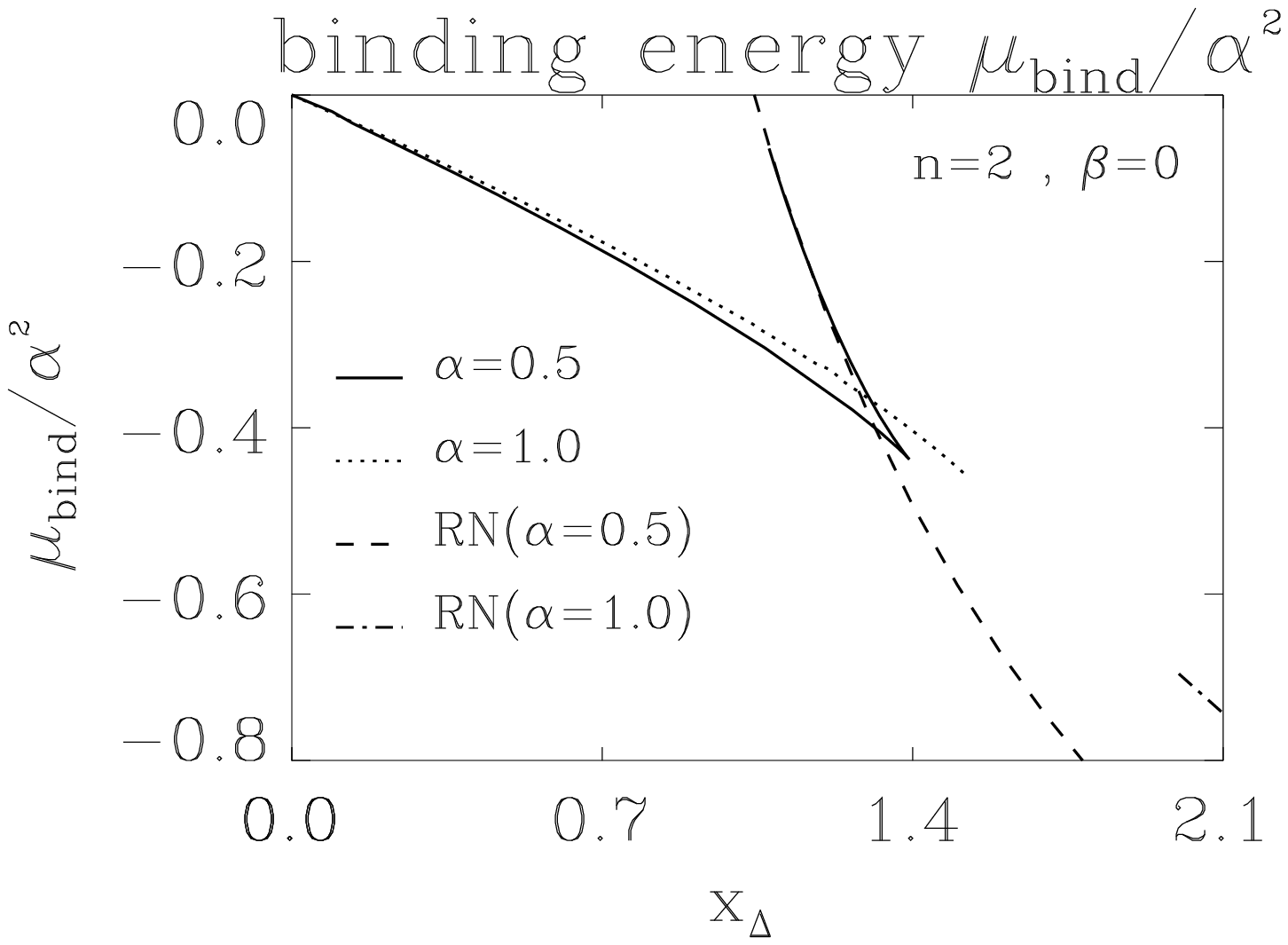}}
\caption{\label{Fig.24}
The dependence of the binding energy $\mu_{\rm bind}/\alpha^2$
of the $n=2$ hairy black hole solutions on the area parameter $x_{\Delta}$ 
is shown in the BPS limit for $\alpha=0.5$ and $\alpha=1$. 
For comparison, the binding energy
of the corresponding RN solutions is also shown. }
\end{figure}
\end{fixy}

\begin{fixy}{0}
\begin{figure}
\centering
\epsfysize=8cm
\mbox{\epsffile{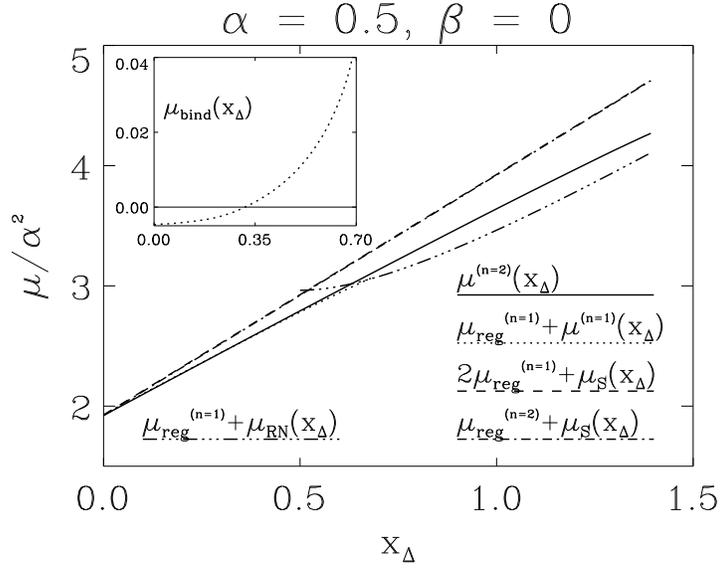}}
\caption{\label{Fig.25a}
The dependence of the mass of the compound system
of a $n=2$ multimonopole and a Schwarzschild black hole,
of two $n=1$ monopoles and a Schwarzschild black hole,
of a $n=1$ monopole and a $n=1$ hairy black hole,
and of a $n=1$ monopole and a RN black hole with unit charge,
on the area parameter $x_{\Delta}$ 
is shown in the BPS limit for $\alpha=0.5$.
For comparison the mass
of the $n=2$ hairy black hole solutions is also shown. 
The inlet shows the binding energy with respect to the compound system
of a $n=1$ monopole and a $n=1$ hairy black hole. }
\end{figure}

\begin{figure}
\centering
\epsfysize=10cm
\mbox{\epsffile{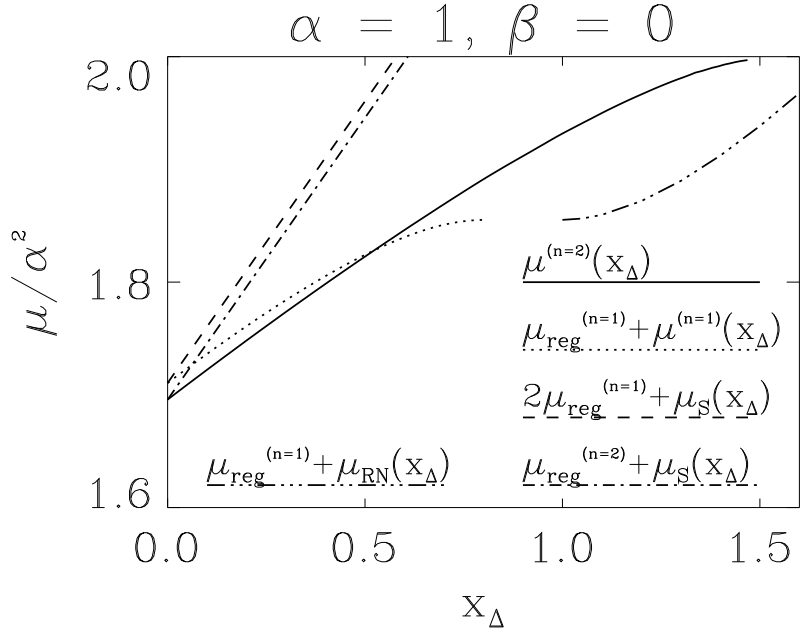}}
\caption{\label{Fig.25b}
The same as Fig. 25a for $\alpha=1$. }
\end{figure}
\end{fixy}

\begin{fixy}{-1}
\begin{figure}
\centering
\epsfysize=10cm
\mbox{\epsffile{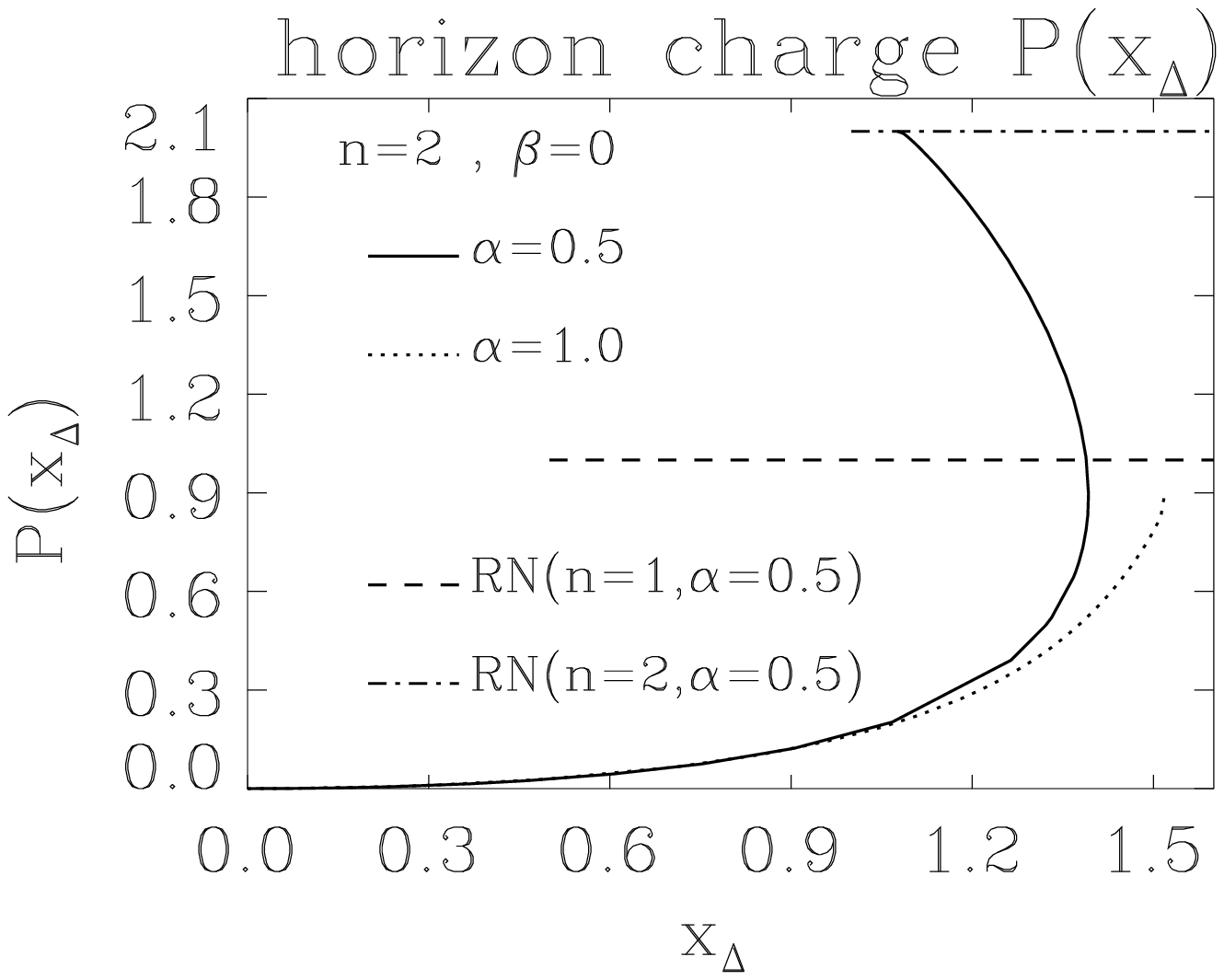}}
\caption{\label{Fig.26} 
The dependence of the non-abelian horizon magnetic charge $P(x_\Delta)$
of the $n=2$ hairy black hole solutions on the area parameter $x_{\Delta}$ 
is shown in the BPS limit for $\alpha=0.5$ and $\alpha=1$.
Also shown is the horizon magnetic charge
of the RN solutions with charge $n=1$ and $n=2$ for $\alpha=0.5$. }
\end{figure}
\end{fixy}


\begin{thebibliography}{000}
\bibitem{foot1}
 For $n=1$ monopoles for large Higgs self-coupling 
 the exterior solution is not of RN-type \cite{lw}.
\bibitem{foot2}
 In the BPS limit and for small values of $\beta$, 
 the branch of monopole solutions with unit charge
 bends backwards at $\alpha_{\rm max}$
 until a critical coupling constant $\alpha_{\rm cr}$ is reached.
 For the multimonopole solutions we do not observe such a
 backbending phenomenon. 
 Already in the BPS limit,
 near $\alpha=\alpha_{\rm max}$ the mass of the non-abelian solutions
 differs from the mass of the extremal RN solution by less than
 $0.01 \%$. Thus either the maximal value $\alpha_{\rm max}$
 coincides with the critical value $\alpha_{\rm cr}$, or both are 
 too close together to extend the branch of multimonopole solutions
 backwards numerically, from $\alpha_{\rm max}$ to $\alpha_{\rm cr}$.
\bibitem{foot3}
 The Lue-Weinberg phenomenon \cite{lw} might also occur for
 axially symmetric multimonopoles, when larger values of $\beta$
 are considered.
\end{thebibliography}
\end{document}